\documentclass[fleqn,usenatbib]{mnras}

\usepackage{newtxtext,newtxmath}

\usepackage[T1]{fontenc}
\usepackage{ae,aecompl}

\DeclareRobustCommand{\VAN}[3]{#2}
\let\VANthebibliography\thebibliography
\def\thebibliography{\DeclareRobustCommand{\VAN}[3]{##3}\VANthebibliography}

\usepackage{graphicx}	% Including figure files
\usepackage{amsmath}	% Advanced maths commands
\usepackage{amssymb}	% Extra maths symbols
\usepackage{longtable}
\usepackage{enumitem} % for listing
\usepackage{array}  % for table-colum width
\usepackage{amssymb}	% Extra maths symbols
\usepackage{pdflscape} % for landscape pages
\usepackage{multicol}  % Multi-column entries in tables
\usepackage{bm}		% Bold maths symbols, including upright Greek
\usepackage{afterpage}
\usepackage{caption}
\usepackage{footnote}
\usepackage{float}
\usepackage{pdflscape}
\usepackage{subcaption}
\usepackage{multirow}
\usepackage[nolist,nohyperlinks]{acronym}
\usepackage{chngcntr}
\usepackage{xcolor}
\usepackage{subcaption}
\usepackage{float}
\usepackage{appendix}
\usepackage{threeparttable}
\newcommand{\angstrom}{\textup{\AA}}

\DeclareUnicodeCharacter{2212}{-} 

\title[ Radio Sources At $z\gtrsim$4]{A Search for Missing Radio Sources At $z\gtrsim$4 Using Lyman Dropouts}
\author[D.Shobhana et al.]{
Devika Shobhana,$^{1}$\thanks{E-mail:devikashobhana@gmail.com}
Ray P. Norris,$^{1,2}$\thanks{raypnorris@gmail.com}      
Miroslav D. Filipovi\'c,$^{1}$ 
Luke A. Barnes,$^{1}$ 
Andrew M. Hopkins, $^3$ 
\newauthor
Isabella Prandoni,$^4$
Michael J. I. Brown,$^5$
Stanislav S. Shabala$^6$
\\ \\
$^{1}$Western Sydney University, Locked Bag 1797, Penrith South DC, NSW 2751, Australia \\
$^{2}$CSIRO Space and Astronomy, Australia Telescope National Facility, PO Box 76, Epping, NSW 1710, Australia \\
$^{3}$Australian Astronomical Optics, Macquarie University, 105 Delhi Rd, North Ryde, NSW 2113, Australia \\
$^{4}$ INAF – Istituto di Radioastronomia, via P. Gobetti 101, 40129 Bologna, Italy\\
$^{5}$ School of Physics and Astronomy, Monash University, Clayton, VIC 3800, Australia\\
$^{6}$ School of Natural Sciences, University of Tasmania, Private Bag 37, Hobart, TAS 7001, Australia\\
}

\date{Accepted XXX. Received YYY; in original form ZZZ}

\pubyear{2022}

\begin{acronym}[AWGN]
\acro{2MASS}{Two Micron All Sky Survey}
\acro{2MASX}{Two Micron All Sky Survey Extended Source Catalogue}
\acro{30Dor}[30 Dor]{30~Doradus}
\acro{AGN}{active galactic nuclei}
\acro{ATCA}{Australia Telescope Compact Array}
\acro{ATESP}{Australia Telescope ESO Slice Project}
\acro{ATNF}{Australia Telescope National Facility}
\acro{ATOA}{Australia Telescope Online Archive}
\acro{AT20G}{Australia Telescope 20 GHz Survey}
\acro{ASKAP}{Australian Square Kilometre Array Pathfinder}
\acro{BETA}{Boolardy Engineering Test Array}
\acro{BL}[BL Lac]{BL Lacertae Objects}
\acro{CASS}{CSIRO Astronomy and Space Science}
\acro{CABB}{Compact Array Broadband Back-end}
\acro{CHII}[{\sc CHii}]{Compact \textsc{Hii}}
\acro{CMB}{Cosmic Microwave Background}
\acro{CSIRO}{Australian Commonwealth Scientific and Industrial Research Organisation}
\acro{CSS}{Compact Steep Spectrum}
\acro{DS9}[\textsc{DS9}]{\textsc{SAOImage DS9}}
\acro{DSS}{Digital Sky Survey}
\acro{EM}{Electromagnetic}
\acro{EMU}{Evolutionary Map of the Universe}
\acro{ev}[eV]{electronvolt\acroextra{: 1 eV $\approx 1.6 \times 10^{-19}$ J}}
\acro{FIRST}{Faint Images of the Radio Sky at Twenty cm}
\acro{FITS}[\textsc{Fits}]{Flexible Image Transport System}
\acro{FRB}{Fast Radio Bursts}
\acro{FSRQ}{Flat Spectrum Radio Quasars}
\acro{FWHM}{Full Width at Half-Maximum}
\acro{FR}{Fanaroff-Riley}
\acro{FRI}{Fanaroff-Riley I}
\acrodefplural{FRI}{Fanaroff-Riley I's}
\acro{FRII}{Fanaroff-Riley II}
\acrodefplural{FRII}{Fanaroff-Riley II's}
\acro{GAMA}{Galaxy And Mass Assembly}
\acro{GLEAM}{GaLactic Extragalactic All-sky MWA}
\acro{GPS}{Gigahertz Peak Spectrum}
\acro{HEMT}{High Electron Mobility Transistor}
\acro{HFP}[HFP]{High Frequency Peaker}
\acro{HPBW}{Half Power Beam Width} 
\acro{HzRG}{High Redshift Radio Galaxies}

\acro{HzRS}{High redshift radio source}
\acrodefplural{HzRS}{High redshift radio sources}
\acro{pHFP}[pHFP]{Potential High Frequency Peaker}
\acro{HI}[H{\sc i}]{Neutral Atomic Hydrogen} 
\acro{HST}{\textit{Hubble Space Telescope}}
\acro{IAU}{International Astronomical Union}
\acro{IFRSs}{Infrared Faint Radio Sources}
\acro{IFRS}{Infrared Faint Radio Source}
\acro{IR}{Infrared}
\acro{IRSA}{Infrared Science Archive}
\acro{JY}[Jy]{Jansky\acroextra{, 1 Jy = $10^{-26} \times \mathrm{W~ m}^{-2}~\mathrm{Hz}^{-1}$}} 
\acro{LLS}{Largest Linear Size}
\acro{LFAA}{Low-Frequency Aperture Array}
\acro{LMC}{Large Magellanic Cloud}
\acro{LSO}[LSOs]{Large Scale Objects}
\acro{MACHO}{Massive Astrophysical Compact Halo Objects}
\acro{MC}[MCs]{Magellanic Clouds}
\acro{mc}[MC]{Magellanic Cloud} 
\acro{MCELS}{Magellanic Cloud Emission Line Survey}
\acro{MW}{Milky Way}
\acro{MIRIAD}[\textsc{Miriad}]{Multichannel Image Reconstruction, Image Analysis and Display}
\acro{MIT}{Massachusetts Institute of Technology}
\acro{MOST}{Molonglo Observatory Synthesis Telescope}
\acro{MQS}{Magellanic Quasars Survey}
\acro{MRC}{Molonglo Reference Catalogue of Radio Sources}
\acro{MWA}{Murchison Widefield Array}
\acro{NRAO}{National Radio Astronomy Observatory}
\acro{NVSS}{NRAO VLA Sky Survey}
\acro{OPAL}{Online Proposal Applications \& Links}
\acro{OVV}{Optically Violent Variable Quasars}   
\acro{PAF}{Phased Array Feed}
\acro{pc}{parsec\acroextra{: 1 pc $\simeq 3.09 \times 10^{16}$ m}}
\acro{PMN}{Parkes-MIT-NRAO}
\acro{PNe}{Planetary Nebulae}
\acro{QSO}{Quasi-Stellar Object}
\acro{RA}{Right Ascension}
\acro{RFI}{Radio-Frequency Interference}
\acro{RMS}[rms]{Root Mean Squared}
\acro{SB}{Starburst}
\acro{SDSS}{Sloan Digital Sky Survey}
\acro{SED}{spectral energy distribution}
\acro{SI}[$\alpha$]{Spectral Index\acroextra{, $S \propto \nu^\alpha$}}
\acro{SKA}{Square Kilometre Array}
\acro{SMBH}{Super Massive Blackhole}
\acro{SMC}{Small Magellanic Cloud}
\acro{SMG}{Submillimeter galaxies}
\acro{SN}{Supernova}
\acro{SNR}{Supernova Remnant}
\acro{SNRs}{Supernova Remnants}
\acro{SF}{Star Formation}
\acro{SFG}{Star-forming Galaxies}
\acro{SFR}{star formation rate}
\acro{SUMSS}{Sydney University Molonglo Sky Survey}
\acro{TOPCAT}[\textsc{Topcat}]{Tool for OPerations on Catalogues And Tables}
\acro{ULIRG}{Ultra Luminous Infrared Galaxy}
\acro{USS}{Ultra Steep Spectrum}
\acro{WBAC}{Wide-Band Analogue Correlator}
\acro{WIFES}[WiFeS]{Wide-Field Spectrograph}
\acro{WISE}{Wide-Field Infrared Survey Explorer}
\acro{VLBI}{Very Long Baseline Interferometry} 
\acro{VLSR}[\textbf{$v_{lsr}$}]{Velocity in the Line of Sight}
\end{acronym}

\begin{document}
\label{firstpage}
\pagerange{\pageref{firstpage}--\pageref{lastpage}}
\maketitle

\begin{abstract}

Using the Lyman Dropout technique, we identify  148 candidate radio sources at $z \gtrsim 4 - 7$ from the 887.5\,MHz \ac{ASKAP} observations of the GAMA23 field.
 About 112 radio sources are currently known beyond redshift $z\sim4$. However, simulations predict that hundreds of thousands of radio sources exist in that redshift range, many of which are probably in existing radio catalogues but do not have measured redshifts, either because their optical emission is too faint or because of the lack of techniques that can identify candidate high-redshift radio sources (HzRSs). Our study addresses these issues  using the Lyman Dropout search technique. This newly built sample probes radio luminosities that are 1-2 orders of magnitude fainter than known radio-\ac{AGN} at similar redshifts, thanks to \ac{ASKAP}'s sensitivity. We investigate the physical origin of radio emission in our sample using a set of diagnostics: (i) radio luminosity at 1.4\,GHz, (ii)  1.4\,GHz-to-3.4\,$\mu$m flux density ratio, (iii) Far-IR detection, (iv) WISE colour, and (v) SED modelling. 
 The radio/IR analysis has shown that the majority of radio emission in the  faint and bright end of our sample's 887.5\,MHz flux density distribution originates from 
\ac{AGN} activity. Furthermore, $\sim10\%$ of our sample are found to have a 250\,$\mu$m detection, suggesting a composite system. This suggests that some
high-$z$ radio-\acp{AGN} are hosted by SB galaxies, in contrast to low-$z$ radio-\acp{AGN}, which are usually hosted by quiescent elliptical galaxies. 
\end{abstract}

%%%%%%%%%%%%%%%%%%%%%%%%%%%%%%%%%%%%%%%%%%%%%%%%%%

%%%%%%%%%%%%%%%%% BODY OF PAPER %%%%%%%%%%%%%%%%%%

\section{Introduction}
\label{sec:intro}

The discovery of quasars beyond $z\sim7$ (eg: \cite{banados2018800,mortlock2011luminous}) poses a crucial question: which cosmic era marks the birth of radio-loud \ac{AGN}? 
Radio loud \ac{AGN}s are amongst the most radio-luminous sources at all cosmic epochs. Their large radio luminosity is attributed to a radio-jet launched by an accreting \ac{SMBH} at their centre. So far, radio-\ac{AGN}s are only known upto $z\sim6$, and the most distant radio quasar is at $z=6.82$ \citep{banados2021discovery}, while the  highest redshift radio galaxy lies at $z=5.72$ \citep{saxena2018discovery} and the highest-redshift blazar is at $z=6.1$ \citep{belladitta2020first}. Of $\sim$200 quasars discovered beyond $z\sim6$, only five are found to be radio sources.

\ac{HzRG}s have been targeted either by looking for ultra steep ($\alpha < -1.3$) radio spectra \citep[USS, ][]{de2000sample}, or by selecting sources with a very faint K-band (2.2\,$\mu$m) counterpart \citep{jarvis2009discovery}. The former technique is the most widely used and is based on the observed steepening of the radio spectrum with redshift. This method has been used to discover almost all known \ac{HzRG}s, including the most distant known source at $z = 5.72$ \citep{saxena2018discovery}. However, some studies \citep{yamashita2020wide,jarvis2009discovery,miley2008distant} have reported the discovery of HzRGs with non-steep spectral index at $z \ge 4$, showing that the USS selection method does not give a complete view of high-redshift radio-\ac{AGN}s.

Simulations of forthcoming radio surveys estimated the source count of radio emitters as a function of redshift \citep{bonaldi2019tiered,wilman2008semi} and it predicts hundreds of thousands of radio sources beyond $z \sim 4$. Although radio emitters such as radio-loud \ac{AGN}s, Starbursts, \ac{SFG}s and radio-quiet \ac{AGN}s contribute to this total radio-source count, simulations demonstrate that $z\ge4$ sky is dominated by radio-AGNs (see \cite[Figure~2]{raccanelli2012cosmological}).
Currently, $\sim$112 published radio-\acp{AGN} are known at redshift $z \gtrsim 4$ (listed in Appendix, Table~\ref{tab:known_radio_AGN}). This number thus indicates that the known radio source population at $z\gtrsim4$ represents a small fraction of the total radio source population.

In some cases, it is unclear whether a  detected high-redshift radio source is a radio galaxy, blazar, quasar, etc., and so we adopt the neutral term ``\ac{HzRS}'' to describe any radio source detected at high redshift ($z\ge4$) .

The mismatch between models and data indicates that known \acp{HzRS} are only the tip of the iceberg. The dearth of radio sources at high redshift can be attributed to the following factors: 
(i) many \acp{HzRS} are probably in existing radio catalogues but their redshifts have not been measured  due to 
%
%ray lack of techniques
their faintness at optical/IR wavelengths, 
and (ii) previous radio surveys were not sensitive enough to detect faint \acp{HzRS}
%, and (iii) selection effects.

 Since the ultimate goal of this series of papers is to establish the \ac{HzRS} count and thus %
%verify the model
test the simulations
\citep{bonaldi2019tiered,wilman2008semi}, we expect that some missing \acp{HzRS} are already in the literature, but are not classified as high-redshift radio sources.  
%To ensure that the census of known radio sources at $z\gtrsim4$ is complete, 
 We demonstrate this by visually cross-matching \ac{SDSS} spectroscopy \citep[DR12,][]{sdss} with the \ac{FIRST} \citep{becker95} and \ac{NVSS} \citep{condon98} catalogues. This search resulted in a further 33 sources at $z \gtrsim 4$, listed in Appendix, Table \ref{tab:new_radio_AGN}. In each case the \ac{SDSS} spectrum has been checked for supporting evidence of the redshift, such as a Lyman break or other spectral features. We note that a further list of candidates is available in the MILLIQUAS \citep{milliquas} catalogue\footnote{\url{https://heasarc.gsfc.nasa.gov/W3Browse/all/milliquas.html}}, but to the best of our knowledge the spectroscopy has not been checked and so that list may include some spurious candidates.

\ac{EMU} is one of the deepest and the largest forthcoming radio continuum surveys \citep{norris2011emu}, to be delivered by the \ac{ASKAP} telescope \citep{johnston2007science}. The \ac{EMU} project started with a series of Early Science observations, followed by the \ac{EMU} Pilot Survey \citep[PS;][]{norris21}. In this paper, we use the \ac{EMU} Early Science Observations of the GAMA23 field (hereafter referred to as ``G23''), which is one of the Galaxy And Mass Assembly (GAMA) survey fields \citep{driver2008galaxy}. 

Motivated by the challenge of finding missing \acp{HzRS}, we make use of a search technique different from conventional radio based techniques, the Lyman Dropout technique  ( a.k.a. Lyman Break Galaxy technique), to identify potential \acp{HzRS} at $ z \gtrsim 4-7$ in the G23 field. The Lyman dropout technique has been a popular technique in optical astronomy over the past two decades for discovering high-redshift galaxies up to  $z\sim11$. However, only one radio galaxy at $z=4.72$ has been identified using the Lyman dropout technique to date \citep{yamashita2020wide}.
Therefore, the primary goal of this study is to test the efficiency of Lyman dropout technique in finding \acp{HzRS}. 
A second goal is to determine the properties of our sample of \acp{HzRS}, 
 a detailed study of which will be discussed in a future paper. 

The Lyman dropout technique looks for the redshifted spectral signature of the \emph{Lyman limit} at 91.2\,nm (Far-UV regime). This is the longest wavelength of light  that can ionise a ground-state hydrogen atom. Light at wavelengths shorter than 91.2\,nm (ie. at higher energies) will be absorbed by sufficiently optically-thick atomic hydrogen present in the galaxy or its circumgalactic medium. This missing radiation creates a break in the observed spectrum. For high-$z$ galaxies, the Lyman break gets redshifted into the optical region, and can be identified using images taken in multiple filters.

The structure of this paper is as follows. In Section~\ref{sec:data} we describe how we select \acp{HzRS} in GAMA23 field from the \ac{ASKAP} 887.5\,MHz radio catalogue using 8-band \textit{ugriZYJK\textsubscript{s}} KiDS/VIKING photometry. In Section~\ref{sec:result}, we present our sample of \ac{HzRS} candidates selected at $z \sim 4$, 5, 6, and 7. In Section~\ref{sec:discuss} we present the analysis of radio and IR properties of our sample. Finally, we summarise our results in Section~\ref{sec:concl}.

This study adopts a $\Lambda$CDM cosmology with $\Omega\textsubscript{m} = 0.3$, $\Omega\textsubscript{$\Lambda$} = 0.7$ and $H_0 = 70~$km~s$^{-1}$Mpc$^{-1}$.

%================================================================================
\section{Data and Methods}
\label{sec:data}

To find \acp{HzRS}, we cross-match the G23 radio observations with the Kilo Degree Survey optical catalogue \citep[KiDS;][]{kuijken2019fourth} and the VIKING DR5 \& CATWISE2020 \citep{marocco2021catwise2020} infrared catalogues. This results in a sample of G23 radio sources with optical and infrared photometry. We then apply the redshift specific Lyman dropout colour cuts \citep{ono2018great,venemans2013discovery} to select the radio source candidates at $z \gtrsim 4$ in the G23 field. This paper is the first in a series describing our search for \acp{HzRS} using the Lyman dropout technique as part of the \ac{EMU} survey.

We use the 887.5\,MHz radio continuum data of the G23 field, produced by \ac{ASKAP} as part of the \ac{EMU} Early Science program in early 2019. \ac{ASKAP} consists of 36 antennas, each of which is equipped with a Phased Array Feed (PAF). It operates in a frequency range from 700 to 1800\,MHz. \ac{ASKAP} data products have been created using the ASKAPsoft pipeline, aided by Selavy software in source extraction. This study examined the following \ac{ASKAP} catalogues from project AS034: (i) selavy-image.i.SB8132.cont.taylor.0.restored.components and (ii) selavy-image.i.SB8137.cont.taylor.0.restored.components, retrieved from the CSIRO Data Access portal
\footnote{\url{https://data.csiro.au/domain/casdaObservation}}.  The observational parameters of G23-ASKAP data are given in Table~\ref{tab:survey_info}.

 A total of 38\,080 radio sources are present in these 2 catalogues, of which 2107 are complex or multi-component (number of components $\ge 2$). In this paper, we focus on simple (or single component) radio sources only, which are fitted by a single Gaussian. 

  For the selection of $z\ge4-6$ radio sources, we used optical data from the complementary Kilo Degree Survey \citep[KiDS,][]{kuijken2019fourth}, in particular we exploited the KiDS DR4.1 multiband source catalogue, featuring Gaussian Aperture and PSF photometry ( GAaP ; see \cite{kuijken2015gravitational} for details) measurements of KiDS-$ugri$ and VIKING-ZYJHK bands for $r$-band detected sources.

To select $z\sim7$ radio sources, we utilized VIKING photometry in the DR5 catalog obtained from the VISTA archive. \footnote{\url{http://horus.roe.ac.uk/vsa/index.html}} 
We converted the Vega magnitudes in the VIKING DR5 catalog to AB magnitudes using the Cambridge Astronomical Survey Unit (CASU) recommendations \footnote{\url{http://casu.ast.cam.ac.uk/surveys-projects/vista/technical/filter-set}}.

The mid-IR (MIR) data used in this paper comes from the Wide-field Infrared Survey Explorer \cite[WISE,][]{wright2010wide}, which is an all-sky survey centred at 3.4, 4.6, 12, and 22\,$\mu$m (referred to as bands W1, W2, W3 and W4), with an angular resolution of 6.1, 6.4, 6.5, and 12.0~arcsec respectively, and typical 5$\sigma$ sensitivity levels of 0.08, 0.11, 1, and 6~mJy/beam. Here, we use data from the CATWISE2020 \citep{marocco2021catwise2020} catalogue.

Far-IR (FIR) observations in the G23 field come from the {\it Herschel} space observatory.  {\it Herschel} carried out observations using two photometric instruments on board, (i) Photodetecting Array Camera and Spectrometer (PACS, \cite{poglitsch2010photodetector} ) and (ii) Spectral and Photometric Imaging Receiver (SPIRE, \cite{griffin2010herschel} ). PACS observations centred at 70\,$\mu$m, 100\,$\mu$m, and 160\,$\mu$m mainly trace the rest-frame mid-IR emission of the high-$z$ ($z>2$) \ac{AGN}. SPIRE observed simultaneously in three wavebands centred at 250\,$\mu$m, 350\,$\mu$m, and 500\,$\mu$m, picking up the starburst emission in high-$z$ \ac{AGN}s \citep{hatziminaoglou2010hermes}. {\it Herschel} ceased operation on 29$^{\rm th}$~April~2013 when the telescope ran out of liquid helium, which is essential for cooling the instruments. This study utilized the SPIRE \citep{schulz2017spire} and PACS point source catalogues \citep{marton2017herschel} available in \ac{IRSA}\footnote{\url{https://irsa.ipac.caltech.edu/}}.

\begin{table}
    \centering
      \caption{ Summary of G23-ASKAP survey.}
    \begin{tabular}{l|c}
      Parameters   & G23-ASKAP \\
         \hline
       Frequency  (MHz)         & 887.5     \\
       Bandwidth  (MHz)         & 288       \\
       Synth. beam size ($\arcsec$) &  10         \\
       RMS ($\mu$Jy/beam)   &    38           \\
       Survey area (deg\textsuperscript{2})  & 50      \\
%      Beam size ($\arcsec$) &              \\
       Astrometric accuracy ($\arcsec$)&      $\sim$ 1    \\
%       accuracy    &                     \\
      \hline
    \end{tabular}
    \label{tab:survey_info}
\end{table}
%The {\it Herschel} space observatory performed an extragalactic sky survey covering northern and southern galactic caps between 70\,$\mu$m and 500\,$\mu$m bands using the (shorter wavelength) PACS and (longer wavelength) SPIRE instruments \citep{poglitsch2010photodetector,griffin2010herschel}.

%%%%%%%%%%%%%%%%%%%%%%%%%%%%%%%%%%%%%%%%%%%%%%%%%%%%%%%%%%%%%%%%%%%%%%%%%%
\subsection{Finding optical and infrared counterparts}
\label{sec:optical_crossmatch}

To find the optical counterparts of radio sources, we use a simple nearest-neighbour technique. We need to choose a search radius that maximises the number of cross-matches while minimising the number of false identifications (hereafter called false-IDs). We achieve this by cross-correlating the \ac{ASKAP} radio catalogue with the KiDS DR4.1 multiband optical (\textit{ugri}) $+$ NIR (ZYJHK\textsubscript{s}) photometry at a range of search radii, measuring the number of cross-matches at each radius. 
We then estimate the false-ID rate by shifting the radio position by 1~arcmin (so that all matches are spurious) and repeating the cross-match at the same set of radii. The false-ID rate is calculated by dividing the number of shifted cross-matches by the number of unshifted cross-matches. The result is shown in Table~\ref{tab:false_id}. Based on this, we have chosen 2~arcsec as the optimal search radius for this study, corresponding to a false-ID rate of 14.64\% and a total cross-match rate of 63.9\%. We reject sources that had multiple matches within 2~arcsec. This reduces the final number of radio-optical cross-matches to 17\,447.

We followed the same procedure to select infrared counterparts to our radio sources. We cross-matched the optical (KiDS) positions of our sample with the CATWISE2020 catalog at a search radius of 2$\arcsec$. This gives a false id rate of 11.9\%.

%%%%%%%%% TABLE 1 %%%%%%%%%%%
\begin{table}
 \centering
 \caption{False-ID rate estimated for \ac{ASKAP}-KiDS cross-match as a function of separation radius, using the single-component source list.}
 \label{tab:false_id}
 \begin{tabular}{c c c c}
 \hline
  Match Radius & No. of Matches & No. of Matches & False-ID Rate \\
  (arcsec) & (unshifted) & (1~arcmin offset) & (\%) \\\hline
  1 & 14\,067 & 833 & 5.92 \\
  2 & 23\,002 & 3\,368 & 14.64 \\
  3 & 29\,350 & 7\,592 & 25.87 \\
  4 & 36\,417 & 13\,410 & 36.82\\
  5 & 44\,884 & 20\,821 & 46.39 \\
  \hline
 \end{tabular}
\end{table}

%%%%%%%%%%%%%%%%%%%%%%%%%%%%%%%%%%%%%%%%%%%%%%%%%%%%%%%%%%%%%%%%%%%%%%%%%%%%%%
\subsection{Selection of radio sources at $z\gtrsim$ 4 -- 6}

 The Lyman dropout technique relies on finding the wavelength or passband at which the Lyman break is detected, which in turn tells us the redshift  of the source, given that rest-frame wavelength of Lyman limit is 91.2\,nm. For example, the Lyman break of a galaxy at $z=4$ will be observed at wavelength 4560\,\AA\,  and hence can be imaged in g-band. Similarly, for higher redshift objects, the Lyman break moves into the $r$ or $i$ or $Z$ bands.
%****************************************************************
\subsubsection{Applying  \textit{g, r, \& i}-band  dropout technique}
\label{sec:initial_sample}

\begin{table}
\caption{KiDS \& VIKING filters, their central wavelength, and their mean 5$\sigma$ limiting magnitude. }
 \centering
 \begin{tabular}{c c c}
 \hline
  Filters & $\lambda$ & 5$\sigma$ Mag. Lim. \\
    & (\AA)    & (AB) \\
    \hline
  $u$ & 3\,550  & 24.23 \\
  $g$ & 4\,775 & 25.12\\
  $r$ & 6\,230  & 25.02 \\
  $i$ & 7\,630  & 23.68 \\
  Z & 8\,770  &23.1 \\
  Y & 10\,200 & 22.3 \\
  J & 12\,520  & 22.1 \\
  H & 16\,450  & 21.5\\
K\textsubscript{s} & 21\,470  & 21.2 \\
\hline
 \end{tabular}
 \label{tab:kiDS_filters}
\end{table}

\begin{table*}
 \begin{threeparttable}
 \caption{
 The criteria used to identify Lyman dropouts at $z\sim$ 4, 5, 6, adopted from \protect \citet{ono2018great} and at $z\sim$ 7, taken from \protect \cite{venemans2013discovery}. The last row is based on our COSMOS tests, to remove low-z interlopers in $z\sim$ 4, 5, 6 sample, as described in subsection \protect \ref{sec:interloper1}.}

 \begin{tabular}{p{2.9cm} | p{2.9cm} | p{2.9cm} | p{2.9cm} |p{3.9cm}}
 \hline
  \multirow{2}{*}{$z\sim4$ (\textit{g} dropouts)} & \multicolumn{2}{c}{$z\sim5$ (\textit{r} dropouts)} & \multirow{2}{*}{$z\sim6$ (\textit{i} dropouts)} &  \multirow{2}{*}{$z\sim7$ (Z dropouts)} \\
  \cline{2-3}
  &  criteria I \tnote{a} &  criteria II \tnote{b} & & \\
  \hline\hline
   S/N(i) $>$5 & S/N(z) $>$5 & S/N(z) $>$5 & S/N(z) $>$5  &  S/N(Y) $>$7\\
   & S/N(g) $<$2  & S/N(g) $<$2 &S/N(g) $<$2 ; S/N(r) $<$2 & Z-Y $\ge$ 1.1  \\
   \textit{g-r $>$ 1.0} &\textit{r-i $>$ 1.2}&\textit{r-i $>$ 1.0} &\textit{i-z $>$ 1.5} & -− 0.5 $<$ Y − J $\le$ 0.5 \\
   \textit{r-i $<$ 1.0} &\textit{i-z $<$ 0.7} &\textit{i-z $<$ 0.5} & \textit{z-Y $<$ 0.5} & Z − Y $>$ Y − J + 0.7 \\
   \textit{g-r $>$1.5 (r-i) + 0.8} & \textit{r-i $>$1.5(i-Z) + 0.8} & \textit{r-i $>$1.5(i-Z) + 0.8} & \textit{i-z $>$2.0 (z-Y) + 1.1} & −0.5 $<$ Y − K $<$ 1.0 \\
  \cline{1-4}
  \multirow{3}{*}{\textit{$i_{AB} > 22.2$}}  & \multirow{3}{*}{\textit{$z_{AB} > 23$}} & \multirow{3}{*}{\textit{$z_{AB} > 23$}}& \multirow{3}{*}{\textit{$z_{AB} > 22$}} & J − K $<$ 0.8 \\
    &  &  & & undetected in \textit{ugri} bands if available\\
   \hline
 \end{tabular}
 \begin{tablenotes}
    \item[a] \cite{ono2018great}
    \item[b]  Our relaxed {\it r} dropout criteria (see text for details.)
    \end{tablenotes}
 \label{tab:dropouts_colour_criteria}
  \end{threeparttable}
\end{table*}

\begin{figure*}
\begin{subfigure}{0.9\textwidth}
 \centering
 % include first image
 \includegraphics[trim={100 290 90 0},width=1.01\textwidth]{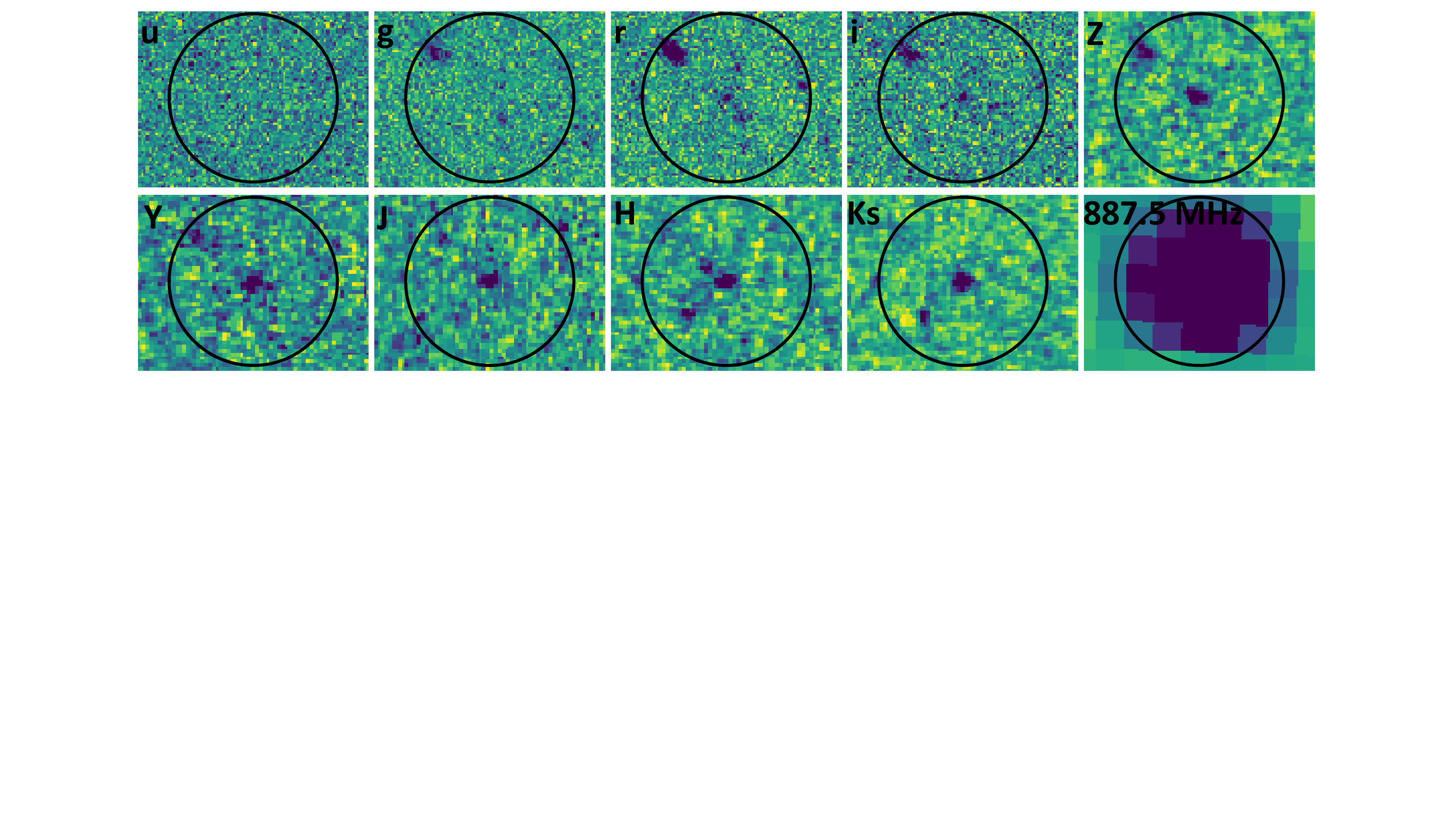}
 \caption{An example of g dropout at $z\sim4$ taken from our final sample:   J223541-311145. The size of the cutouts is $\sim20\arcsec\times15\arcsec$ and the circle is of radius $\sim7.4\arcsec$ which represents the apparent size of the \ac{ASKAP} detected radio source.}
\end{subfigure}
\begin{subfigure}{0.9\textwidth}
 \centering
 % include first image
 \includegraphics[trim={100 290 90 0},width=\textwidth]{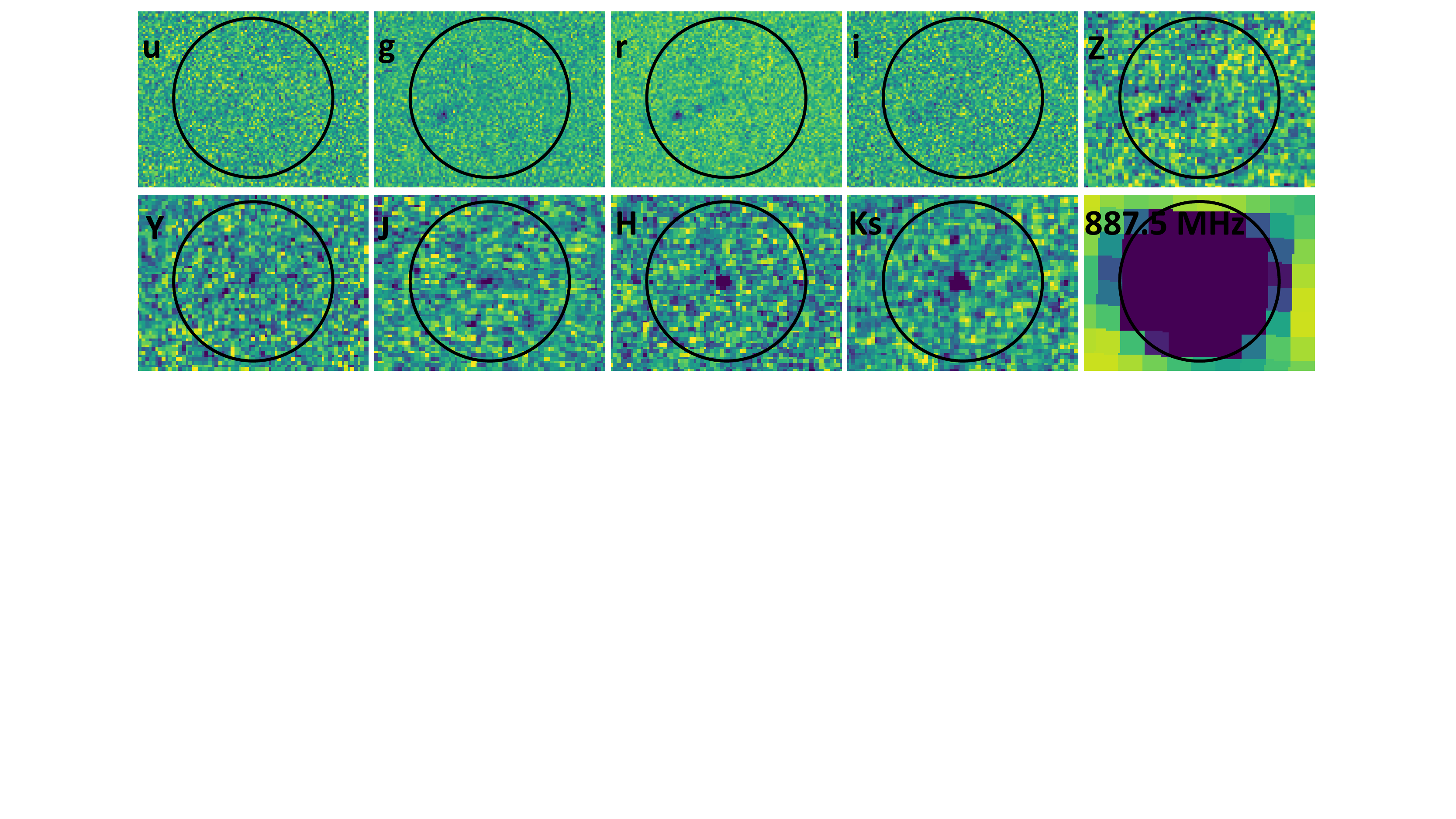}
 \caption{An example of r dropout at $z\sim5$ taken from our final sample:   J230551-343338. The size of the cutouts is $\sim24\arcsec\times18\arcsec$ and the circle is of radius $\sim8\arcsec$ which represents the apparent size of the \ac{ASKAP} detected radio source.}
\end{subfigure}
\begin{subfigure}{0.9\textwidth}
 \centering
 % include first image
 \includegraphics[trim={100 290 90 0},width=\textwidth]{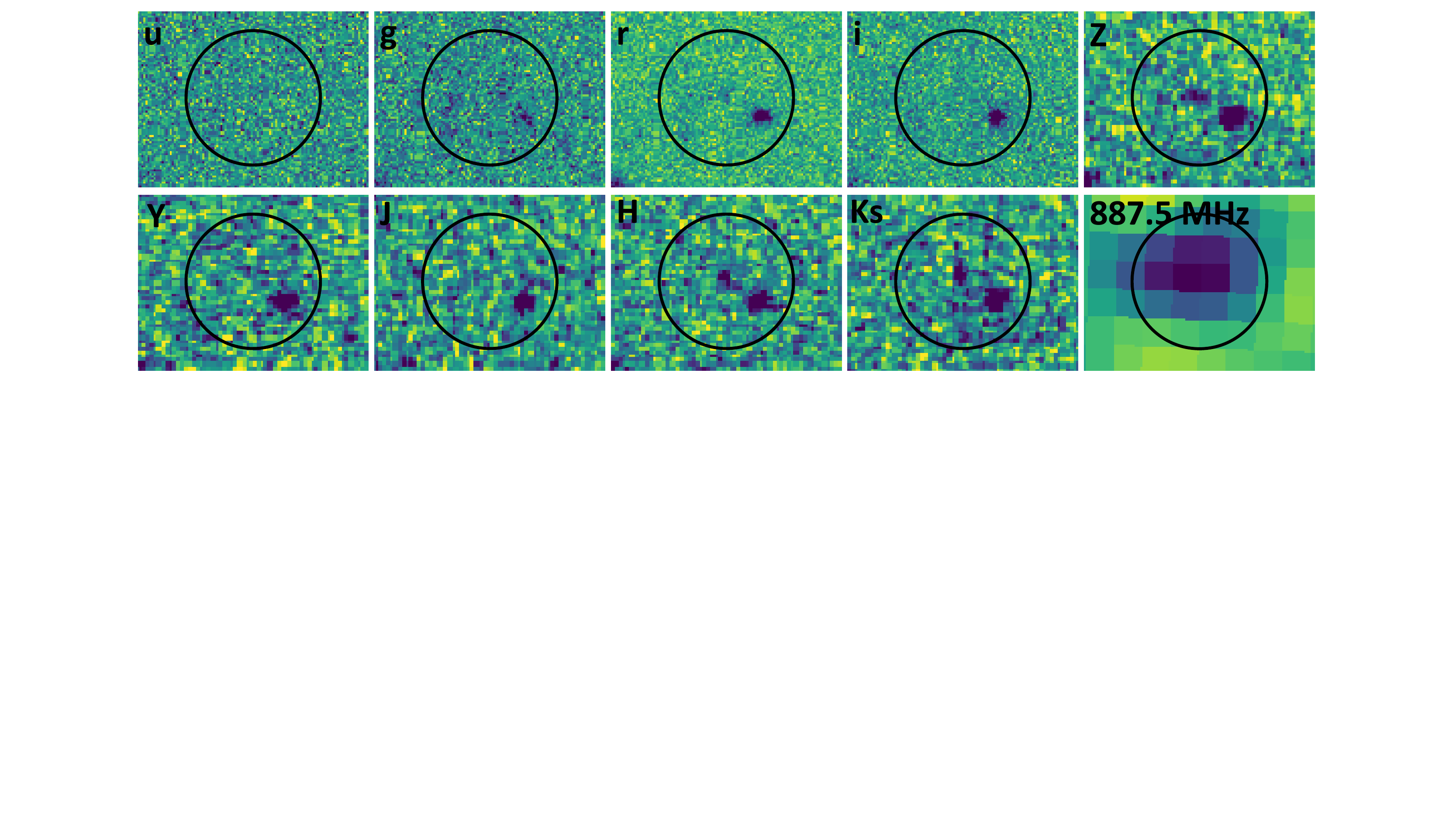}
 \caption{An example of i dropout at $z\sim6$ taken from our final sample:     J230246-293923. The size of the cutouts is $\sim20\arcsec\times15\arcsec$ and the circle is of radius $\sim6\arcsec$ which represents the apparent size of the \ac{ASKAP} detected radio source.}
\end{subfigure}

\caption{KiDS-\textit{ugri} and ViKING-ZYJHKs images of Lyman dropouts at each redshift are illustrated above in the order of increasing wavelength. The Lyman break can be identified by combining photometry in three consecutive bands. At $z \sim 4$, the Lyman limit ($\lambda\textsubscript{rest}=912\angstrom$) falls in the g-band and shorter wavelengths are significantly absorbed, as indicated by faint $u$ and $g$ bands. Similarly for $z \sim 5$ and 6 sources, the Lyman limit falls in the $r$ and $i$-bands respectively.}
\label{fig:example_dropouts}
\end{figure*}

To find \acp{HzRS}, we adopt the dropout criteria of \cite{ono2018great}, shown in Table~\ref{tab:dropouts_colour_criteria}.   The spectroscopically confirmed redshift ranges covered by each dropout, adopted from \citet[Figure~6]{ono2018great}, are as follows: (i) g-dropout: $3\le z \le 4.5$ (ii) r-dropout: $4.3\le z \le 5.4$ (iii) i-dropout: $5.6\le z \le 6.2$.  To keep it simple, we use redshifts, $z\sim4$, 5 and 6 to represent $g$, $r$, and $i$-band dropouts respectively. We use photometry from the KiDS DR4.1 multi-band catalogue (shown in Table~\ref{tab:kiDS_filters}), based on the GAaP magnitudes corrected for both zero-point and Galactic extinction. The 5$\sigma$ limiting magnitude for the $g$-band was used in the \textit{$g - r$} colour if objects were undetected ( i.e. no entry in the catalogue) in $g$-band. Similarly, a 5$\sigma$ limiting magnitude for the $r$ and $i$-bands were used to estimate \textit{$r - i$} and \textit{$i - z$} colours for $r$ and $i$ dropouts if objects were undetected in $r$ and $i$ band respectively.  In Table~\ref{tab:total_rg-dropout}, we describe the photometric selection and number of sources remaining after applying $z \sim 4$, 5, and 6 colour cuts. We present examples for all three dropouts, taken from our final sample, in Figure~\ref{fig:example_dropouts}, showing their cutouts at each of the \textit{ugri}ZYJHKs bands.

 Ideally, the signal-to-noise-ratio (SNR) of the sources should be used to 
%
%ray ensure
measure
their detection or non-detection in a given band. Since such information is not present in the KiDS catalogue, we utilized the mean 5$\sigma$ limiting magnitude of each passband to define detection and undetection. Given that the 5$\sigma$ limiting magnitude follows a continuous distribution \citep{kuijken2019fourth}, this could  result in some dropouts being missed.

\begin{table*}
    \centering
    \begin{threeparttable}
     \caption{ Our initial sample: the number of radio sources remaining after each criterion. Sources lacking detection in a given dropout (no entry in the catalog) band is replaced with their respective mean 5$\sigma$ limiting magnitude.}
    \begin{tabular}{c|c|c}
    \hline
       Sample & Criteria & Radio source \\
              & &  Count        \\
       \hline
       G23-ASKAP &-- & 35,973 \\
                                & cross-match with KiDS at 2\arcsec & 23\,002 \\
                                & After removing multiple matches & 17\,447 \\
       & imaflags\_iso = 0 \& &  \multirow{2}{*}{17\,396}  \\    
       & nimaflags\_iso = 0  & \\
       & flag\_gaap\_$gri$ = 0  & 17\,396 \\
       & flag\_gaap\_$griZ$ = 0  & 17\,376 \\
                 \hline
        & col~1,Table~\ref{tab:dropouts_colour_criteria} colour cuts \& & \multirow{4}{*}{229}\\
        $z\sim4$ sample  & Mag\_gaap\_i < 23.8 \& & \\
        ($g$ detected)& Mag\_gaap\_u > 24.23 or undetected\\
                 \hline
         & col~1,Table~\ref{tab:dropouts_colour_criteria} colour cuts \& & \multirow{4}{*}{6}\\
                          & Mag\_gaap\_g = 25.12 \& &\\
         $z\sim4$ sample  & Mag\_gaap\_i < 23.8 \tnote{a} \,\& & \\
         ($g$ undetected) &  Mag\_gaap\_u > 24.23 or undetected & \\
                         
                 \hline
         &   col~3,Table~\ref{tab:dropouts_colour_criteria} colour cuts\& & \multirow{4}{*}{58}\\
        $z\sim5$ sample  & Mag\_gaap\_Z < 23.6 \tnote{b} \,\& & \\
        ($r$ detected)& Mag\_gaap\_u > 24.23 or undetected \& & \\
                      & Mag\_gaap\_g > 25.12 or undetected \\
                 \hline
      &  col~4,Table~\ref{tab:dropouts_colour_criteria} colour cuts \& & \multirow{5}{*}{6}\\
        $z\sim6$ sample  & Mag\_gaap\_Z < 23.6\tnote{b} \,\& & \\
        ($i$ detected)& Mag\_gaap\_u > 24.23 or undetected\\
                      & Mag\_gaap\_g > 25.12 or undetected \\
                      & Mag\_gaap\_r > 23.68 or undetected \\
                 \hline
        & 
        col~4,Table~\ref{tab:dropouts_colour_criteria} colour cuts \& & \multirow{5}{*}{9}\\
        $z\sim6$ sample  & Mag\_gaap\_Z < 23.6\tnote{b} \,\& & \\
        ($i$ undetected)& Mag\_gaap\_u > 24.23 or undetected\\
                        & Mag\_gaap\_g > 25.12 or undetected \\
                        & Mag\_gaap\_r > 23.68 or undetected \\
                 \hline         

    \end{tabular}
    \begin{tablenotes}
    \item[a]  $i$-band $5\sigma$ limiting magnitude (AB) correspond to the end of the peak of the distribution \citep[Figure~3]{kuijken2019fourth}.
    \item[b]  $Z$ band $5\sigma$ limiting magnitude (AB) distribution correspond to the end of the peak of the distribution, assuming a spread of $\pm0.5\sigma$, given that the nominal $5\sigma$ $Z_{lim} = 23.1$. 
    \end{tablenotes}
    \label{tab:total_rg-dropout}
    \end{threeparttable}
\end{table*}

\begin{figure*}
    \centering
    \includegraphics[trim={0 370 0 25 clip=true},width=\textwidth]{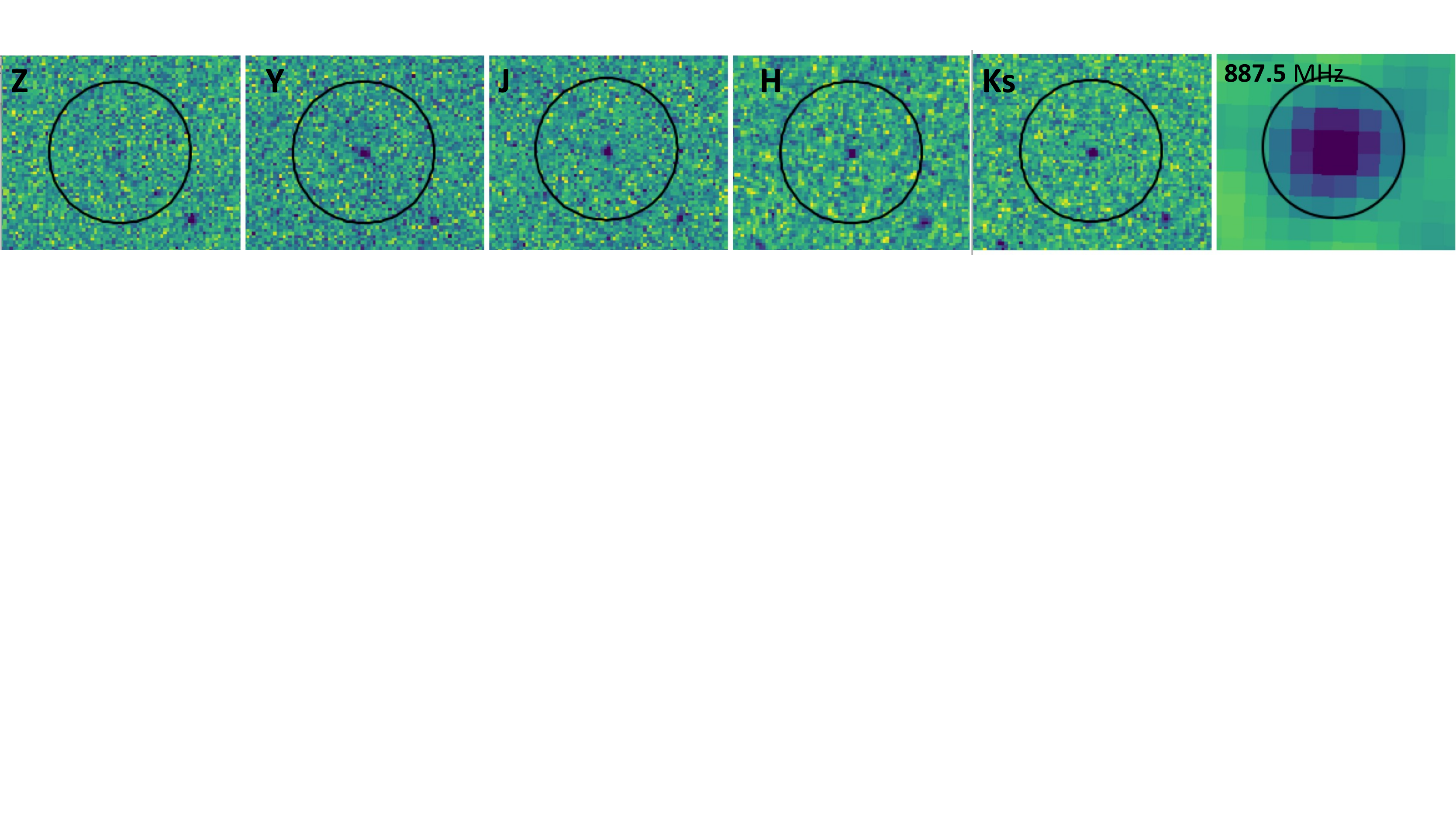}
    \caption{VIKING bands of a Z dropout at $z\sim7$ taken from our final sample.  The size of the cutouts is $\sim25\arcsec\times20\arcsec$ and the circle is of radius $\sim8\arcsec$ which represents the apparent size of the \ac{ASKAP} detected radio source. }
    \label{fig:my_label}
\end{figure*}

%%%%%%%%%%%%%%%%%%%%%%%%%%%%%%%%%%%%%%%%%%%%%%%%%%%%%%%%%%%%%%%%%%%%%

\subsubsection{Removing low-$z$ interlopers from our $z \gtrsim 4-6$ sample}
\label{sec:interloper1}

Low-$z$ objects such as dwarf stars and passive galaxies can enter the selection region defined by Lyman dropout colour cuts due to photometric errors, even though intrinsically they do not enter the colour selection window. We evaluate the contamination rate in the $z \sim 4$, 5 and 6 selection window by testing the Lyman dropout technique on a set of objects with known spectroscopic redshifts. We chose the COSMOS field \citep{scoville2007cosmic}, a well-studied region of the sky where both broadband optical photometry and spectroscopic redshifts are available.  We used the following catalogues of the COSMOS field, available in the public \ac{IRSA} domain, to test $g$, and $r \&$ $i$-band dropout techniques respectively ; (i) COSMOS Photometry Catalogue January 2006 (hereafter, COSMOS2006 catalog; \cite{capak2007first})  and (ii) COSMOS2015 Catalog \citep{laigle2016cosmos2015}. The spectroscopy for COSMOS sources was obtained from COSMOS DEIMOS Catalogue \citep{hasinger2018deimos}.  We utilized the  Subaru-\textit{griz} photometry in the COSMOS catalogues to test the colour cuts.  Furthermore, in this paper, we follow lowercase and uppercase notation in the literature for the Subaru {\it z} filter and the VIKING Z filter respectively.

 We crossmatched the COSMOS 2006 catalogue and the COSMOS DEIMOS Catalogue at 1\arcsec, resulting in 8906 matches. We further excluded multiple matches and apply the following criteria as recommended in \cite{capak2007first} to obtain a cleanest sample: blend$\_$mask = 0, i$\_$mask = 0, b$\_$mask = 0, and v$\_$mask = 0. This results in 6\,787 unique sources that have been deblended and are without any photometric flags to test the $g$-band dropout colour cuts. We further corrected the $g$, $r$, and $i$ magnitudes for Galactic extinction following the recommendations in \cite{capak2007first}.

 Similarly, we crossmatched the COSMOS2015 catalogue and the COSMOS DEIMOS Catalogue at 1\arcsec, resulting 7640 matches. The sources with multiple matches were excluded and the following criteria were applied as per \cite{laigle2016cosmos2015}: $flag\_HJMCC=0 \& flag\_cosmos=1 \& flag\_peter=0$, giving a source count of 7\,502. We finally applied respective Galactic extinction corrections to $r$, $i$, $z$, and $Y$ bands as per \cite{laigle2016cosmos2015}. 

\begin{figure*}
\begin{subfigure}{.45\textwidth}
 \centering
 \includegraphics[width=\textwidth]{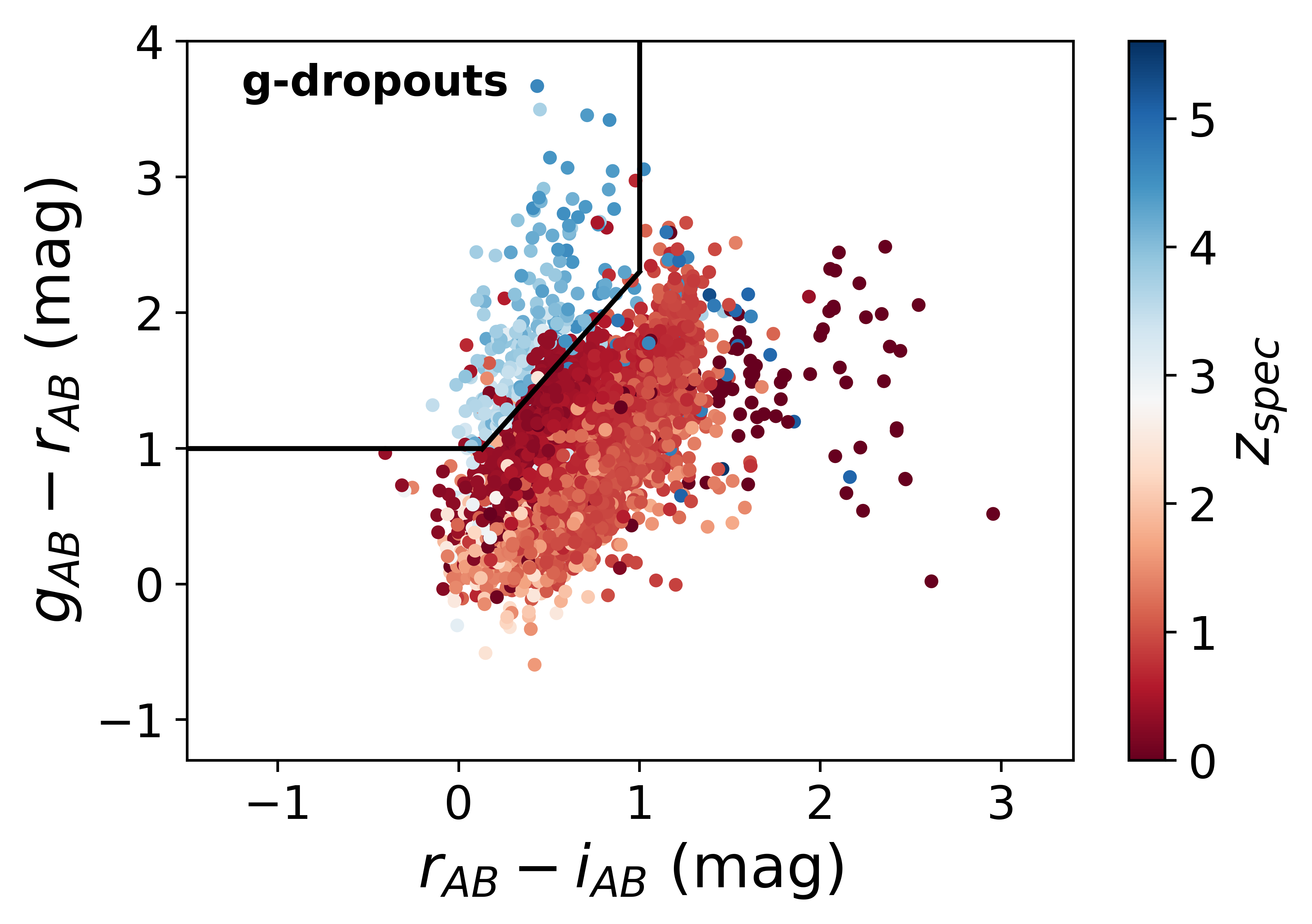} 
\end{subfigure}\hspace{0.5cm}
\begin{subfigure}{.45\textwidth}
 \centering
 \includegraphics[width=\textwidth]{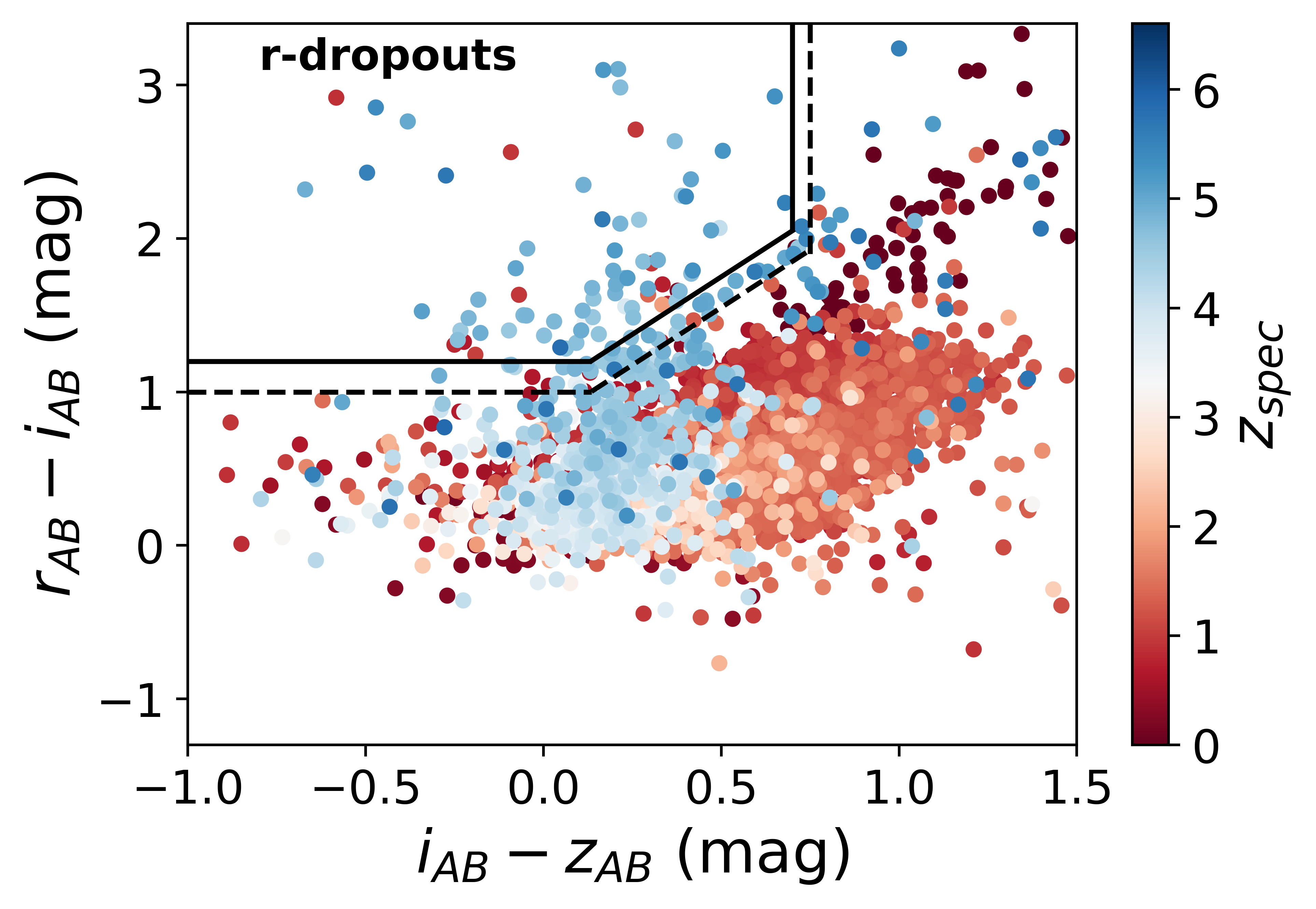} 
\end{subfigure}
\begin{subfigure}{.45\textwidth}
 \centering
 \includegraphics[width=\textwidth]{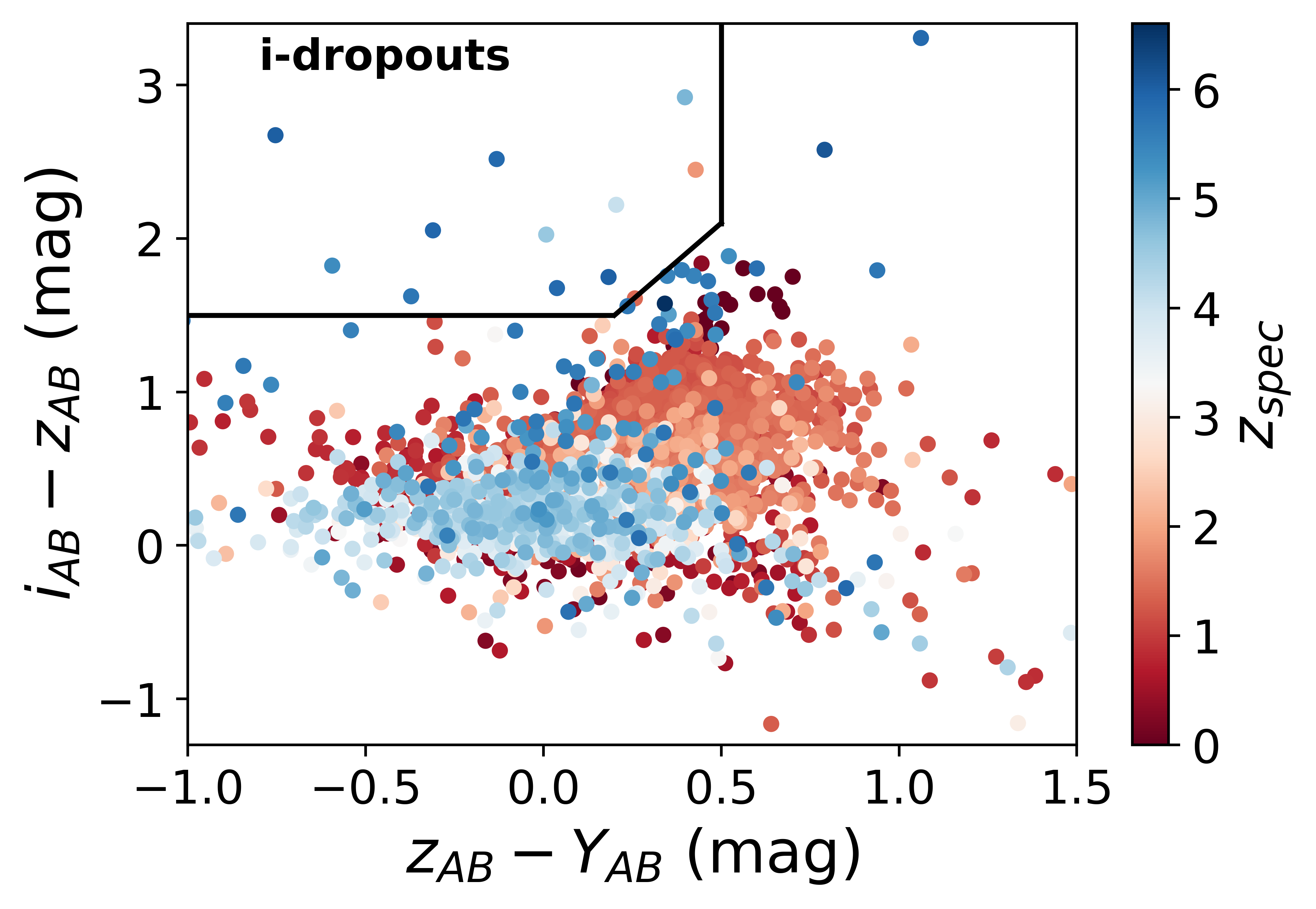} 
\end{subfigure}
\caption{\textbf{(Top Left)} 
Selection of $z\sim4$ sources by means of $g$-band dropout technique using $gri$ broadband filters: \textit{(g-r)} vs. \textit{(r-i)} colour-colour diagram of COSMOS sources, colour coded according to their spectroscopic redshift. The black box represents the $g$ dropout selection criteria in \protect\cite{ono2018great}. It encompasses almost all $z\sim4$ sources, but with a small contamination from low-$z$ sources too. \textbf{(Top Right)} Selection of $z\sim5$ sources by means of $r$-band dropout technique using $riz$ broadband filters: \textit{(r-i)} vs. \textit{(i-z)} colour-colour diagram of COSMOS sources, colour coded according to their spectroscopic redshift. The black box (solid lines) represents the $r$ dropout selection criteria in \protect\cite{ono2018great}.  The black box bordered by dashed black lines represent the colour locus from relaxed $r$ band colour-cuts.  %The selection window misses a significant fraction of $z>5$ sources. 
 \textbf{(Bottom)} Selection of $z\sim6$ sources by means of $i$-band dropout technique using $izY$ broadband filters: \textit{(i-z)} vs. \textit{(z-Y)} colour-colour diagram of COSMOS sources, colour coded according to their spectroscopic redshift. The black box represents the $i$ dropout selection criteria in \protect\cite{ono2018great}. }

\label{fig:cc_plot_cosmos}
\end{figure*}

Table~\ref{tab:dropouts_colour_criteria} shows that $z \sim 4$, 5, and 6 galaxy candidates can be selected based on their \textit{gri}, \textit{riz}, \textit{izy} colours respectively. We demonstrate this in Figure~\ref{fig:cc_plot_cosmos} by plotting the spectroscopic redshift distribution of COSMOS sources in \textit{g$-$r} vs. \textit{r$-$i}, \textit{r$-$i} vs. \textit{i$-$z} and \textit{i$-$z} vs. \textit{z$-$Y} colour-colour space. 
 To test the $z \sim 7$ or Z-band dropout colour cuts, deep spectroscopic data ($z_{spec}>6.4$) is needed, which is not available in the COSMOS DEIMOS catalog. It is evident that the photometric selection window of g-dropouts (the black box) encompasses almost all $z \sim 4$ sources, with a small contribution from low-$z$ ``interloper'' sources. By contrast, the r-dropout selection window misses a significant fraction of $z \sim 5$ sources.   Therefore, we relaxed the $r$ dropout colour cuts as follows:

\begin{equation} \label{eq:1}
 r-i > 1.0; \\
i-z < 0.75 ; \\
r-i > 1.5*(i-z)+0.8
\end{equation}

The resulting colour locus is indicated by black dashed lines, showing 
 that more $z \sim 5$ sources get included than low-$z$ ones.

\begin{figure*}
\begin{subfigure}{.45\textwidth}
 \centering
 % include first image
 \includegraphics[width=\textwidth]{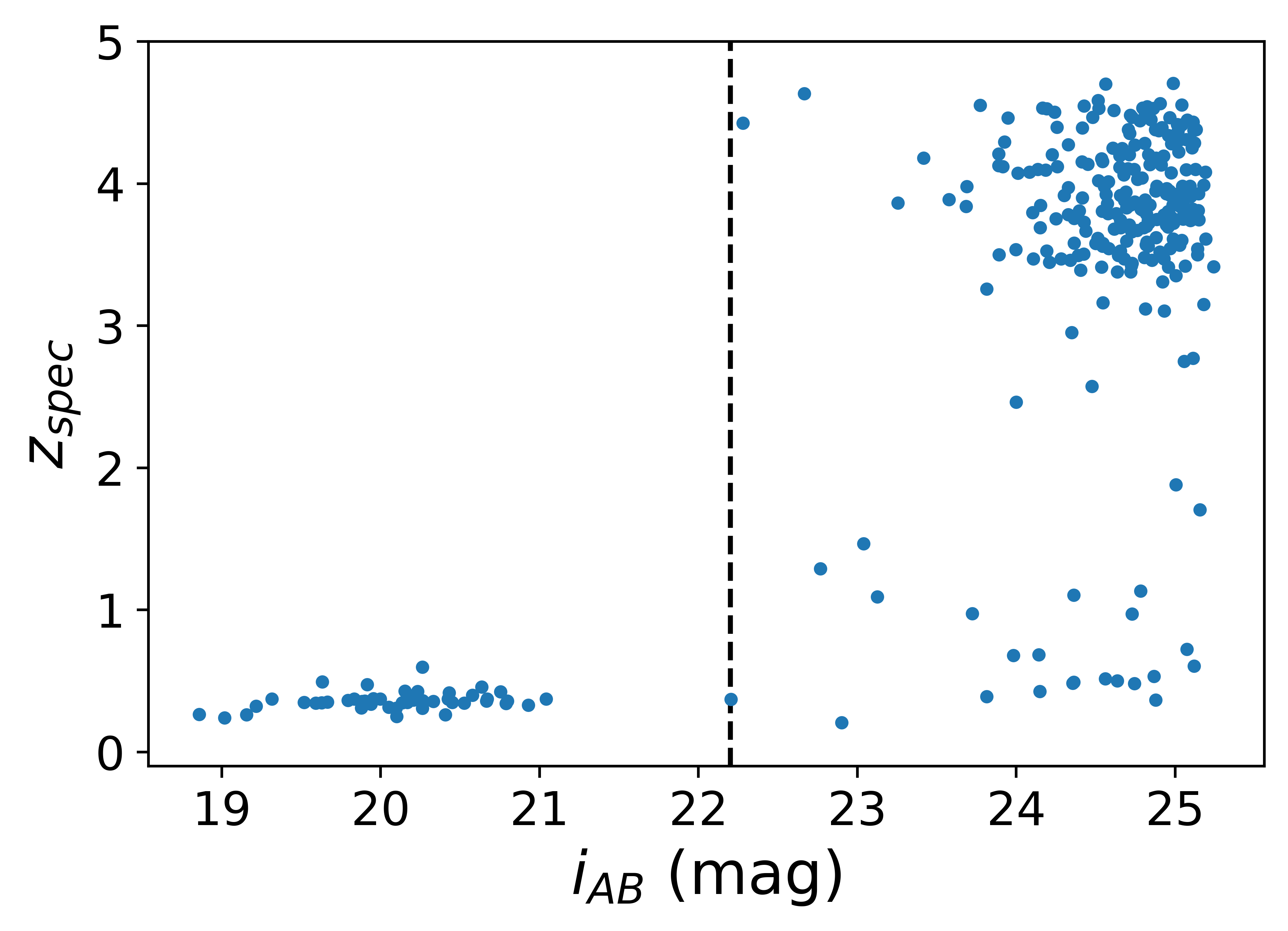} 
\end{subfigure}\hspace{0.5cm}
\begin{subfigure}{.45\textwidth}
 \centering
 % include first image
 \includegraphics[width=\textwidth]{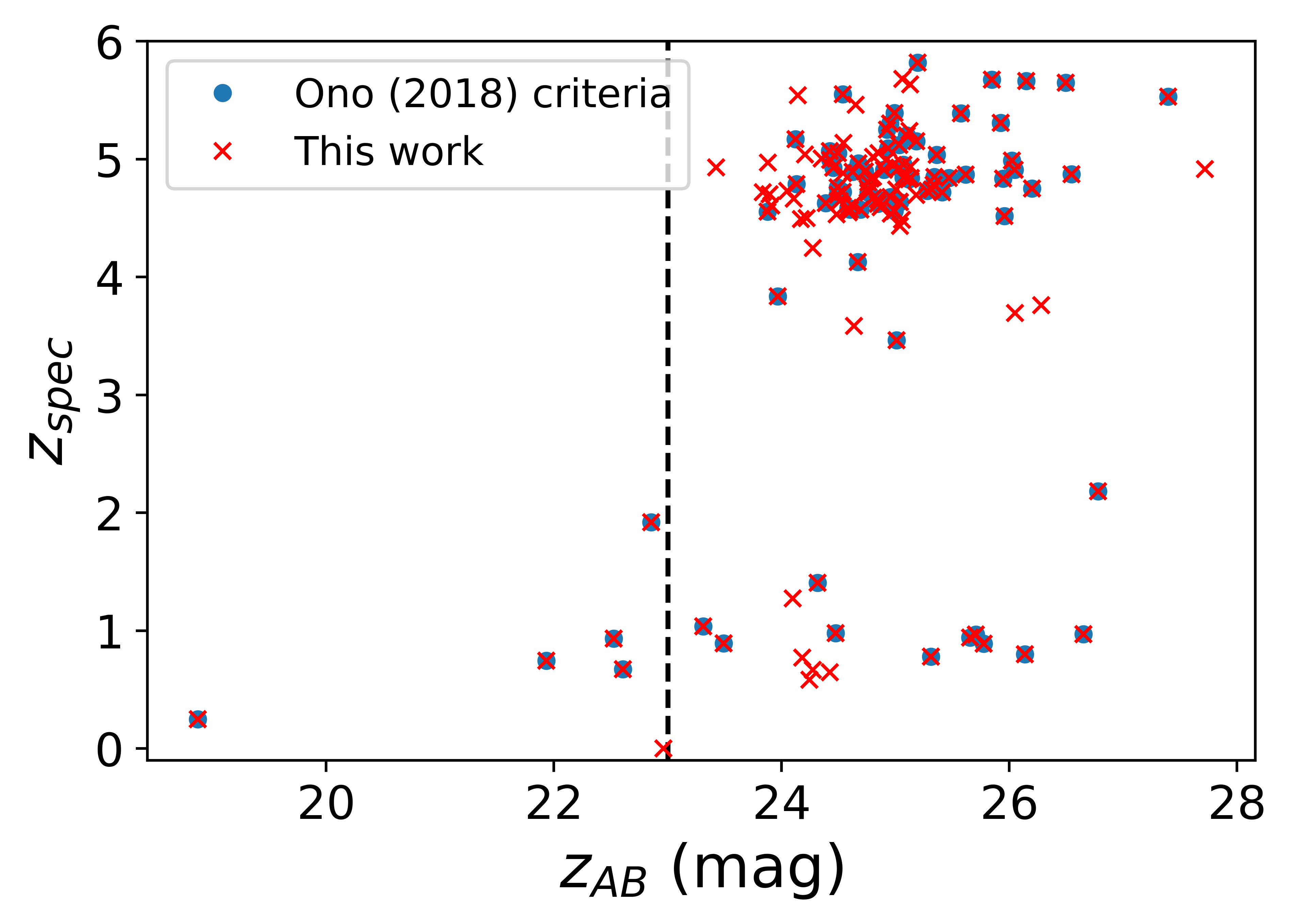} 
\end{subfigure}
\begin{subfigure}{.45\textwidth}
 \centering
 % include first image
 \includegraphics[width=\textwidth]{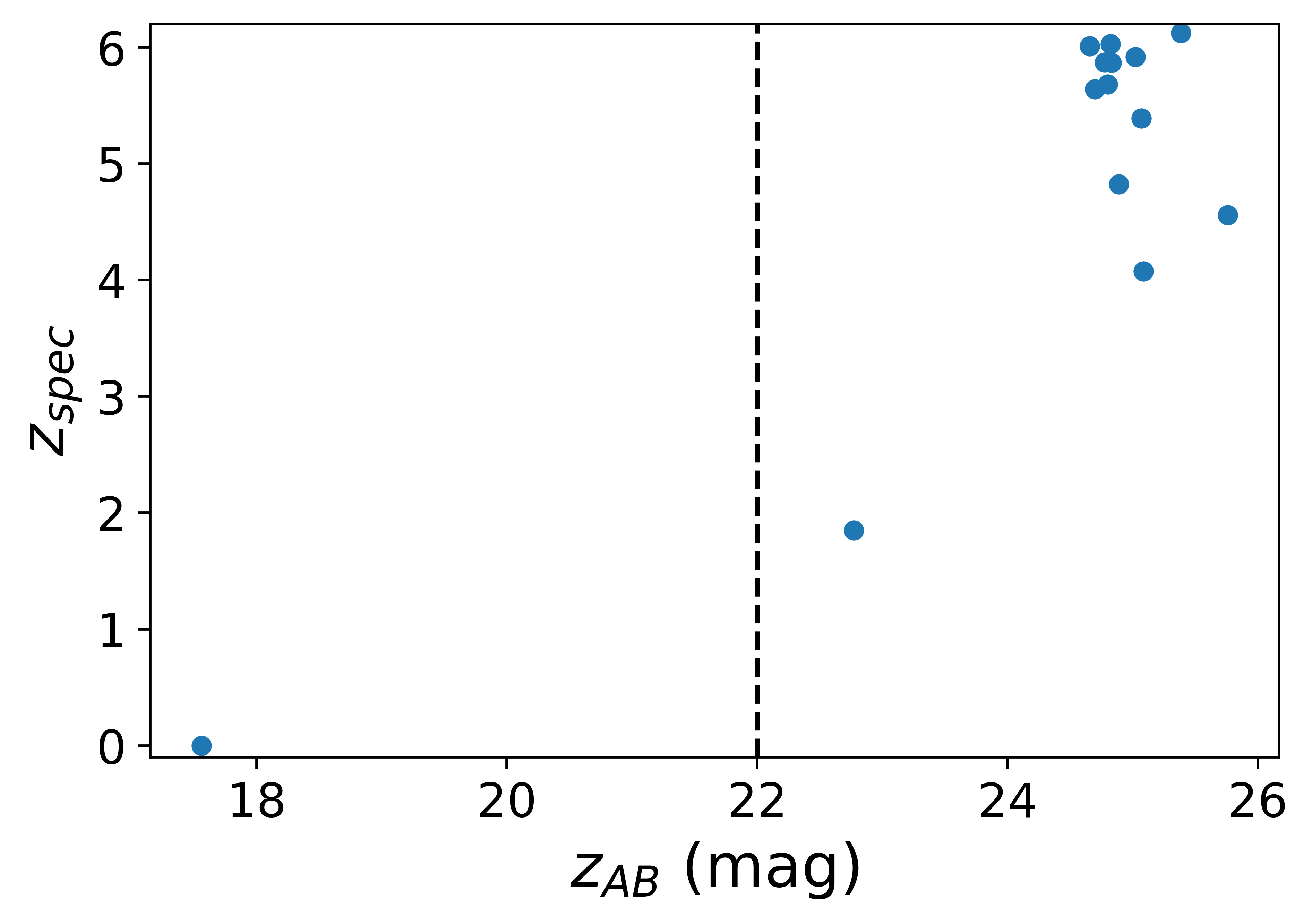} 
\end{subfigure}
\caption{Selecting low-$z$ interlopers in $g$ and $r$ dropout samples in COSMOS field by utilising the respective detection bands of each dropout: \textbf{(Top Left)} $i$ magnitude of $g$ dropouts together with their spectroscopic redshifts. \textbf{(Top Right)} $z$ magnitude of $r$ dropouts together with their spectroscopic redshifts.   Blue solid circles represent the $r$ dropouts selected using \protect\cite{ono2018great} criteria and red crosses, the $r$ dropouts selected using our relaxed colour-cuts (Equation~\ref{eq:1}).    (Bottom:) $z$ magnitude of $i$ dropouts together with their spectroscopic redshifts. } 
\label{fig:mag_redshift_cosmos}
\end{figure*}

In Figure~\ref{fig:mag_redshift_cosmos}, we demonstrate that $g$ and $r$ dropouts suffer significant contamination from low-$z$ sources at the bright end. Based on this, we introduce further selection criteria in the form of a magnitude cut-off in $i$-band (i\textsubscript{AB} $>$ 22.2) for $g$-dropouts and in $z$-band  (z\textsubscript{AB} $>$ 23) for $r$-dropouts, which helps to reduce the contribution from low-z sources (see bottom row of Table~\ref{tab:dropouts_colour_criteria}). 

 We further note that our sample size of $i$-band dropouts is 
%
%ray significantly less 
too small
to draw any conclusion. The $Z$ band magnitude cut-off ($Z_{AB}>24$) implied by the figure~\ref{fig:mag_redshift_cosmos} is too high and the VIKING data is also not deep enough to apply this cut-off.  Therefore, we performed a literature search to to find the $Z$-band magnitude of known $z\sim6$ quasars. Based on the spectroscopically confirmed $i$ - band dropouts in \citet[Table~3]{venemans2015first}, we applied a magnitude cut-off, Z\textsubscript{AB} $>$ 22 for $i$-dropouts.

\begin{figure*}
\begin{subfigure}{0.35\textwidth}
 \centering
 % include first image
 \includegraphics[width=\textwidth]{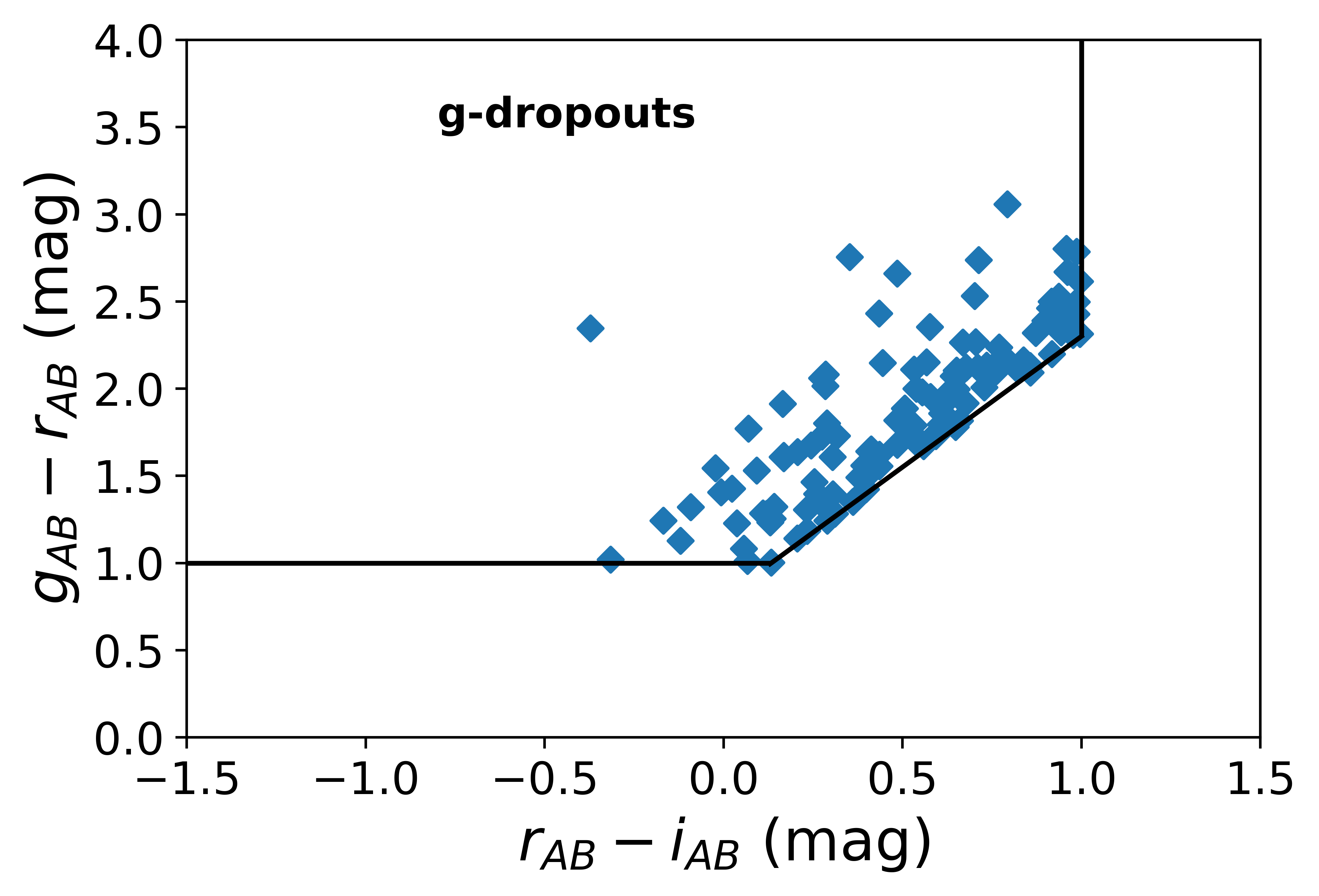}
\end{subfigure} \hspace{1 cm}
\begin{subfigure}{0.35\textwidth}
 \centering
 % include first image
 \includegraphics[width=\textwidth]{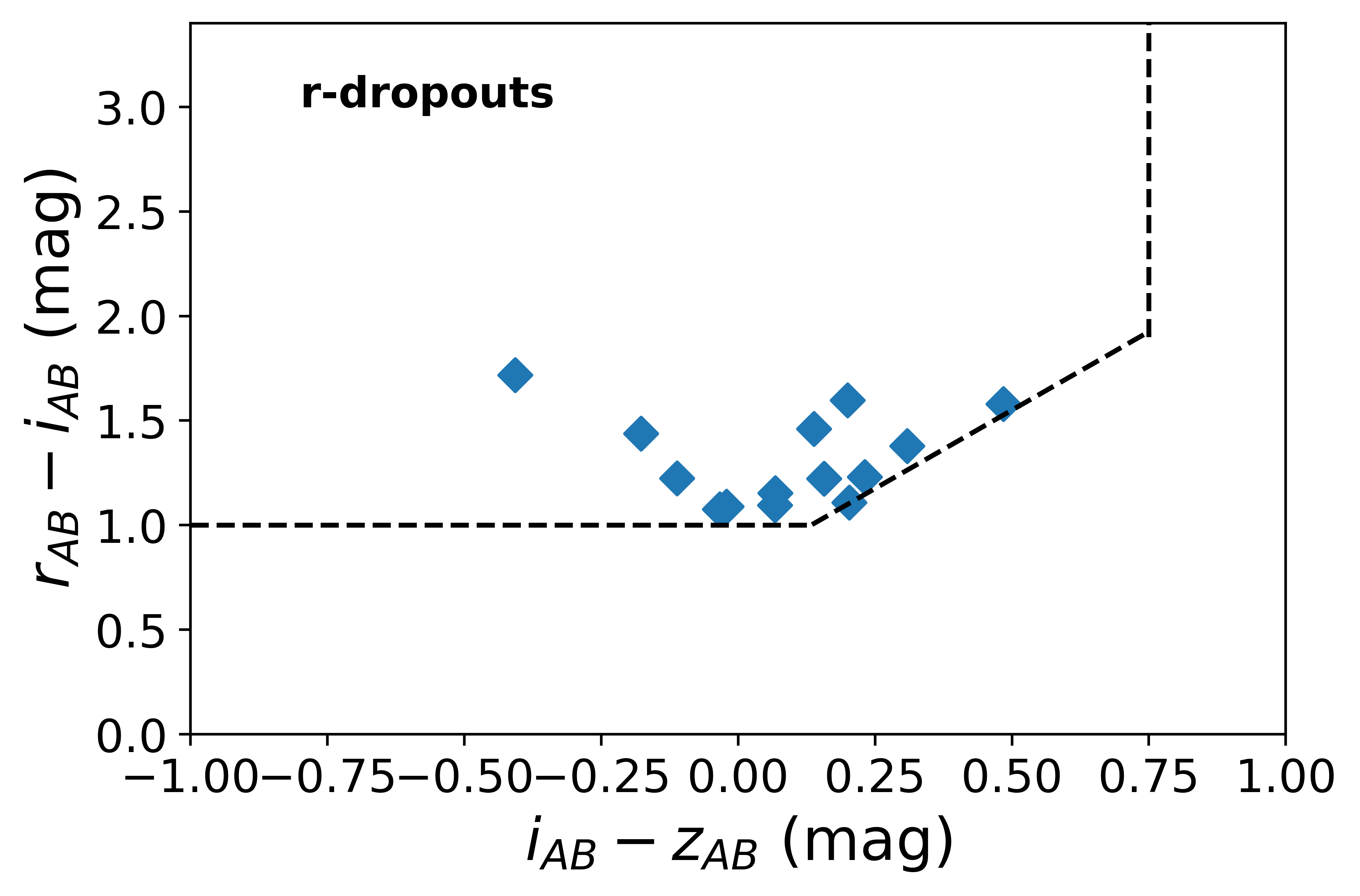}
\end{subfigure}
\begin{subfigure}{0.35\textwidth}
 \centering
 % include first image
 \includegraphics[width=\textwidth]{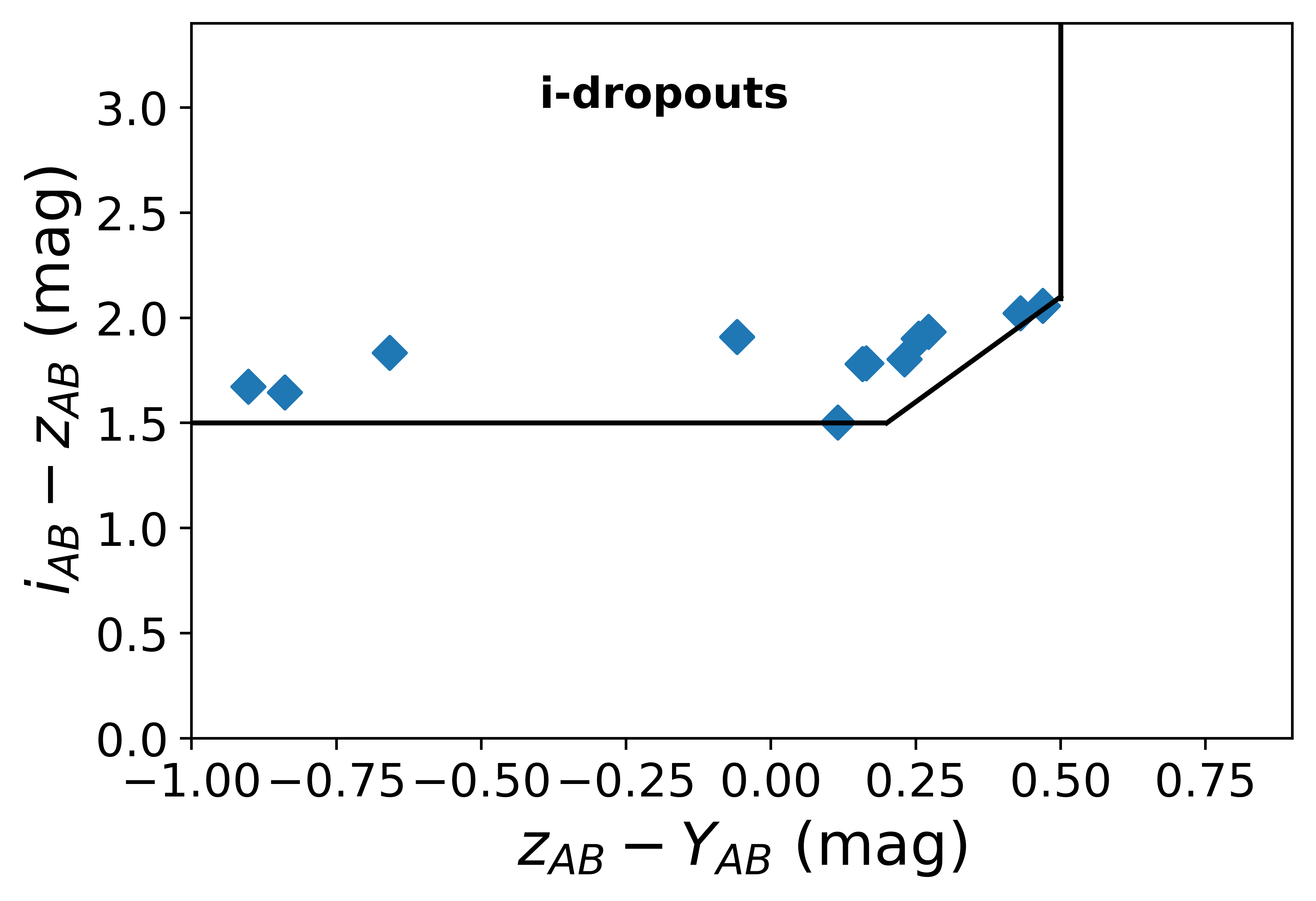}
\end{subfigure}\hspace{1 cm}
\begin{subfigure}{0.35\textwidth}
 \centering
 % include first image
 \includegraphics[width=\textwidth]{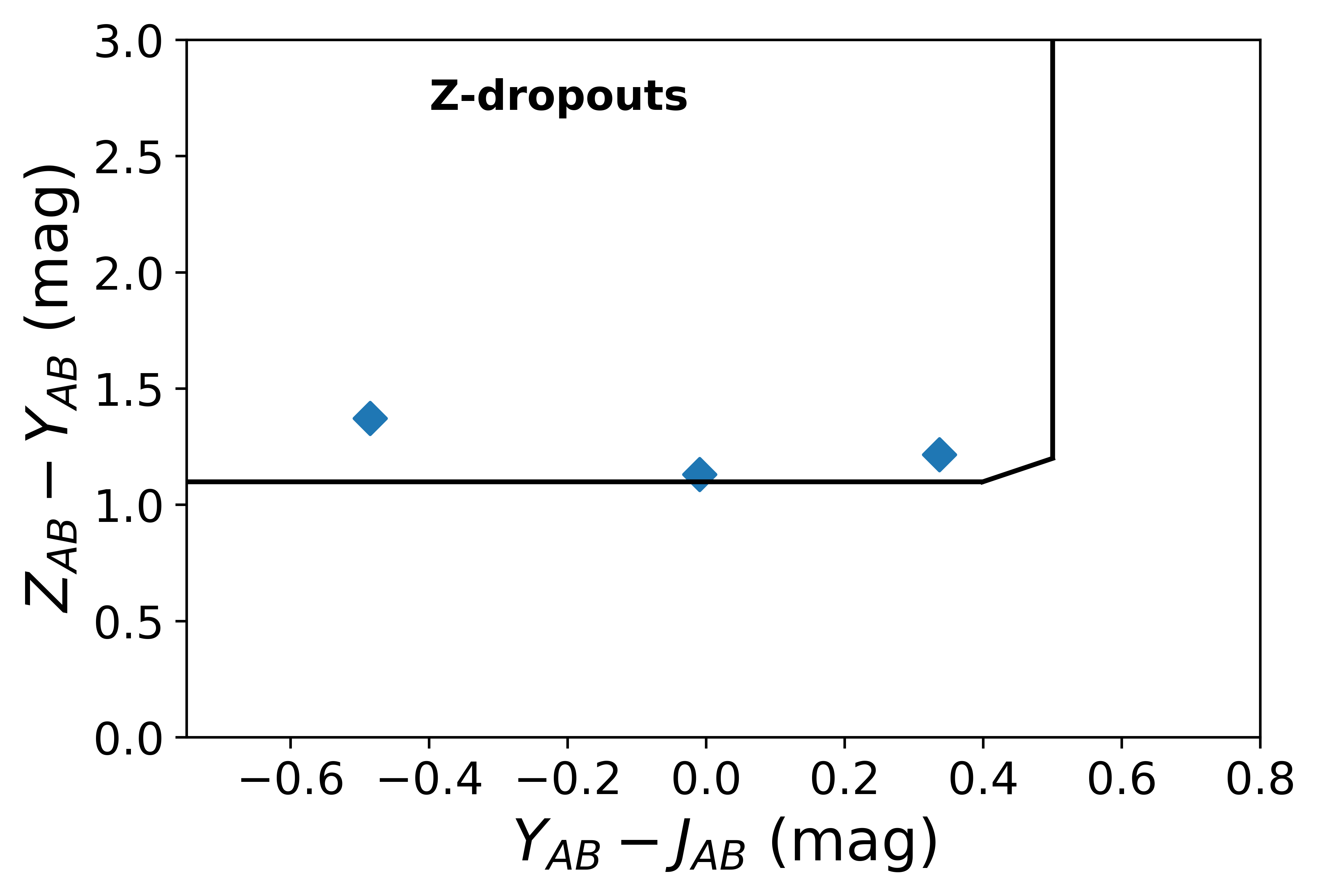}
\end{subfigure}
\caption{ Colour-colour diagram of our final sample of \acp{HzRS} at $z\gtrsim$4-7 selected via the Lyman dropout technique. The black box shows the selection criteria of $g$, $r$, $i$, and $Z$ dropouts in $gir$, $riZ$, $iZY$, and $ZYJ$ colour-colour plane respectively.}
\label{fig:cc_dropouts}
\end{figure*}

%%%%%%%%%%%%%%%%%%%%%%%%%%%%%%%%%%%%%%%%%%%%%%%%%%%%%%%%%%%%%%%%%%%%%

\subsection{Selection of radio sources at $z\sim7$}

We utilised VIKING photometry to search for radio sources in the redshift range, $6.44\le z \le 7.44$. \cite{venemans2013discovery} demonstrated that Lyman dropouts at $z\sim7$ (a.k.a Z dropouts) can be selected using  ZYJ near-IR filters. We cross-matched G23-ASKAP radio catalogue and VIKING DR5 catalogue at a search radius of 2$\arcsec$,  which  gives a false ID rate of 7.9\% . Using the selection method in \cite{venemans2013discovery} (see Table~\ref{tab:dropouts_colour_criteria}), we select a sample of 3 radio source candidates at $z\sim7$, of which one is a known radio-quasar, VIK J2318−3113, at $z=6.44$ \citep{ighina2021radio}. Scattering of foreground galaxies into the Z-dropout selection region in the ZYJ color-color space is minimised by selecting point sources only by applying the criterion: pGalaxy$<0.95$ ( see \cite{venemans2013discovery} and reference therein for details), where pGalaxy is the probability that the source is a galaxy.   

%%%%%%%%%%%%%%%%%%%%%%%%%%%%%%%%%%%%%%%%%%%%%%%%%%%%%%%%%
\subsection{Estimate of Reliability}
\label{reliability}
Finally, we estimate the success rate of our magnitude cutoff by counting the fraction of low-$z$ sources remaining after applying the $i$-band and $z$-band cutoff to $g$ and $r$ \& $i$ dropouts respectively. 
 A sample of  294 sources were identified from the COSMOS catalog as satisfying the $z \sim 4$ colour cuts (Table~\ref{tab:dropouts_colour_criteria}), of which  75 sources have $z_\textrm{spec}$ below 3.0, resulting in a contamination rate of 25.5\%. The further application of the i\textsubscript{AB} $>$ 22.2 criterion reduces the low-z source count to 29 and the total sample size to 248. This gives a final contamination rate of  11.7\%.

 A total of 78 sources were identified as $r$ dropouts using \cite{ono2018great} criteria, 
%
%ray where 19 0f which 
of which 19
have $z_\textrm{spec} < 4.3$ giving a contamination rate of $\sim24.4\%$. On the other hand, relaxed $r$ band colour cuts (Equation~\ref{eq:1}) selected 127 sources in total, of which 29 have $z_\textrm{spec} < 4.3$ resulting a contamination rate of $\sim22.8\%$. This shows that relaxed $r$ band colour cuts select $\sim1.7$ times more $z_\textrm{spec} > 4.3$ sources than the \cite{ono2018great} criteria while the contamination rate remains almost constant. We further reduce the contamination rate to 19\% by applying a  z\textsubscript{AB} $>$ 23 cutoff. 

 Due to the dearth of deep spectroscopic data, only 14 sources were selected as $i$ dropouts, 6 of which have a $z_\textrm{spec}<5.6$ classifying as interlopers. This gives an initial contaminant rate of $\sim 43 \%$ and the further application of magnitude cut-off, $Z_{AB} > 22$, results a final contaminants rate of $\sim 38.4\%$.

 This shows that the $z \sim 4$ sample is  $\sim88\%$ reliable while the $z \sim 5$ sample is around 81\% reliable. 
%
%ray The $z \sim 6$ sample must be atleast $\sim62\%$ reliable, given that sample size was smaller. 
The $z \sim 6$ sample is $\sim62\%$ reliable, which is likely to be a lower limit given that the sample size was smaller.
There is no robust way of estimating the reliability for the $z\sim7$ sample, as there is a lack of deep spectroscopic data in COSMOS. 

 We applied these magnitude cut-offs to the sources in Table~\ref{tab:total_rg-dropout} to select our final sample of radio source candidates at $z\gtrsim4-6$ in the G23-\ac{ASKAP} field. The resulting source count of each dropout is shown in Table~\ref{tab:final}. As a further check on reliability, we cross-matched our sample with the GAMA spectroscopic redshift \citep{baldry18} catalogue and the Gaia Early Data Release 3 \citep[EDR3,][]{brown2020gaia}. The Gaia satellite measures parallaxes and proper motions for nearby stars in our Milky Way, and allow us to identify cool dwarf stars, if any, in our final sample. Gaia EDR3 is the latest release, providing information for about 1.8 billion objects.

None of our sources were found to have either GAMA spectroscopic or GAIA counterparts out to a search radius of 4~arcsec. The lack of GAMA spectroscopic counterparts is consistent with a high redshift, as the GAMA spectroscopic survey is complete to $z\sim0.4$. Similarly, the lack of GAIA counterparts confirms that (i) no Milky Way stars are present in our sample, and (ii) no low-z quasars that are bright enough to be detected by GAIA are in our final sample.

%%%%%%%%%%%%%%%%%%%%%%%%%%%%%%%%%%%%%%%%%%%%%%%%%%%%%%%%%%%%%%
\section{Results}
\label{sec:result}

  Using the Lyman Dropout photometric technique and our additional magnitude cut-offs (Table~\ref{tab:dropouts_colour_criteria}) to reduce low-$z$ interlopers, we select our final sample of 148 radio source candidates at $z \gtrsim 4-7$, where (i) {\it g} \& {\it i} dropouts criteria come from \cite{ono2018great} (ii) {\it r } dropout criteria from this study (Equation~\ref{eq:1}) and (iii) Z dropout criteria from \cite{venemans2013discovery}. 
 
  The colour-colour plot of our final sample is shown in Figure~\ref{fig:cc_dropouts}. Furthermore, we identify a known radio quasar at $z=6.44$ and 2 radio source candidates at $z\sim7$ using the Z-dropout selection technique from \cite{venemans2013discovery}.  Thus our final sample of 149 radio sources in 50\,deg\textsuperscript{2} implies a sky density of $\sim3$ per $deg^{2}$ for $S_{888}\gtrsim0.1$\,mJy radio sources at $z\gtrsim4$. For comparison, \cite{norris21} suggested a sky density of $\sim5$ per $deg^{2}$ beyond $z\sim4$ at EMU flux limit,
 based on simulations.  We present the catalogues of $z \gtrsim 4-6$ and $z\sim7$ sources, including KiDS/VIKING and WISE W1 photometry, in Table~\ref{tab:catalog} and Table~\ref{tab:z7_sample} respectively.

%%%%%%%%%%%%%% Table 5 %%%%%%%%%%%%%%%%%%%
\begin{table}
 \caption{Our final sample of  $ z\gtrsim4-7$ candidates in G23 field selected using Lyman dropout colour cuts (Table~\ref{tab:dropouts_colour_criteria}) together with our magnitude cutoffs from Figure~\ref{fig:mag_redshift_cosmos}}.
 \centering
 \begin{tabular}{c|c}
 \hline
  Redshift ($z$) & Sample Size \\
  \hline
  $z\sim4$ &  117\\
  $z\sim5$ &  14 \\
  $z\sim6$ &  15\\
  $z\sim7$ & 3\\
  \hline
 \end{tabular}
 \label{tab:final}
\end{table}
%%%%%%%%%%%%%%%%%%%

%%%%%%%%%%%%%% Table 6 %%%%%%%%%%%%%%%%%%%
\clearpage
\onecolumn
\begin{longtable}[c]{ccccccccccc}
 \caption{Our final $z\gtrsim4-6$ sample, showing radio properties from \ac{ASKAP} and optical/ IR photometry from KiDS and CATWISE catalogues respectively. Missing optical magnitudes indicate non-detection in that respective filter. The missing WISE magnitude (W1) indicate no counterpart within 2~arcsec of the KiDS position.   KiDS/VIKING magnitudes are given in the AB system and the WISE magnitudes in the Vega unit. }\\
 \hline
 Index & \ac{ASKAP} &RA & DEC & S\textsubscript{887.5}&Mag\_u &Mag\_g& Mag\_r & Mag\_i & Mag\_Z & W1 \\
  & name & (deg) & (deg) & (mJy) & & & & & & (mag)  \\
  \hline
 \endfirsthead
 \hline
  Index & \ac{ASKAP} &RA & DEC & S\textsubscript{887.5}&Mag\_u &Mag\_g& Mag\_r & Mag\_i & Mag\_Z & W1 \\
  & name & (deg) & (deg) & (mJy) & & & & & & (mag)  \\
  \hline
  \endhead
  \multicolumn{11}{c}{$z\sim4$ Sample ($g$ dropouts) } \\
  \hline

1&  J225745-311209 & 344.437  & -31.203   & 0.32     & 25.067 & 25.779  & 24.108  & 23.549  & 22.928  &        \\
2&  J225802-345205 & 344.511  & -34.868    & 0.948    &            & 25.954  & 23.293  & 22.808  & 21.720  & 17.588 \\
3&  J225835-325509 & 344.649  & -32.919   & 0.237    &            & 25.220  & 23.268  & 22.689  & 21.573  & 15.633 \\
4&  J225915-314116 & 344.815   & -31.688   & 0.57     &            & 24.348  & 22.954  & 22.649  & 21.796  & 16.055 \\
5&  J230041-322125 & 345.172  & -32.357   & 0.247    &            & 25.854  & 23.887  & 23.248  & 21.810  & 15.95  \\
6&  J230122-294717 & 345.342  & -29.788   & 0.254    & 24.800 & 25.084   & 23.678  & 23.685  & 23.143  &        \\
7&  J230156-331615 & 345.484  & -33.271   & 0.261    & 24.993  & 25.791  & 23.483  & 22.506  & 21.295   & 15.556 \\
8&  J230223-294300 & 345.599  & -29.716   & 0.584    & 24.999  & 25.341   & 23.511  & 22.884   & 22.187 & 17.686 \\
9&  J230234-335010 & 345.645  & -33.836   & 0.246    &            & 25.306  & 23.750 & 23.314  & 23.198 &        \\
10&  J230249-322437 & 345.708  & -32.411   & 0.233    &            & 26.273  & 24.163   & 23.632   & 22.896  & 16.641 \\
11&  J230306-333335 & 345.778  & -33.559   & 0.282    &            &            & 23.845  & 23.724   &            &        \\
12&  J230317-345556 & 345.824  & -34.932   & 2.428    &            & 25.179  & 23.003 & 22.217 & 21.372 & 16.771 \\
13&  J230432-341722 & 346.133   & -34.289   & 6.517    & 25.132 & 24.487  & 23.360   & 23.481  & 23.491  &        \\
14&  J230504-292342 & 346.267  & -29.395   & 0.646    &            & 25.831  & 23.852  & 23.297  & 22.074  & 17.746 \\
15&  J230516-331612 & 346.317  & -33.270   & 0.421    &            & 25.096  & 23.024   & 22.381   & 22.067   &        \\
16&  J230544-351610 & 346.434  & -35.269   & 0.302    &            & 24.140  & 23.057  & 23.002  & 22.886  &        \\
17&  J230547-293327 & 346.449  & -29.557   & 2.886    & 24.393   &            & 23.511  & 23.343  & 22.727   &        \\
18&  J230718-310125 & 346.826   & -31.024   & 0.364    &            & 24.013  & 22.692  & 22.784  & 22.728  &        \\
19&  J230831-350015 & 347.133  & -35.004   & 0.203    &            & 24.927  & 23.111   & 22.451  & 21.681  & 16.86  \\
20&  J230832-340644 & 347.137   & -34.112   & 8.009    & 25.594  & 24.677  & 23.069  & 22.766  & 22.104  & 16.219 \\
21&  J230905-334318 & 347.272  & -33.722   & 0.204    &            & 25.661   & 23.539  & 22.862  & 21.849  &        \\
22&  J230937-323421 & 347.405  & -32.573   & 0.272    &            & 25.599   & 23.829  & 23.760   & 22.196  & 16.971 \\
23&  J230940-335143 & 347.418  & -33.862     & 0.298    &            & 24.765  & 23.522  & 23.691  & 24.487  &        \\
24&  J230942-335049 & 347.427  & -33.847   & 0.304    &            &            & 23.385  & 22.792  & 21.804  & 16.54  \\
25&  J231057-294135 & 347.740   & -29.693   & 0.431    &            & 25.159  & 23.852  & 23.621  & 22.309  & 16.91  \\
26&  J231117-323952 & 347.822   & -32.664   & 0.247    &            & 25.543  & 24.052 & 23.674  & 23.917  &        \\
27&  J231148-311231 & 347.950  & -31.209   & 0.323    & 25.134  & 24.726  & 22.381  & 22.753  & 22.904  & 18.202 \\
28&  J231149-304758 & 347.958  & -30.799   & 0.292    & 24.414   & 24.550   & 23.228   & 23.087  & 22.582   & 16.862 \\
29&  J231210-332436 & 348.045  & -33.410   & 0.26     & 24.598  & 25.670  & 24.111  & 23.718  & 23.112  & 17.172 \\
30&  J231421-344141 & 348.587   & -34.695   & 0.749    & 24.784   & 24.396  & 22.718  & 22.233  & 22.000  & 16.423 \\
31&  J231444-291949 & 348.684  & -29.330   & 0.276    &            & 26.165   & 23.711  & 22.739  & 21.711  & 16.301 \\
32&  J231449-293938 & 348.706  & -29.661   & 1.205    & 24.960  & 25.635  & 23.636  & 23.097  & 21.851  & 16.812 \\
33&  J231508-312105 & 348.784  & -31.351   & 0.244    &            & 25.858  & 24.062  & 23.455  & 22.462   & 17.192 \\
34&  J231508-341955 & 348.787  & -34.332    & 98.088   &            & 25.778   & 23.454   & 22.511   & 21.470  & 15.897 \\
35&  J231555-311458 & 348.979  & -31.249   & 0.472    &            & 26.288  & 23.857  & 23.423   & 22.709   & 17.889 \\
36&  J231604-324740 & 349.020  & -32.795   & 0.151    &            & 26.329  & 24.171  & 23.403  & 22.477  & 16.847 \\
37&  J231617-303200 & 349.073  & -30.533   & 3.779    &            & 25.763  & 23.302   & 22.391  & 21.255  & 15.878 \\
38&  J231632-331953 & 349.134  & -33.332   & 1.342    & 24.946   & 24.309  & 22.669  & 22.263  & 21.888  & 17.111 \\
39&  J231645-301948 & 349.189  & -30.330  & 0.763    &            & 24.803  & 23.166  & 22.959  & 21.994   & 18.09  \\
40&  J231648-303629 & 349.201  & -30.608   & 0.356    &            & 25.818  & 23.713  & 23.062  & 22.658  & 16.939 \\
41&  J231719-315344 & 349.332  & -31.896   & 0.205    & 25.291  & 24.586  & 23.584  & 23.451    & 23.077   &        \\
42&  J231723-301556 & 349.348  & -30.266   & 0.464    &            & 24.145   & 22.748  & 22.487  & 22.705  & 18.284 \\
43&  J231737-313228 & 349.407  & -31.541   & 0.259    & 24.763  & 26.235  & 23.565  & 22.605   & 21.880   & 16.663 \\
44&  J231752-311151 & 349.468  & -31.197   & 0.324    &            & 24.784  & 23.772  & 23.705  & 22.739  & 16.736 \\
45&  J231826-323746 & 349.610  & -32.629    & 0.374    & 24.867  & 25.198  & 23.653  & 23.676  & 22.280  & 16.01  \\
46&  J231828-310408 & 349.618  & -31.069   & 0.197    &            & 24.853  & 23.426   & 23.403  & 22.139  & 17.125 \\
47&  J231850-303818 & 349.708  & -30.638   & 0.2      &            & 25.686  & 22.949  & 22.236  & 21.583  & 16.557 \\
48&  J231853-293420 & 349.722  & -29.572   & 0.361    & 24.365   & 26.231  & 24.235  & 23.583  & 22.366  & 16.635 \\
49&  J232019-294205 & 350.081  & -29.702  & 0.572    &            & 25.674  & 24.177  & 23.777  & 22.864  & 15.97  \\
50&  J232020-343818 & 350.084  & -34.638   & 0.257    &            & 25.217  & 23.124  & 22.267  & 21.1660  & 16.308 \\
51&  J232034-305242 & 350.143  & -30.879   & 0.506    &            & 26.007  & 23.508  & 22.591  & 21.225   & 16.13  \\
52&  J232043-320457 & 350.182  & -32.083   & 3.006    & 25.217  & 25.887  & 23.793   & 23.129  & 22.256  & 16.822 \\
53&  J232130-320208 & 350.377  & -32.036   & 0.294    & 25.261  & 24.562  & 23.421  & 23.215  & 21.940  & 17.03  \\
54&  J232135-320117 & 350.399  & -32.022   & 0.266    &            & 24.778   & 22.919   & 22.309  & 21.839  & 16.359 \\
55&  J232140-311522 & 350.417  & -31.256   & 0.47     &            & 25.638  & 23.374  & 22.705   & 21.997  & 17.304 \\
56&  J232246-331448 & 350.691  & -33.247     & 0.277    & 24.752  & 26.199  & 23.584  & 22.589  & 21.546  & 16.676 \\
57&  J232331-345632 & 350.882  & -34.942   & 0.639    &            & 26.101  & 23.673  & 22.688  & 21.728  & 15.694 \\
58&  J232338-350353 & 350.910  & -35.065   & 0.988    &            & 24.471   & 23.451  & 23.767  & 22.534  & 18.065 \\
59&  J232347-344625 & 350.948   & -34.774   & 0.726    &            & 25.682  & 23.363   & 22.491  & 21.452   & 15.994 \\
60&  J232413-303039 & 351.057  & -30.511   & 0.457    &            & 26.281  & 23.783   & 22.797  & 21.864  & 16.669 \\
61&  J232515-295957 & 351.314  & -29.999   & 1.317    &            & 24.565  & 22.944  & 22.509   & 22.097  & 17.487 \\
62&  J232515-314448 & 351.314  & -31.747   & 1.431    & 24.848  & 26.250  & 23.447  & 22.489  & 21.277  & 16.122 \\
63& J222915-334311  & 337.314  & -33.719   & 1.992    &            & 25.057  & 23.829   & 23.792   & 23.329    &        \\
64& J223018-304213  & 337.576  & -30.704   & 2.954    &            & 25.314  & 23.629  & 23.089   & 22.573  & 17.405 \\
65& J223054-322552  & 337.727 & -32.431    & 75.829   &            & 25.564  & 23.444  & 22.627  & 21.861  & 17.538 \\
66& J223215-301935  & 338.066   & -30.327   & 10.809   &            & 26.243  & 23.488   & 23.135   & 22.872  & 18.29  \\
67& J223508-295809  & 338.785  & -29.969   & 1.081    &            & 24.139  & 22.897   & 22.607  & 22.438  &        \\
68& J223531-291912  & 338.879  & -29.320   & 0.744    & 24.549  & 25.323   & 23.572  & 23.050   & 22.325   & 16.968 \\
69& J223533-343305  & 338.889  & -34.552   & 1.942    & 24.406  & 24.286    & 23.031  & 22.894   & 21.702  & 17.279 \\
70& J223541-311145  & 338.922  & -31.196   & 0.689    &            & 26.473  & 24.012  & 23.065  & 21.906  & 16.757 \\
71& J223710-305549  & 339.294  & -30.930   & 2.54     &            & 25.052  & 23.322  & 23.006  & 22.722  & 16.801 \\
72& J223729-325838  & 339.374  & -32.977  & 0.337    & 24.749  & 25.078   & 23.796  & 23.485  & 22.158   & 16.236 \\
73& J223743-333305  & 339.432  & -33.551   & 1.05     &            & 24.747  & 23.514   & 23.384   & 22.600  & 16.724 \\
74& J223831-330728  & 339.629  & -33.124   & 0.249    &            & 25.445  & 23.981  & 23.727  & 24.024   &        \\
75& J223936-295303  & 339.900  & -29.884    & 2.703    &            & 25.995  & 23.209 & 22.222  & 21.617   & 16.493 \\
76& J224017-325128  & 340.073  & -32.858   & 0.398    & 24.662  & 25.188  & 23.179  & 22.451   & 22.441 & 16.653 \\
77& J224040-295600  & 340.168  & -29.933    & 2.673    & 24.283  & 24.770  & 22.952   & 22.466  & 20.989  & 16.221 \\
78& J224047-331331  & 340.197  & -33.225   & 0.646    &            & 25.308  & 23.958  & 23.689   & 23.193  &        \\
79& J224138-331346  & 340.411  & -33.229    & 0.242    &            & 25.162  & 23.246  & 22.570  & 21.842  & 17.368 \\
80& J224145-340622  & 340.439   & -34.106    & 66.186   &            & 26.2801  & 23.891   & 22.992  & 21.688  & 16.254 \\
81& J224203-333606  & 340.513  & -33.602   & 0.769    & 25.436   & 25.747  & 23.598  & 23.154  & 24.637   & 16.188 \\
82& J224246-335928  & 340.692  & -33.991   & 1.751    &            & 25.566  & 23.892  & 23.647  & 23.481  &        \\
83& J224249-333814  & 340.706  & -33.637   & 0.55     &            & 25.751  & 23.515   & 22.745  & 21.976  & 17.565 \\
84& J224308-313622  & 340.784  & -31.606   & 0.662    & 24.432 & 25.657  & 24.005  & 23.593  & 22.933  & 17.425 \\
85& J224311-332250  & 340.796  & -33.381   & 4.432    &            & 25.643  & 23.491  & 22.924  & 21.404   & 16.056 \\
86& J224354-333900  & 340.976  & -33.650   & 0.272    & 25.384  & 25.028  & 23.303   & 23.028  & 21.557  & 16.502 \\
87& J224401-315431  & 341.0056   & -31.909   & 11.74    &            & 26.160  & 23.101   & 22.309  & 21.1800  & 15.864 \\
88& J224417-350540  & 341.072  & -35.094   & 0.622    & 24.869  & 25.819  & 23.907  & 23.741  & 22.267   & 17.523 \\
89& J224445-314547  & 341.188 & -31.763   & 4.325    &            & 25.334  & 23.252     & 22.967  & 21.969  &        \\
90& J224519-342942  & 341.329  & -34.495   & 0.535    &            & 25.644  & 23.446  & 22.529  & 21.990  & 16.562 \\
91& J224540-313747  & 341.418  & -31.629   & 0.221    & 25.059  & 24.642  & 22.851   & 22.319  & 21.532  & 16.265 \\
92& J224540-330700  & 341.419  & -33.117   & 1.679    & 24.517   & 25.802  & 23.269  & 22.568  & 21.588   & 16.067 \\
93& J224552-312733  & 341.468  & -31.4592   & 0.409    &            & 26.263  & 23.615  & 22.636  & 22.315   & 16.989 \\
94& J224601-310331  & 341.508 & -31.059   & 0.685    & 25.250  & 26.278  & 24.116  & 23.278  & 22.031  & 16.936 \\
95& J224615-285257  & 341.565  & -28.883   & 1.177    &            & 24.328  & 22.526  & 22.237  & 22.035  &        \\
96& J224623-311641  & 341.597  & -31.278     & 0.206    & 24.845   & 25.129  & 23.845  & 23.735  & 22.802  & 18.402 \\
97& J224629-323825  & 341.622 & -32.640  & 0.824    &            & 25.533  & 23.647   & 23.141  & 22.331  & 16.925 \\
98& J224630-343853  & 341.628  & -34.648   & 0.599    & 24.742  & 26.381  & 24.066   & 23.071   & 21.513  & 16.402 \\
99& J224642-311527  & 341.677 & -31.258   & 0.189    & 24.730  & 25.230   & 23.807   & 23.410  & 22.712  &        \\
100& J224710-321939  & 341.792   & -32.328   & 0.476    & 24.249  &            & 23.011   & 22.249   & 21.336  & 16.425 \\
101& J224716-310058  & 341.817  & -31.016   & 0.32     &            & 26.009  & 23.856  & 23.021  & 21.859  & 17.402 \\
102& J224812-335035  & 342.051  & -33.843   & 0.36     & 25.272  & 26.052  & 24.037  & 23.753  & 21.790  & 16.874 \\
103& J224918-314457  & 342.327  & -31.749   & 4.265    &            & 25.877  & 23.815  & 23.540  & 24.035  &        \\
104& J224955-294536  & 342.479  & -29.760     & 0.294    & 24.707  & 26.023  & 23.892  & 23.036  & 22.479  & 16.346 \\
105& J225029-350452  & 342.622  & -35.0811   & 0.663    &            & 25.287  & 23.936  & 23.575  & 22.385  & 16.907 \\
106& J225052-331301  & 342.719  & -33.217   & 0.49     & 24.649  & 25.228  & 23.448  & 22.800  & 21.656  & 16.539 \\
107& J225214-310846  & 343.061  & -31.146   & 2.607    &            & 26.549  & 24.020   & 23.083  & 22.204  & 16.635 \\
108& J225240-302302  & 343.166  & -30.384   & 0.384    &            & 26.053  & 23.786   & 23.082  & 21.805  & 16.802 \\
109& J225257-351113  & 343.238  & -35.187   & 0.275    & 25.510  & 25.626  & 23.263  & 22.320  & 21.177  & 16.124 \\
110& J225300-302322  & 343.253  & -30.389   & 0.49     &            & 24.503  & 23.148  & 22.869  & 22.856  &        \\
111& J225312-305344  & 343.303  & -30.896   & 0.188    & 24.612  & 25.492  & 23.359  & 22.624 & 22.381  & 16.758 \\
112& J225314-295139  & 343.311  & -29.861   & 0.365    &            & 25.334  & 23.804  & 23.712  & 24.634  &        \\
113& J225332-313319  & 343.386  & -31.555   & 0.242    & 24.613  & 25.953  & 23.599   & 23.023   & 21.794  & 16.304 \\
114& J225343-313305  & 343.429  & -31.552  & 0.367    &            & 25.312  & 23.191  & 22.481  & 21.788   & 16.938 \\
115& J225647-285027  & 344.196  & -28.841   & 2.049    &            & 24.798  & 23.615   & 23.382   & 22.680  & 17.25  \\
116& J225733-313857  & 344.388  & -31.649   & 0.249    &            & 25.377  & 23.456  & 22.837  & 21.364  & 16.001 \\
117& J225827-342715  & 344.614  & -34.454   & 0.286    & 25.139  & 26.231  & 24.271  & 23.650  & 22.228   & 16.669 \\

\hline
\multicolumn{11}{c}{$z\sim5$ Sample  ($r$ dropouts)} \\
\hline

1&  J231423-331509 & 348.596  & -33.253    & 0.232    &            & 26.248  & 24.738 & 23.642  & 23.575  & 16.47  \\
2& J224105-345956  & 340.271  & -34.999   & 0.261    &            & 25.664  & 24.804  & 23.343  & 23.205  & 16.581 \\
3&  J232503-340057 & 351.264  & -34.016    & 0.293    & 25.392  &            & 24.578 & 23.471  & 23.267  & 17.989 \\
4&  J231919-320058 & 349.831  & -32.016  & 0.294    &            &            & 25.372  & 23.793  & 23.308  & 17.586 \\
5& J224820-301317  & 342.086  & -30.221    & 0.295    & 24.808  & 25.333  & 24.632   & 23.402    & 23.171 & 17.793 \\
6&  J230855-335352 & 347.229  & -33.898   & 0.323    &            &            & 24.749  & 23.372  & 23.063 & 17.508 \\
7& J224858-343550  & 342.242  & -34.597   & 0.356    & 25.101   & 26.083  & 24.681  & 23.457  & 23.569     &        \\
8&  J232021-315838 & 350.090  & -31.977   & 0.382    &            & 25.374  & 24.706  & 23.483   & 23.327  & 17.392 \\
9&  J230235-292456 & 345.648  & -29.415    & 0.443    &            & 25.676  & 24.440  & 23.003 & 23.181   & 16.789 \\
10&  J232035-314851 & 350.149  & -31.814   & 0.453    & 25.137  & 25.209  & 24.388  & 23.235    & 23.167  &        \\
11& J225340-300626  & 343.417  & -30.107   & 0.604    & 24.703  & 26.143  & 24.253   & 23.164  & 23.186  & 16.329 \\
12&  J230551-343338 & 346.463  & -34.561   & 1.248    &            & 26.491  & 25.365  & 23.768  & 23.568   & 15.901 \\
13&  J230946-314050 & 347.446  & -31.681   & 1.925    & 25.160  & 25.341  & 24.632   & 23.556  & 23.590   & 16.995 \\
14&  J230301-305405 & 345.755  & -30.902   & 48.28    & 24.559  & 25.334  & 24.356  & 22.638  & 23.045 & 17.474 \\

\hline
\multicolumn{11}{c}{$z\sim6$ Sample  ($i$ dropouts)} \\
\hline

1& J223445-332701  & 338.687  & -33.450   & 0.418    & 25.773  &            & 24.113  & 24.442   & 22.824  & 18.457 \\
2& J223719-333857  & 339.330  & -33.649   & 0.253    & 24.446   & 25.272  & 24.209 & 24.738  & 22.934  & 16.201 \\
3& J224115-303408  & 340.313  & -30.569   & 1.534    &            & 25.440   & 24.481  & 24.176  & 22.531  & 17.176 \\
4& J224130-302918  & 340.377 & -30.488   & 0.414    & 24.261 & 25.51  & 24.133  & 23.502  & 22.001  & 16.495 \\
5& J224652-340238  & 341.717  & -34.044   & 0.364    &            & 25.277  & 24.919  & 24.649  & 22.740  & 17.127 \\
6& J224957-332626  & 342.487  & -33.441   & 0.348    &            &            & 24.518  & 24.346  & 22.288  & 17.592 \\
7&  J230246-293923 & 345.693  & -29.656   & 0.255    & 24.531  & 25.785  & 24.542  & 24.67  & 23.152  & 17.14  \\
8&  J230423-322732 & 346.098  & -32.459    & 0.323    &            & 25.833  & 24.599  & 24.326  & 22.393  & 17.412 \\
9&  J230617-335108 & 346.574  & -33.852   & 1.338    & 24.927  & 26.045    & 24.540 & 24.729  & 22.707   & 17.383 \\
10&  J231051-334628 & 347.713  & -33.775   & 2.471    &            & 25.472  & 24.659  & 25.338 & 23.557 & 18.277 \\
11&  J231125-342657 & 347.857  & -34.449   & 0.212    & 25.349   & 25.270   & 24.544  & 24.442  & 22.541  & 16.731 \\
12&  J231245-291817 & 348.188  & -29.305   & 0.447    &            & 25.961  & 24.485  & 24.084 & 22.301  & 15.885 \\
13&  J231004-334153 & 347.519  & -33.698   & 0.287    &            & 25.541  & 25.099  & 24.757 & 23.125  &        \\
14& J223430-311836  & 338.627  & -31.310   & 1.23     & 25.057  & 25.216  & 24.858 & 25.361  & 23.528 &        \\
15&  J232457-322915 & 351.241  & -32.488   & 0.353    & 25.209    & 25.133  & 24.645   & 24.775   & 23.102  &        \\

 \hline
 \label{tab:catalog}
\end{longtable}
\twocolumn

\begin{table*}
    \centering
    \caption{  Our final $z\sim7$ sample  ($Z$ dropouts), showing radio properties from \ac{ASKAP} and IR photometry from VIKING and CATWISE catalogues respectively. Missing WISE magnitudes (W1/W2) indicate no counterpart within 2~arcsec of the VIKING position. pGalaxy indicates the probability that the source is a galaxy, obtained from the VIKING DR5 catalogue.}
    \begin{tabular}{cccccccccccc}
    \hline
  Index & \ac{ASKAP} &RA & DEC & S\textsubscript{887.5} & Mag\_Z & Mag\_Y & Mag\_J & Mag\_Ks & W1\_mag & W2\_mag &pGalaxy \\
  & name & (deg) & (deg) & (mJy) & (AB) &(AB) & (AB)&(AB) &(Vega) & (Vega) &  \\
         \hline
1 &  J230535--341213 & 346.399453 & -34.203674 & 0.251 & 22.53 &21.16  &21.65  &23.08 & -- &-- & $9.5\times10^{-6}$  \\
2 &  J223833--320822 & 339.63998 & -32.139507 & 0.335 & 21.41 &20.19 &19.86 &21.99 &17.09 &16.10 &0.00017  \\ 
3 &  J231818--311345 & 349.576502 & -31.229435 & 0.662 &21.91  &20.78 &20.79 &22.35 &18.02 &18.11 &0.000171 \\ 
\hline
    \end{tabular}
    \label{tab:z7_sample}
\end{table*}

%*******************************************************************************
\section{Discussion}
\label{sec:discuss}

\subsection{Radio flux density}
\label{sec:radio_f_de}

\begin{figure}
 \centering
 % include first image
 \includegraphics[width=0.45\textwidth]{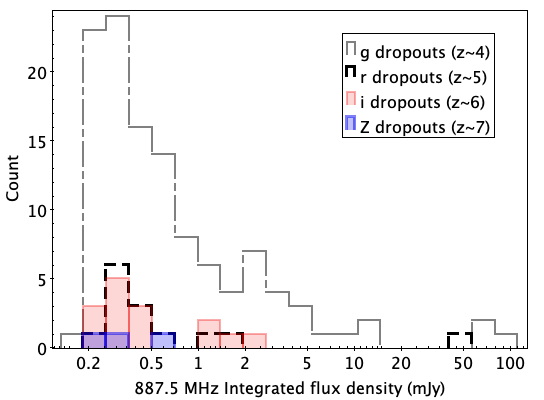} 
 \caption{887.5\,MHz Total radio flux density distribution of our final sample of   148 \ac{HzRS} candidates and a known radio-quasar (selected as a Z-dropout).}
 \label{fig:fluxD}
\end{figure}
%***********************
\begin{figure*}
\begin{subfigure}{.45\textwidth}
 \centering
 % include first image
 \includegraphics[width=\textwidth]{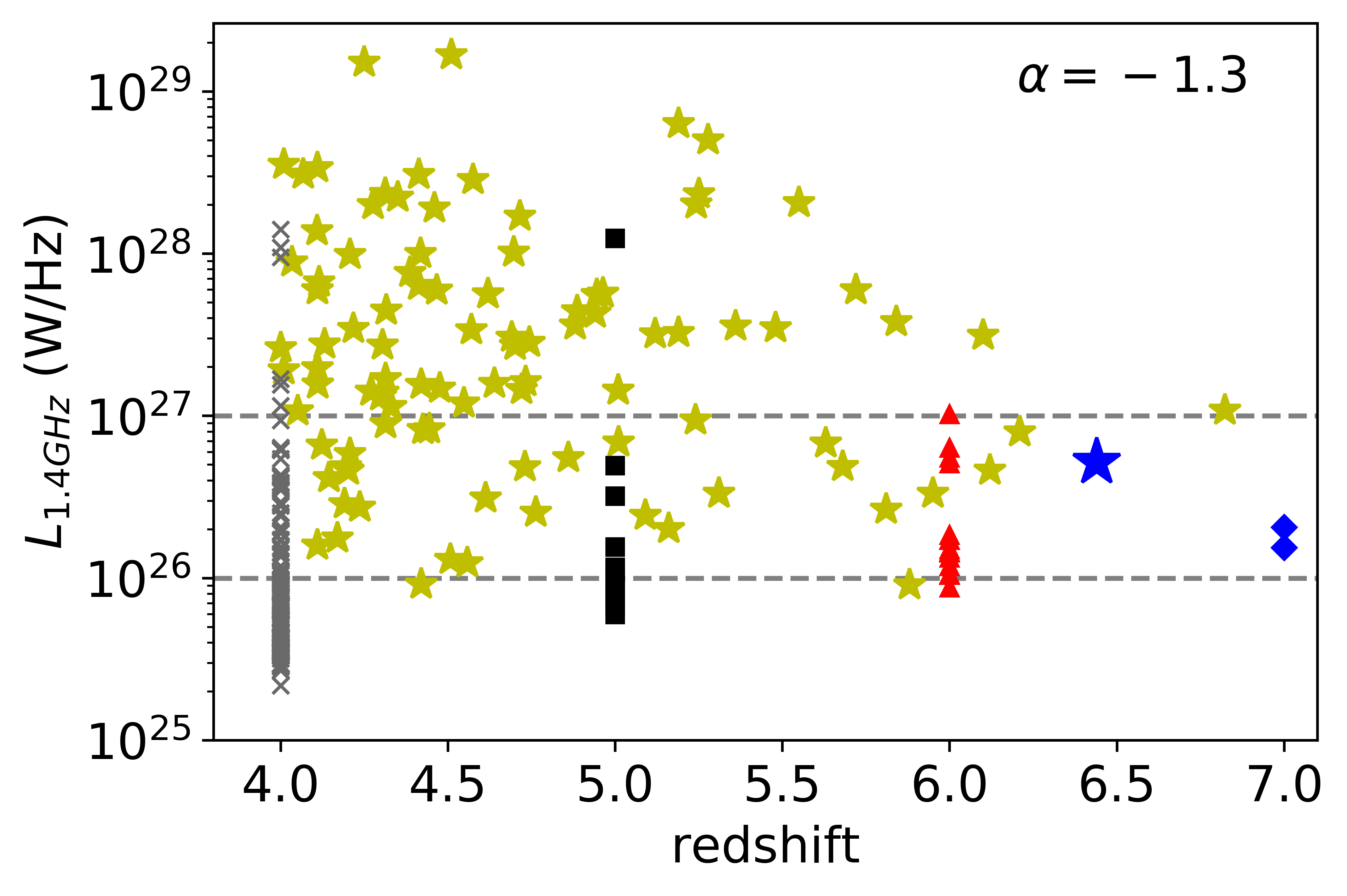} 
\end{subfigure}
\hspace{0.3cm}
\begin{subfigure}{.45\textwidth}
 \centering
 % include first image
 \includegraphics[width=\textwidth]{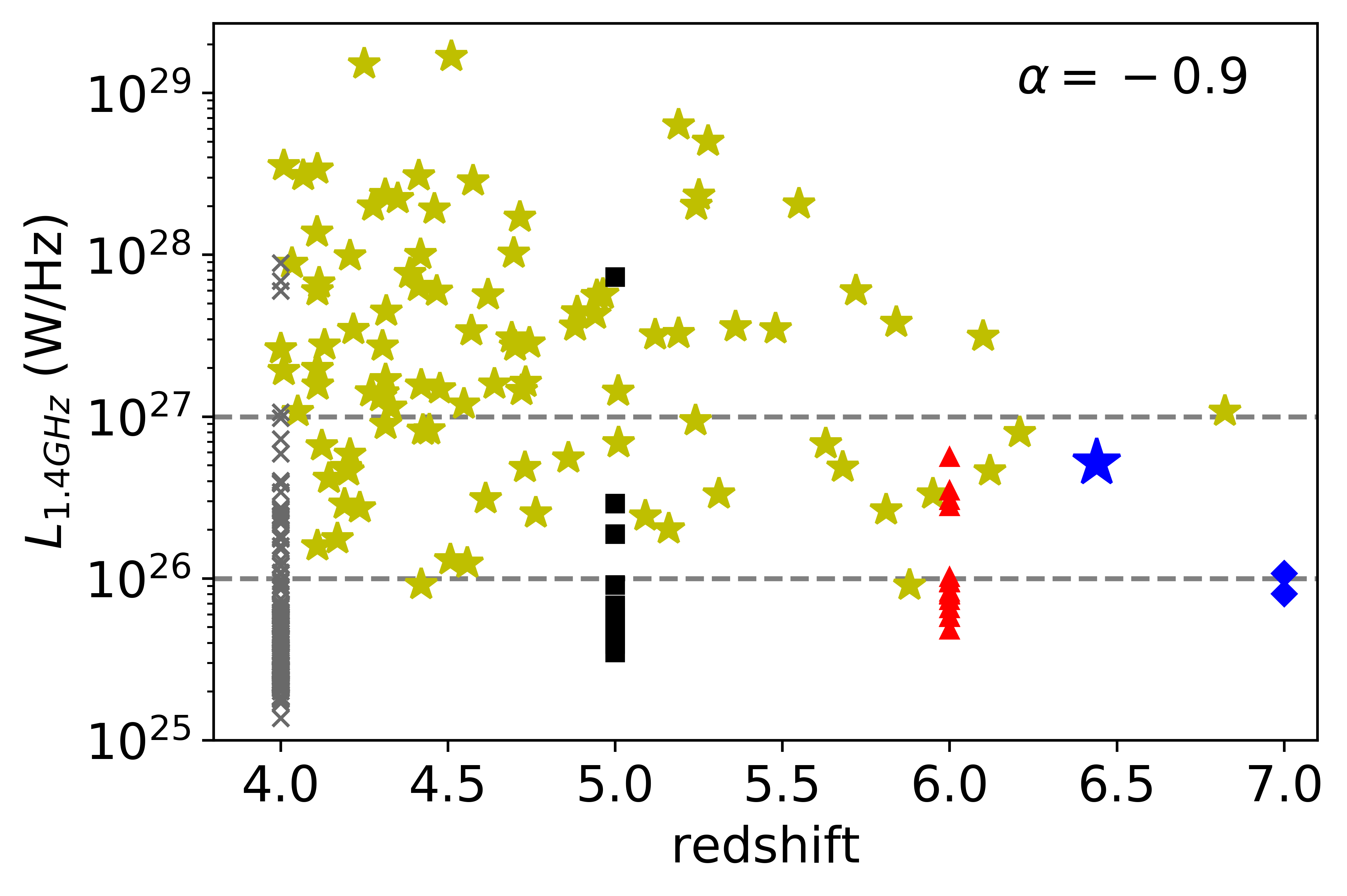} 
\end{subfigure}
\hspace{0.3cm}
\begin{subfigure}{.45\textwidth}
 \centering
 % include first image
 \includegraphics[width=\textwidth]{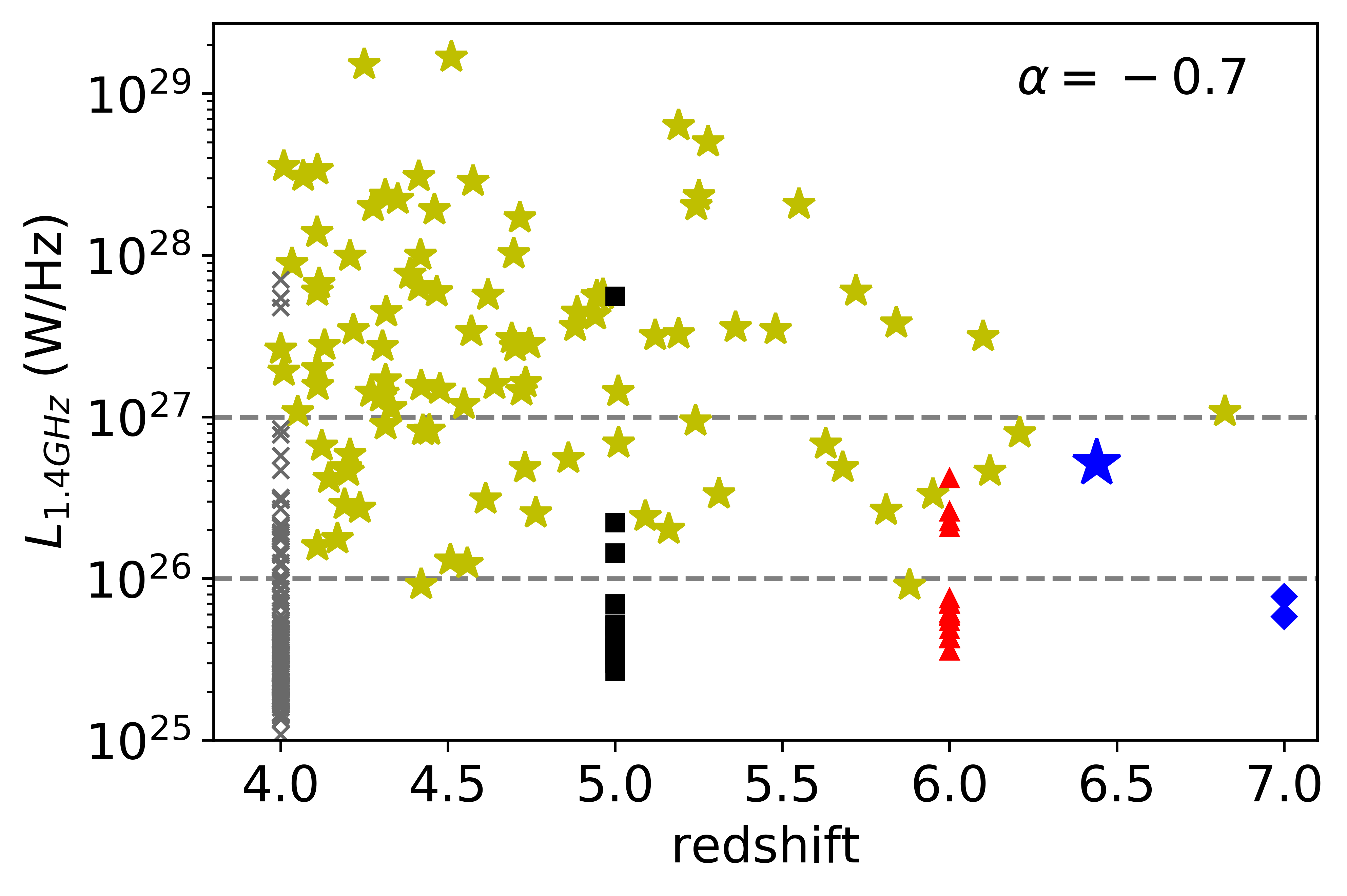} 
\end{subfigure}
\hspace{0.3cm}
\begin{subfigure}{.45\textwidth}
 \centering
 % include first image
 \includegraphics[width=\textwidth]{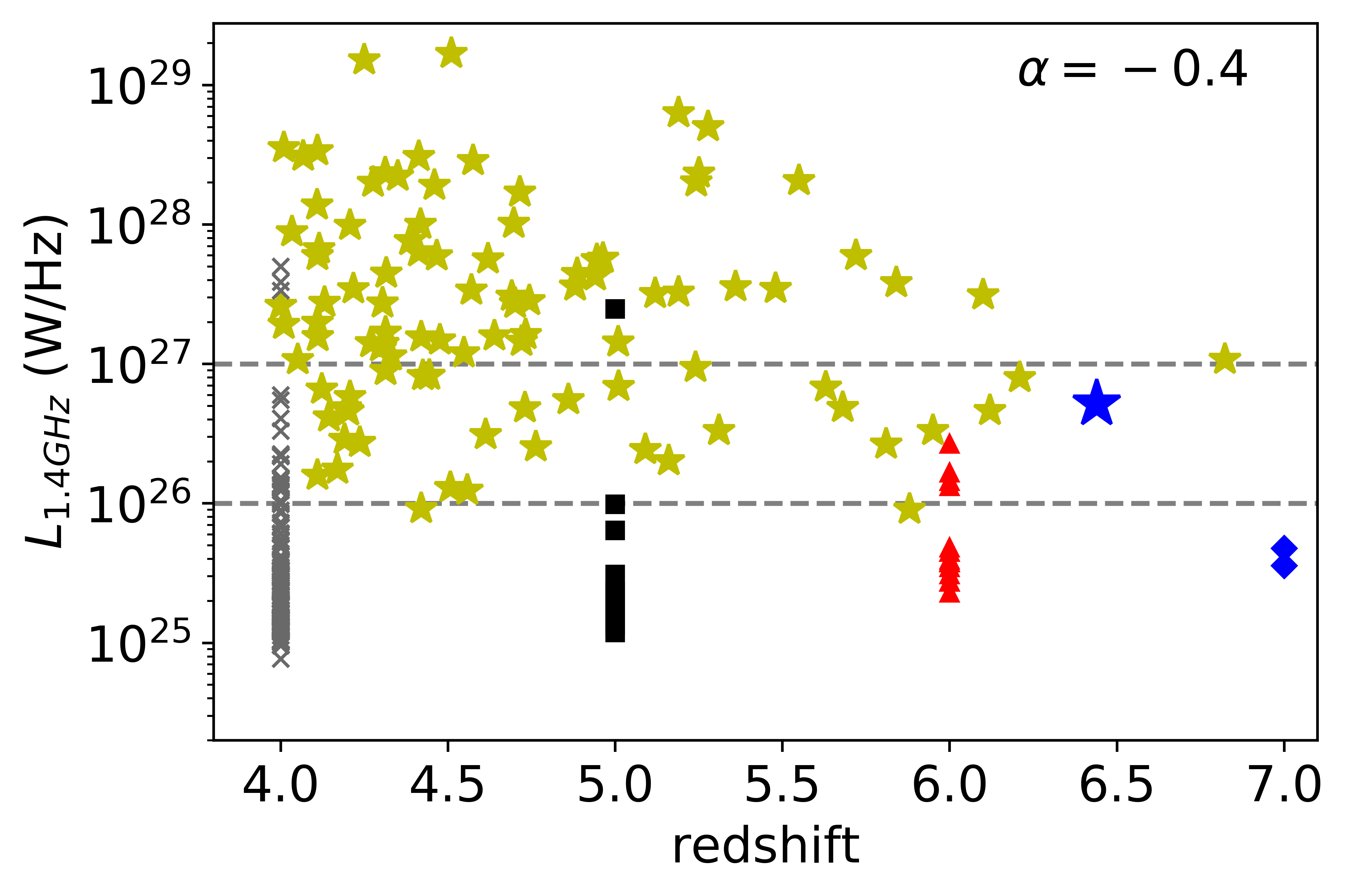} 
\end{subfigure}\\
\vspace{0.3cm}
\begin{subfigure}{.55\textwidth}
 \centering
 % include first image
 \includegraphics[width=\textwidth]{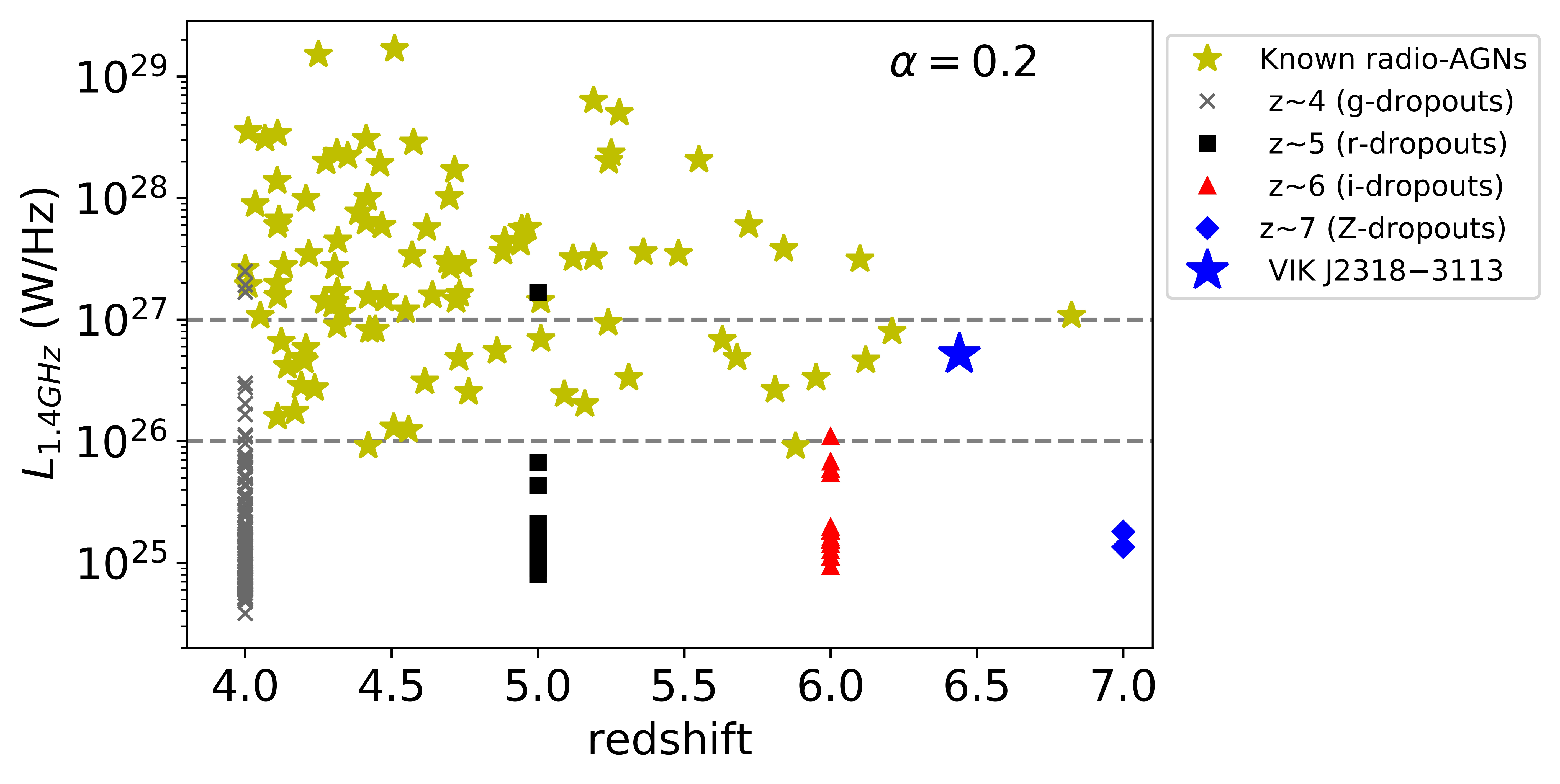} 
\end{subfigure}\vspace{-0.2cm}
\caption{ The 1.4\,GHz radio luminosity distribution of our sample estimated for a set of spectral indices and known radio-\ac{AGN}s (Appendix~\ref{A}, yellow stars) as a function of redshift. Irrespective of spectral index, our sample is 1-2 orders of magnitude less luminous than known \ac{AGN}s at the similar redshift. }
\label{fig:FD}
\end{figure*}

The observed 887.5\,MHz flux density distribution of our final sample is shown in Figure~\ref{fig:fluxD}. Our sample is selected from the G23 \ac{ASKAP} radio catalogue and no additional radio properties are used in the selection.
The distribution of flux densities of such a flux-limited sample is expected to peak at low flux densities, and this is seen in our \acp{HzRS} sample, which peaks at 0.2--0.4\,mJy.

 Although the $z \sim 5$ and $z \sim 6$ sample  are smaller, they also peak at low flux densities and follow a distribution similar to that of $z\sim4$. We  confirmed the statistical significance of this similarity by performing a Kolmogorov-Smirnov (KS) test, comparing the $z \sim 5$ and $z \sim 6$ samples with the $z \sim 4$ sample. Using the KS test, the probability that the $z \sim 5$, $z \sim 6$, and $z\sim7$ flux density distributions are drawn from the same population as the $z \sim 4$ sample is 0.44, 0.31, and 0.65. Thus, the test suggests that the $z \sim 4$, 5, 6, and 7 sources are most likely to represent a similar type of radio source population.

 %We further excluded the highest flux density points ($\ge5$\,mJy) in the $z\sim4$ sample and repeated the KS test to check whether the difference is solely caused by the lack of high flux density sources in the $z\sim6$ sample. This gives a probability of 0.002, confirming that the difference between the $z \sim 4$ and $z \sim 6$ samples is statistically significant. 

We calculate the rest-frame 1.4\,GHz luminosity of our sources from the total flux density observed at 887.5\,MHz, using Equation~\ref{eq:lum_1.4GHz}, by assuming a power-law ($S_\nu \propto \nu^\alpha$) radio spectral energy distribution (SED) and adopting a  set of radio spectral indices $\alpha =\{-1.3, -0.9, -0.7, -0.4, 0.2 \}$, as indicative of our complete \acp{HzRS} sample. We assumed redshift, $z=$ 4, 5, 6, and 7 for \textit{g, r, i,} and Z dropouts respectively in the Equation~\ref{eq:lum_1.4GHz}. The luminosity distance (D\textsubscript{L}) is calculated using the online cosmology calculator \citep{wright2006cosmology}. In other words:

\begin{align}
  L_{\nu_{1}} &= \frac{4\pi D_\textrm{L}^{2}}{(1+z)^{(1+\alpha)}} \left ( \frac{\nu_{1}}{\nu_{2}}\right)^{\alpha}S_{\nu_{2}} (\textrm {W/Hz}),
  \label{eq:lum_1.4GHz}
\end{align}
where L\textsubscript{$\nu$\textsubscript{1}} is the radio luminosity at rest-frame $\nu\textsubscript{1}$ derived from the observed flux density S\textsubscript{$\nu$\textsubscript{2}} at $\nu\textsubscript{2}$ and  (1+z)\textsuperscript{-($1+\alpha$)} denotes the standard radio K-correction.

 The resulting 1.4\,GHz luminosity of our sources for a given spectral index, shown in Figure~\ref{fig:FD},  as a function of redshift indicates that our sample probes less powerful \acp{HzRS} compared to known radio sources (listed in the Appendix) at the same redshift. This is expected because of the greater sensitivity of the \ac{ASKAP} survey. Very little is known about the properties of \acp{HzRS} with $L_{1.4} < 10^{26}$\,W/Hz as the known \acp{HzRS} are mostly brighter. 

We therefore used the following diagnostics to investigate the physical origin of radio emission in our sample: (i) radio luminosity at 1.4\,GHz, (ii) 1.4\,GHz-to-3.4\,$\mu$m flux density ratio, (iii) WISE colour, (iv) FIR detection at 250\,$\mu$m, and (v) SED modelling.

%============================**********************++++++++++++

\subsection{1.4~GHz luminosity diagnostic }

If we assume that all the radio emission is caused by star formation processes, we can calculate the \ac{SFR} following \citet{bell03},
 \begin{align}
 SFR(L\textsubscript{1.4})& = L\textsubscript{1.4} \times 5.52 \times 10^{-22} M\textsubscript{$\odot$}/yr.
 \label{eq:sfr}
\end{align}
We show the results in Table~\ref{tab:sfr}.

Among extragalactic objects, \acp{SMG} have the highest reported \ac{SFR}: $\sim 6000~$M\textsubscript{$\odot$}/yr \citep{barger2014there}. By comparison, most Lyman dropouts do not have as high \ac{SFR} as SMGs, with a maximum of $\sim 300~$M\textsubscript{$\odot$}/yr \citep{barger2014there}. Furthermore,  \citet[Figure~23]{barger2014there} demonstrated a turn-down in the observed \ac{SFR} distribution function beyond 2000\,M\textsubscript{$\odot$}/yr. In Table~\ref{tab:sfr}, we present the \acp{SFR} estimated  for each dropout at the following two flux densities at a given spectral index: (i) the minimum of 887.5\,MHz flux density distribution for each dropout (ii) flux density which marks the  exceeding of \ac{SFR} beyond the highest known limit (6000\,M\textsubscript{$\odot$}/yr); \cite{barger2014there}. In some cases, the minimum of 887.5\,MHz flux density distribution itself results in an unphysical SFR. 
Thus, most of the \ac{SFR} shown in Table~\ref{tab:sfr} exceed by far the highest-known \ac{SFR} of a galaxy \citep{barger2014there}, and we conclude that most of the radio emission from these galaxies is unlikely to be primarily generated by star formation,  implying the presence of a radio-\ac{AGN} as well.

This finding is also consistent with the simulation \citep{bonaldi2019tiered,wilman2008semi} of the redshift evolution of radio sources, which predicts that star forming galaxies (SFGs) are a negligible fraction of the observed radio sources beyond redshift $z > 2$ at the flux limit of 0.1\,mJy. 
  We acknowledge that \textit{radio}-\ac{SFR} of the faintest end ($S_{887}\le0.2$\,mJy; 5 sources) of our $z\sim4$ sample lie in the range $\sim$4000-8000\,M\textsubscript{$\odot$}/yr,  which is very high but in priciple possible given that estimated space density of \acp{SMG} with SFRs above 2000\,M\textsubscript{$\odot$}/yr is  $1.4\times10^{-5}$\,M\textsubscript{$\odot$}/yr/Mpc\textsuperscript{3} \citep{fu2013rapid}. Therefore, it is likely that faintest end of $z\sim4$ sample  have a \ac{SB} component or be pure \acp{SB}.

\begin{table}
 \centering
 \caption{ 
 \ac{SFR} estimated for different flux densities at $z \sim 4$, 5, 6 and 7, assuming all radio emission is powered by star formation. \ac{SFR} is calculated using Equations \ref {eq:lum_1.4GHz} at spectral index, $\alpha = -0.8$, -0.7, -0.6 given that mean spectral index of \acp{SFG} is $\sim0.75$ }
 \begin{tabular}{c|c|c|c|c|c}
 \hline
 Redshift & D\textsubscript{L} & S\textsubscript{887.5}& S\textsubscript{1.4}  &L\textsubscript{1.4} & SFR\textsubscript{1.4}\\
  $z$ & (10\textsuperscript{3}Mpc) & (mJy) & (mJy) &  (W~Hz$^{-1}$) & (10\textsuperscript{3}M\textsubscript{$\odot$}/yr)\\
  \hline 
 
  %%%%%%%%%%%%%%%%%%%%%%%%%%%%%%%%%%%%%%%%%

   \multicolumn{6}{c}{$\alpha=-0.8$} \\
 \hline
 4 & 36.6 & 0.15 & 0.10 &1.211$\times 10^{25}$ & 6.67 \\

 5 & 47.6   &0.2 &0.14 & 2.63$\times 10^{25}$ &14.55 \\

 6 & 58.98  & 0.2 &0.14 &3.92$\times 10^{25}$  &21.62 \\
 
7 & 70.54   & 0.2 &0.14 &5.45$\times 10^{25}$  &30.11 \\

  \hline
%%%%%%%%%%%%%%%%%%%%%%%%%%%%%%%%%%%%%%%%%%%%%%  
  
   \multicolumn{6}{c}{$\alpha=-0.7$} \\
 \hline
  \multirow{2}{*}{4} & \multirow{2}{*}{36.6} & 0.15 & 0.11 &1.07$\times 10^{25}$ & 5.95 \\
       &       & 0.2 & 0.15  &1.44$\times 10^{25}$ & 7.93\\
\hline
  5 & 47.6  &0.2 &0.15 & 2.31$\times 10^{25}$ &12.73 \\

  6 & 58.98  & 0.2 &0.15 &3.37$\times 10^{25}$  &18.63 \\
 
7 & 70.54  & 0.2 &0.15 &4.64$\times 10^{25}$  &25.53 \\
  \hline
  
%%%%%%%%%%%%%%%%%%%%%%%%%%%%%%%%%%%%%%%%%%%%%%%  
   \multicolumn{6}{c}{$\alpha=-0.6$} \\
 \hline
  \multirow{2}{*}{4} & \multirow{2}{*}{36.6} & 0.15 & 0.11 &6.9$\times 10^{24}$ & 3.84 \\
       &       & 0.24 & 0.18  &1.11$\times 10^{25}$ & 6.15\\
\hline

 5 & 47.6   &0.2 &0.15 & 1.41$\times 10^{25}$ &7.78 \\

 6 & 58.98  & 0.2 &0.15 &1.97$\times 10^{25}$  &10.87 \\
 
7 & 70.54   & 0.2 &0.15 &2.6$\times 10^{25}$  &14.36 \\

  \hline
 \end{tabular}
 \label{tab:sfr}
\end{table}

%============================================================%
\subsection{Radio-to-3.4\,$\mu$m flux density ratio}
\label{sec:NIR}

\citet[][Figure 4]{norris2011deep} presented a new measure for radio-loudness of a source, 1.4\,GHz-to-3.6\,$\mu$m flux density ratios. They compared the 1.4\,GHz-to-3.6\,$\mu$m flux density ratios for different galaxy populations  namely, \acp{HzRG}, \acp{SMG}, \acp{SB}, RL \& RQ quasars and \acp{ULIRG}, as a function of redshift.  They showed that galaxies powered primarily by star forming activity have a lower ratio in contrast to radio-AGNs at all redshifts. 
 We use 3.4\,$\mu$m photometry from CATWISE2020 catalogue \citep{marocco2021catwise2020} since there is no substantial difference between 3.4 \& 3.6\,$\mu$m passbands, and 3.4\,$\mu$m magnitude (W1) is converted to flux density \citep{cutri2012explanatory} using,
\begin{align}
  S\textsubscript{3.4\,$\mu$m} = 306.682 \times 10^{-W1/2.5} ~ \textrm{Jy.}
\end{align}.

\begin{figure*}
\begin{subfigure}{.32\textwidth}
 \centering
 % include first image
 \includegraphics[width=\textwidth]{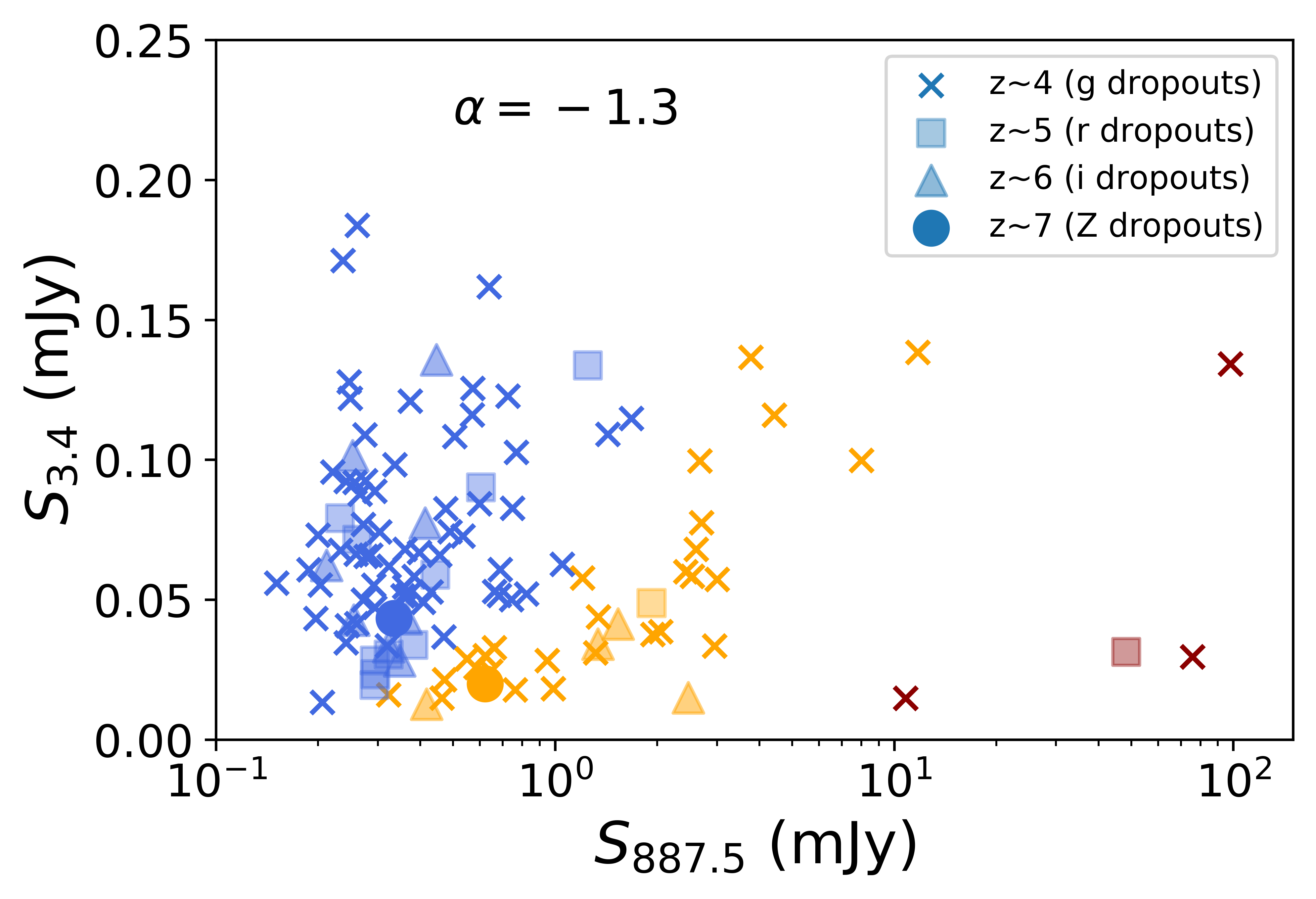} 
\end{subfigure}
\hspace{0.1cm}
\begin{subfigure}{.32\textwidth}
 \centering
 % include first image
 \includegraphics[width=\textwidth]{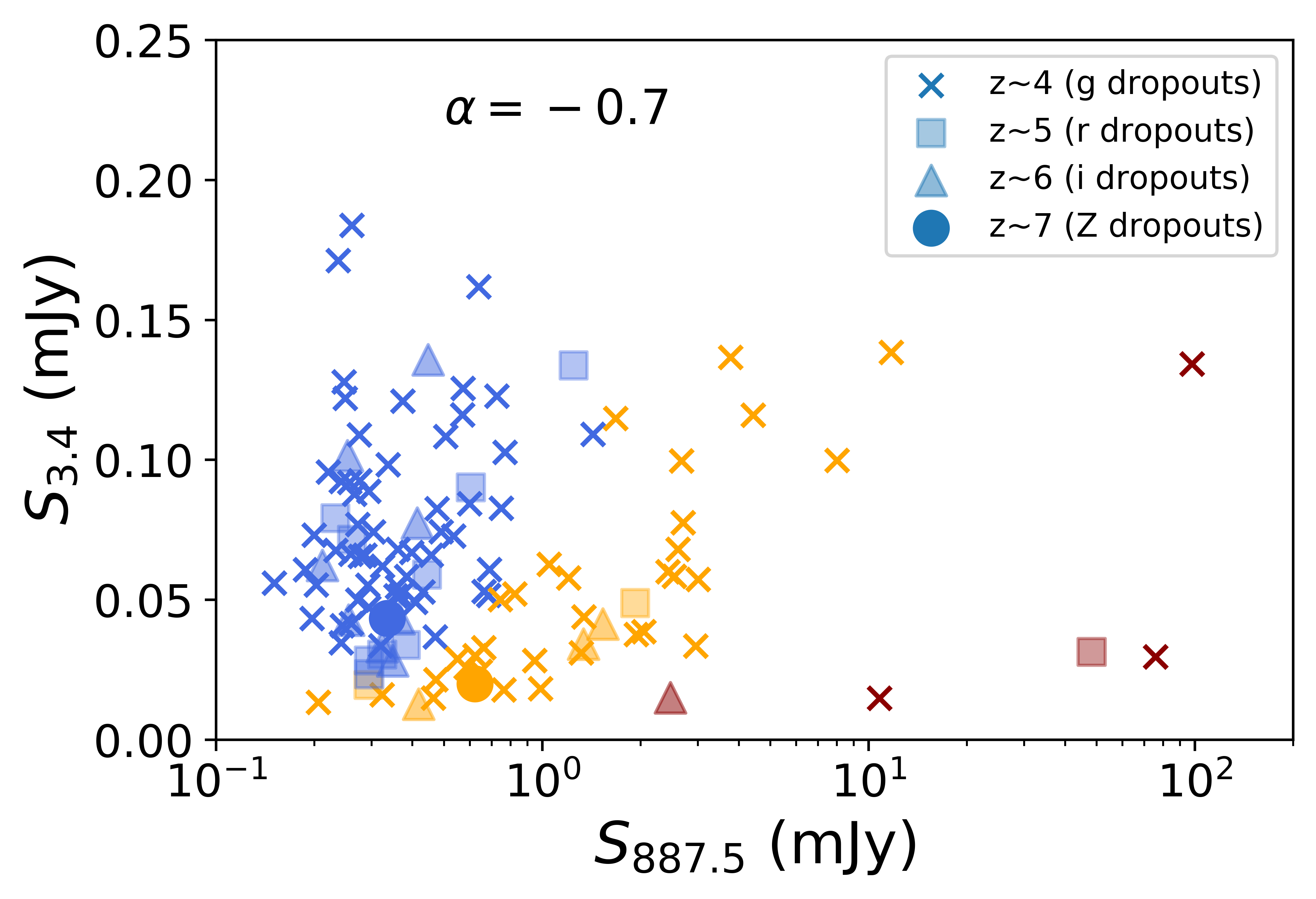} 
\end{subfigure}
\hspace{0.1cm}
\begin{subfigure}{.32\textwidth}
 \centering
 % include first image
 \includegraphics[width=\textwidth]{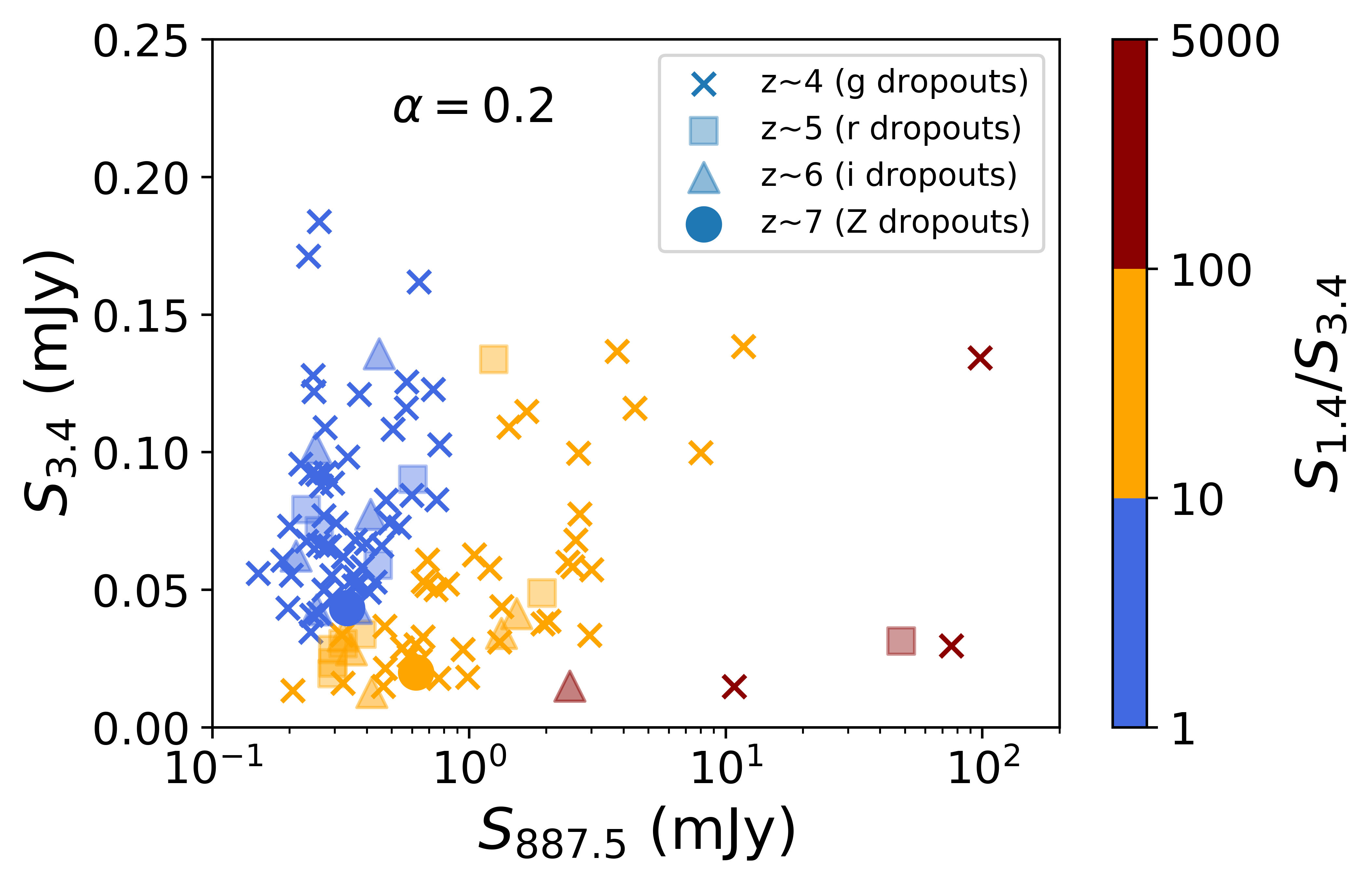} 
\end{subfigure}

 \caption{ 887.5\,MHz flux density vs. 3.4\,$\mu$m flux density for $z\gtrsim4$ sample, colour coded according to their 1.4\,GHz-to-3.4\,$\mu$m flux density ratio. 1.4\,GHz flux density is estimated from observed flux density at 887.5\,MHz assuming spectral indices, $\alpha = -1.3$, -- 0.7, 0.2.  }
 \label{fig:radio_ir_ratio}
\end{figure*}

We compare \cite{norris2011deep} model with the 1.4\,GHz-to-3.4\,$\mu$m flux density ratio of our sample at redshifts, $ z \sim 4$, 5, and 6 in Figure~\ref{fig:radio_ir_ratio}.  The measured flux density at 887.5\,MHz was used to estimate the 1.4\,GHz flux density, assuming spectral indices,  $\alpha$ of -1.3, -0.7, and 0.2 as indicative of our entire \ac{HzRS} sample.
According to that model, radio-\acp{AGN} have a radio-to-3.6\,$\mu$m flux density ratio above 100 at all redshifts. However, \cite{maini2016infrared} showed that the ratio for Type 1 \& 2 \ac{AGN}/QSO and high power radio galaxies start to decrease (between 10 and 100) beyond $z\sim4$ and $z\sim5$ . We contend that reason for this difference is that the \cite{norris2011deep} model refers to \emph{powerful} radio-\ac{AGN}s only, at all redshifts.  About $55\%$ of our sample have a 1.4\,GHz-to-3.4\,$\mu$m flux density ratio between 1 and 10 and $\sim41\%$ between 10 and 100. Only 5 sources in our sample have a radio-to-3.4\,$\mu$m ratio greater than 100.

It therefore seems that high-$z$ radio-\ac{AGN}s can extend to even lower values ($<10$) of radio/IR ratios than previously thought.  The following are some possible explanations for their low ratio: (i) radio-\ac{AGN} is accompanied by SB activity, and the optical emission being redshifted to W1 band causing an increase in the 3.4\,$\mu$m flux density (ii) the inherently low radio powers of high-$z$ radio-\acp{AGN}, as predicted by simulation in \cite{saxena2017modelling} (iii) optical emission from un-obscured quasars at $z\ge4$ being redshifted to W1 band. We discuss this further in Section \ref{sec:FIR}.

%%%%%%%%%%%%%%%%%%%%-----------------------------------%%%%%%%%%%%%%%%%%%%

\subsection{FIR detection}
Dust heated by stars or \ac{AGN} generally re-radiates the absorbed energy at FIR wavelengths. Each galaxy will have a distribution of dust temperatures which is determined by the size and distribution of dust with respect to the heating source (\ac{SF} or \ac{AGN}). Cool ($T\sim20\,K$) dust arises from the diffuse interstellar medium (ISM), warm ($T\sim40\,K$) dust from SF regions, and even warmer $T\sim60-100\,K$ dust results from \ac{AGN} activity. 

We queried the {\it Herschel} database to check whether any of our sources have been detected at FIR wavelengths. We searched the SPIRE point source catalogues at 250\,$\mu$m, 350\,$\mu$m, and 500\,$\mu$m \citep{schulz2017spire} separately in a 5~arcsec cone around the WISE position of each of our HzRS sample,  which yielded a false-id rate of $\sim16\%$. Several of our sources have a SPIRE counterpart, but none was found in the PACS point source catalogues.  A total of 14 and 3 sources from our $z\sim4$ and $z\sim6$ sample respectively were found to have a SPIRE counterpart, representing only 10\% of our entire \ac{HzRS} sample.
 
We present the results in Table~\ref{tab:spire_sources}.  The SPIRE 250\,$\mu$m band probes a rest wavelength $\sim$45 - 63\,$\mu$m  at $z=3-4.3$ ($g$ dropouts) and $\sim$34 - 38\,$\mu$m  at $z=5.6-6.2$ ($i$ dropouts). At wavelength $<50$\,$\mu$m in the rest-frame, an \ac{AGN} heated dust torus 
%
%ray emission is not negligible.
may contribute significantly to the FIR emission.
Therefore the SPIRE detection of our radio sources suggests that the observed 250\,$\mu$m emission may arise from either (i) a dust torus heated by an \ac{AGN} or (ii) \ac{SB} activity or (iii) a combination of both \ac{SF} and \ac{AGN}.

\begin{table}
    \centering
     \caption{Our \acp{HzRS} with a FIR detection in either of {\it Herschel}-SPIRE 250\,$\mu$m, 350\,$\mu$m $\&$ 500\,$\mu$m bands, selected from a 5~arcsec search around their WISE position at which false-ID rate is $\sim16\%$.  The probability of a source's measured SPIRE flux being contaminated by a nearby neighbour is about 1/14.}
    \begin{tabular}{c|c|c|c}
    \hline
        Source ID & S\textsubscript{250} & S\textsubscript{350} & S\textsubscript{500}  \\
         &  (mJy) & (mJy) & (mJy)   \\
        
        \hline
\multicolumn{4}{c}{$z\sim4$ candidates}        \\
\hline

J224308-313622 & 59.3 &	 &  \\
J223533-343305 & 67.4 & 40.7 &  --    \\
J231210-332436 & 49.9 &	32.3 &   --   \\
J224203-333606 & 59.5 &	68.5 &  --    \\
J231826-323746 & 39.9 & -- & -- \\
J223729-325838 & 34.8 & -- & -- \\
J224017-325128 & 66.2 & -- & -- \\
J224552-312733 & 51.2 &--   &--   \\
J225343-313305 & 65.4     & --  & --  \\
J231648-303629 & 65.6     & --  &--   \\
J232019-294205 & 93.6     &77.2   &--   \\
J232135-320117 & 44.5     & --   &   \\
J225915-314116 & --& 78.1 & 62.9 \\
J224955-294536 & -- & -- & 49 \\

\hline
\multicolumn{4}{c}{$z\sim6$ candidates}  \\
\hline 
J224652-340238 & 92.9 & 62.4 & 47.9 \\
J231245-291817 & 98.5 & 86.8 & 48.9 \\
J223719-333857 & 65.6 & 52.5 & -- \\
\hline
    \end{tabular}
    \label{tab:spire_sources}
\end{table}

%%%%%%%%%%%%%%%%%%%%%%%%%%%%%%%%%%%%%%%%%%%%%%%%%%%%%%%%%%%%%%%%%%%%%%%%%%
\subsubsection{Radio-FIR relation}

In this section, we explore the 250\,$\mu$m-to-1.4\,GHz luminosity ratio of our sources, as defined by Equation~\ref{eq:q250},  which is a tracer of star formation activity. Here we are only interested in order-of-magnitude estimates and hence we neglect evolution. Another paper in which we model evolutionary effects is in preparation. 
$q\textsubscript{250}$ is defined as:
\begin{align}
  q_{250} = \log_{10}\left (\frac{L_{250}}{L_{1.4}}\right ) ~,
  \label{eq:q250}
\end{align}
where $L\textsubscript{250}$ and $L\textsubscript{1.4}$ are the rest-frame luminosities at 250\,$\mu$n and 1.4\,GHz respectively.

To estimate rest-frame $L\textsubscript{250}$, we must first estimate the 
K-correction (k\textsubscript{corr}) defined as 
\begin{align}
  k_{corr} = \frac{S_{rest}}{S_{obs}} ~,
  \label{eq:k_corr}
\end{align}
where $S_{\rm obs}$ and $S_{\rm rest}$ are the observed and the rest-frame flux densities respectively. 

Assuming a greybody thermal emission, the spectral flux density at a given frequency and temperature, $S_{\nu}(T)$, would be a modified Planck's radiation law $B(\nu,T)$,
\begin{align}
 S_{\nu}(T) &\propto \nu^{\beta}B(\nu,T) ~, \label{eq:S_nu}\\  \intertext{where}
 B(\nu,T) &= \frac{2h}{c^{2}}\frac{\nu^{3}}{e^{\frac{h\nu}{k_{B}T}}-1}
 \label{eq:planck}
\end{align}
 and $\beta$ represents the dust emissivity index, for which we assume a value of  1.5 \citep{kirkpatrick2015role}.

We calculate the K-correction using Equations~\ref{eq:k_corr}, \ref{eq:S_nu} and \ref{eq:planck}, assuming 2 dust temperatures, 45\, K, $\&$ 80\,K, corresponding to warm and warmer dust components  based on \cite{kirkpatrick2015role} study, and present the results in Table~\ref{tab:k_corr}.
\begin{table}
 \centering
 \caption{K-correction estimated for the observed frame at 250\,$\mu$m,  assuming the dust emission either dominated by a warmer dust ($T_{w} \sim$ 80\,K) or a warm dust ($T_{w} \sim$ 45\,K).}
 %or a cold dust ($T_{c} \sim$ 20\,K)}.}
 \begin{tabular}{c|c|c|c|c}
 \hline
  redshift & $\nu$\textsubscript{obs} & $\lambda$\textsubscript{rest} & $\nu$\textsubscript{rest}  & k\textsubscript{corr} \\
  ($z$) & (Hz) & ($\mu$m) & (Hz)  & \\
\hline
  \multicolumn{5}{c}{$T_{dust} \sim 45\,K$} \\
  \hline
   4 &\multirow{3}{*}{$1.2\times10^{12}$} &50 & $5.9\times10^{12}$ &  6.22 \\
  5 & & 42 & $7.2\times10^{12}$  & 3.8 \\
  6 & &36 & $8.4\times10^{12}$  &  2.11\\
  \hline
  \multicolumn{5}{c}{$T_{dust} \sim 80\,K$} \\
  \hline
  4 &\multirow{3}{*}{$1.2\times10^{12}$} &50 & $5.9\times10^{12}$ &  40.77 \\
  5 & & 42 & $7.2\times10^{12}$  & 44.93 \\
  6 & &36 & $8.4\times10^{12}$  & 43.46 \\
  \hline
 \end{tabular}
 \label{tab:k_corr}
\end{table}

We then calculate the luminosity $L_{250}$,
\begin{align}
  L_{250} = \frac {4\pi S_{250}K_{cor} D_\textrm{L}^{2}}{1+z} ~, 
  \label{eq:L250}
\end{align}
where $S_{250}$ is the observed flux densities of our sources in the SPIRE database and D\textsubscript{L} is the luminosity distance.

\begin{figure*}
\begin{subfigure}{.45\textwidth}
 \centering
 % include first image
 \includegraphics[width=\textwidth]{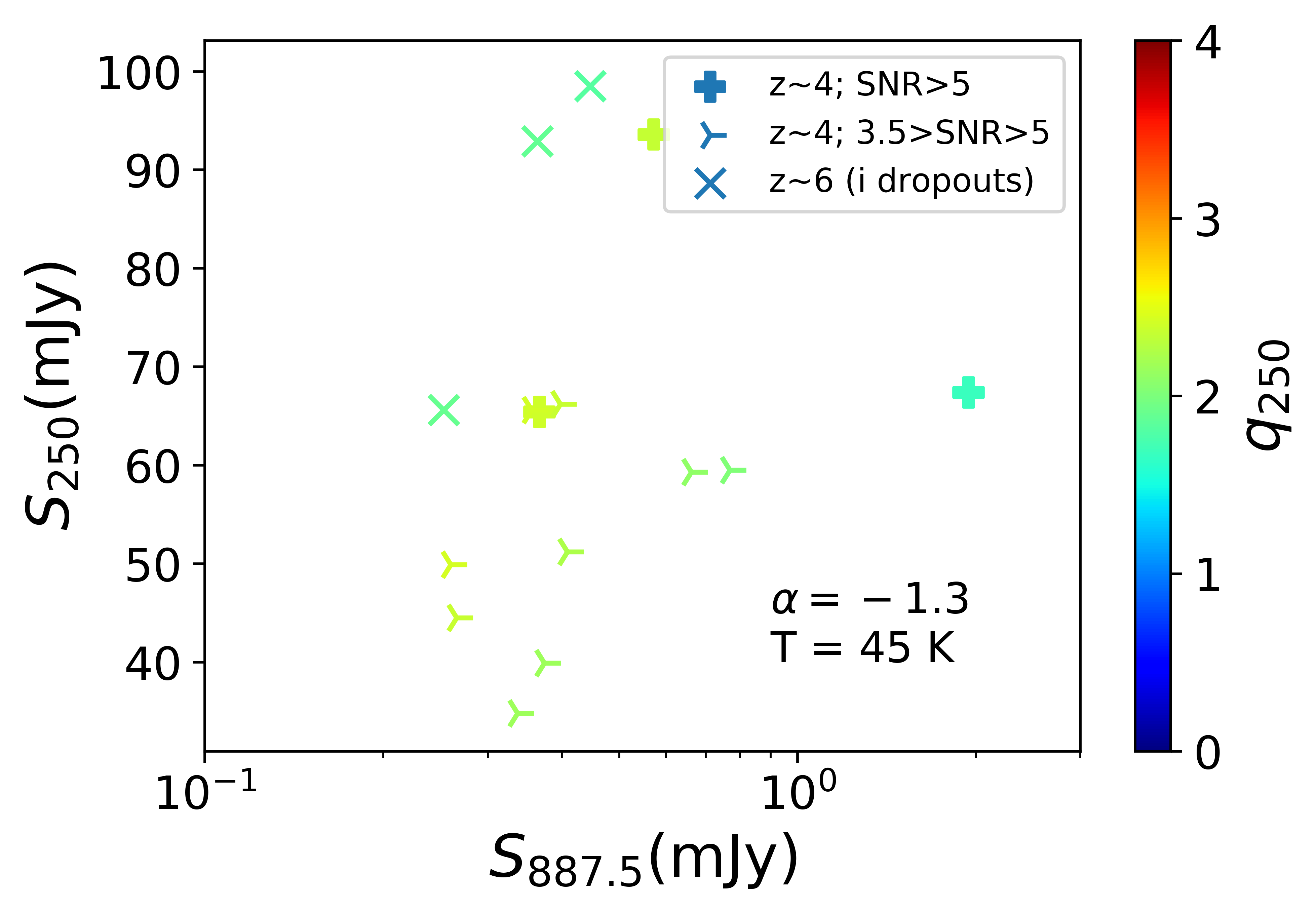} 
\end{subfigure}
\hspace{0.3cm}
\begin{subfigure}{.45\textwidth}
 \centering
 % include first image
 \includegraphics[width=\textwidth]{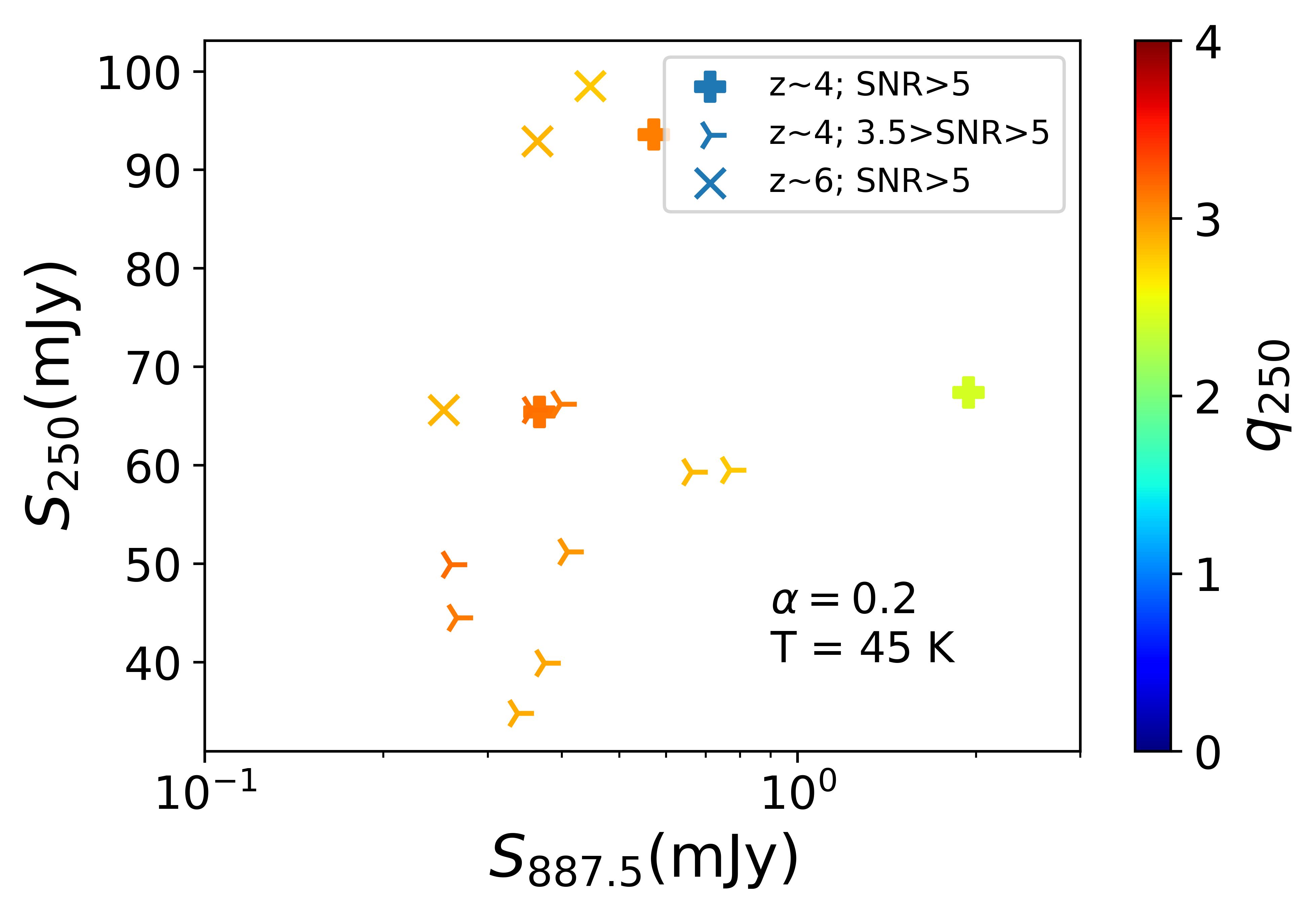} 
\end{subfigure}
\begin{subfigure}{.45\textwidth}
 \centering
 % include first image
 \includegraphics[width=\textwidth]{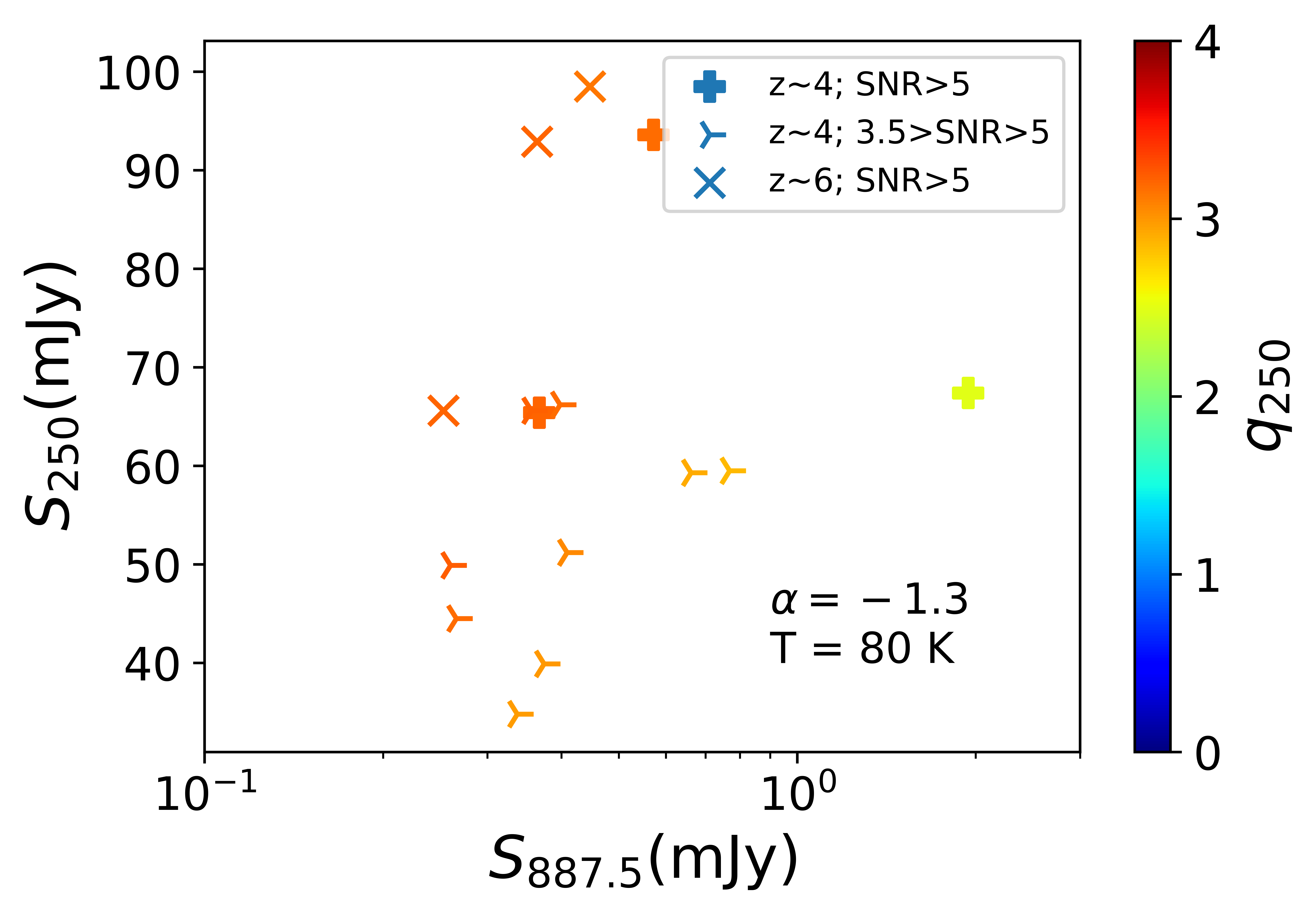} 
\end{subfigure}
\begin{subfigure}{.45\textwidth}
 \centering
 % include first image
 \includegraphics[width=\textwidth]{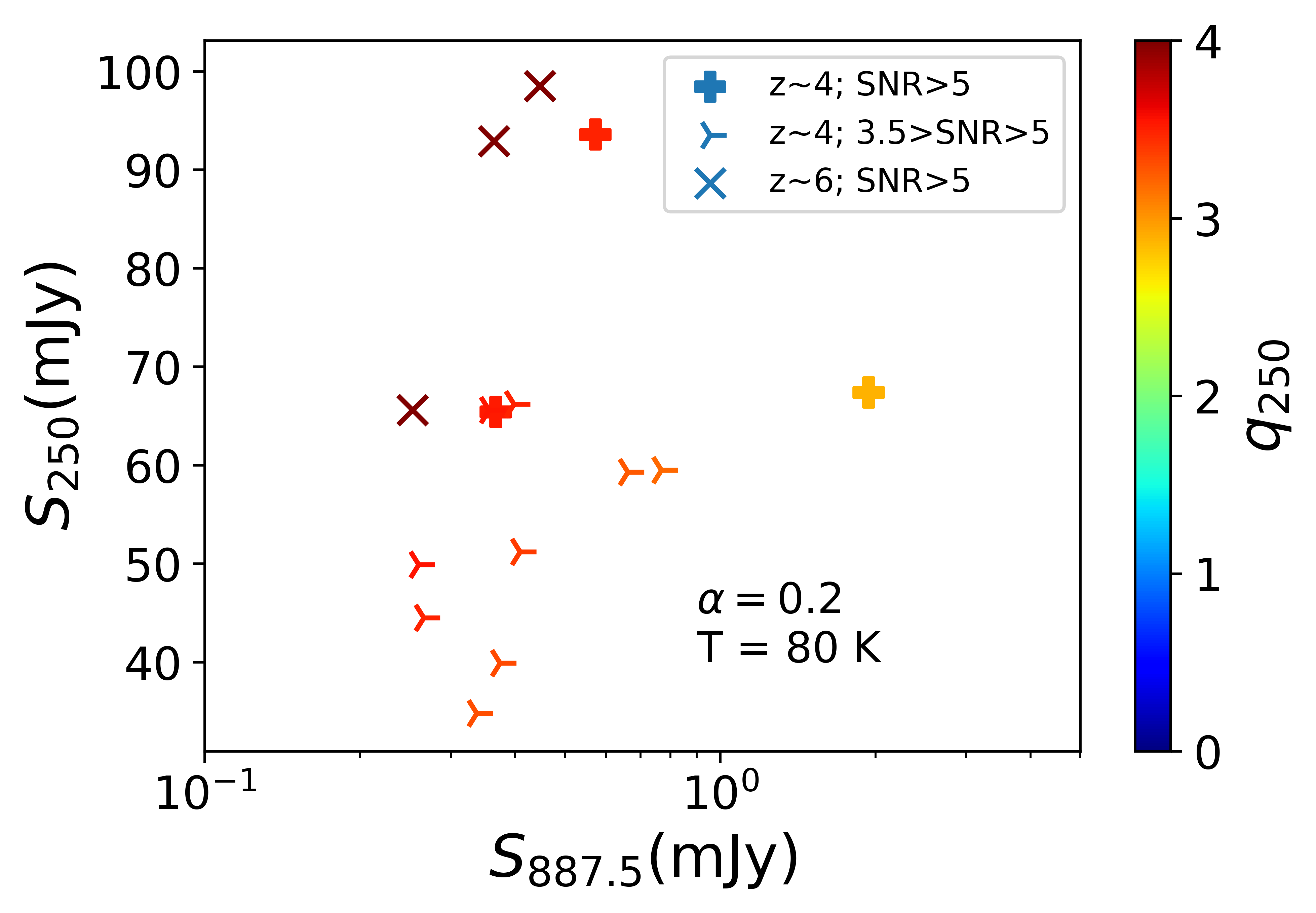} 
\end{subfigure}
 \caption{ 887.5\,MHz flux density vs. 250\,$\mu$m flux density for $z\gtrsim4$ sample, colour coded according to their q\textsubscript{250} measurements.  }
 \label{fig:q250}
\end{figure*}

 The resulting $q\textsubscript{250}$ estimated for two extremes of assumed radio spectral indices (see Section~\ref{sec:radio_f_de}) is shown in Figure~\ref{fig:q250}. All sources with a reliable detection as indicated by SNR>5 and 3>SNR>5 have q$_{250}$ exceeding the threshold of 1.3 \citep{jarvis2010herschel, virdee2013herschel} indicating luminous SFG.   The high values of $q\textsubscript{250}$,  irrespective of radio spectral index, suggest the presence of an active \ac{SB} component in SPIRE detected sources but on the other hand our K-correction is very uncertain because of our lack of knowledge of the FIR SED, and so we do not consider this a definitive indicator of SF. Furthermore, SPIRE detected sources have $S_{887.5} > 0.2$\,mJy implying the presence of a radio-\ac{AGN} component according to Table~\ref{tab:sfr}.   Therefore, we attempt to estimate \ac{SFR} (see Section~\ref{sec:FIR}), from the observed 250\,$\mu$m emission to verify whether the resulting SFR is sufficient enough to generate the observed radio luminosity. We also note  that q\textsubscript{250} criterion is based on the study of powerful radio-\acp{AGN} in the literature, whereas this study probe low power radio-\acp{AGN}, suggesting q$_{250}>1.3$ may not be applicable in this case.

%%%%%%%%%%%%%%%%%%%%%%%%%%%%%%%%%%%%%%%%%%%%%%%%%%%%%%%%%%%%%%%
\subsubsection{\ac{SFR} from S\textsubscript{250}}
\label{sec:FIR}

We estimate the SFR from S\textsubscript{250} via the rest-frame 24\,$\mu$m luminosity, $L_{24\,\mu m}$ (in erg/s), using the following equations from \citet{brown2017calibration},
\begin{align}
 \log L_{24_{\mu m}} &= 40.93+1.3(\log L_{H\alpha} - 40)
 \label{eq:Halpha}\\
 SFR &= L_{H\alpha}\times5.5\times10^{-42}~\rm{M_\odot /yr}~.
  \label{eq:sfr_ha}
\end{align}

For the calculation of $\text{L}_{24\,\mu m}$, we consider 4 different spectral energy distribution (SED) models corresponding to: (i) an ultra luminous infrared galaxy (Arp~220), (ii) a starburst (M\,82),  (iii) Mrk\,231 (luminous infrared \ac{AGN}), and (iv) a composite system (IRAS F00183--7111), in which the optical/IR SED is dominated by the starburst surrounding the \ac{AGN}, with the \ac{AGN} emerging only at radio wavelengths \citep{norris12}. In each case, we interpolate the SED using flux density measurements obtained via NASA/IPAC Extragalactic Database (NED).

We perform the following calculations to estimate $\text{L}_{24\,\mu m}$ of our sources:

\begin{enumerate}
 \item We calculate the luminosity of our sources corresponding to their emitted wavelengths from their observed flux density at 250\,$\mu$m using the equation,
 
 \( L_{\lambda} = \frac{4\pi D_\textrm{L}^{2}S_{250}}{1+z}\), \\
 where $\lambda$ represents the respective emitted (or rest-frame) wavelength at $z\sim4,5,6$ as shown in column 3 of  Table~\ref{tab:k_corr}.
 \item For each model source (Arp~220, M\,82, Mrk\,231, IRAS F00183--7111) we assume that the ratio $L_{24\mu m}/L_{\lambda}$ for our sources is the same as that of the model source as measured from their SED:

  \(\left(\frac{L_{24\mu m}}{L_{\lambda}}\right)_{our source}\) =
  \(\left(\frac{S_{24\mu m}}{S_{\lambda}}\right)_{model}\).
 
\end{enumerate}

\begin{figure*}
    \centering
    \includegraphics[trim={15 210 15 0},width=\textwidth]{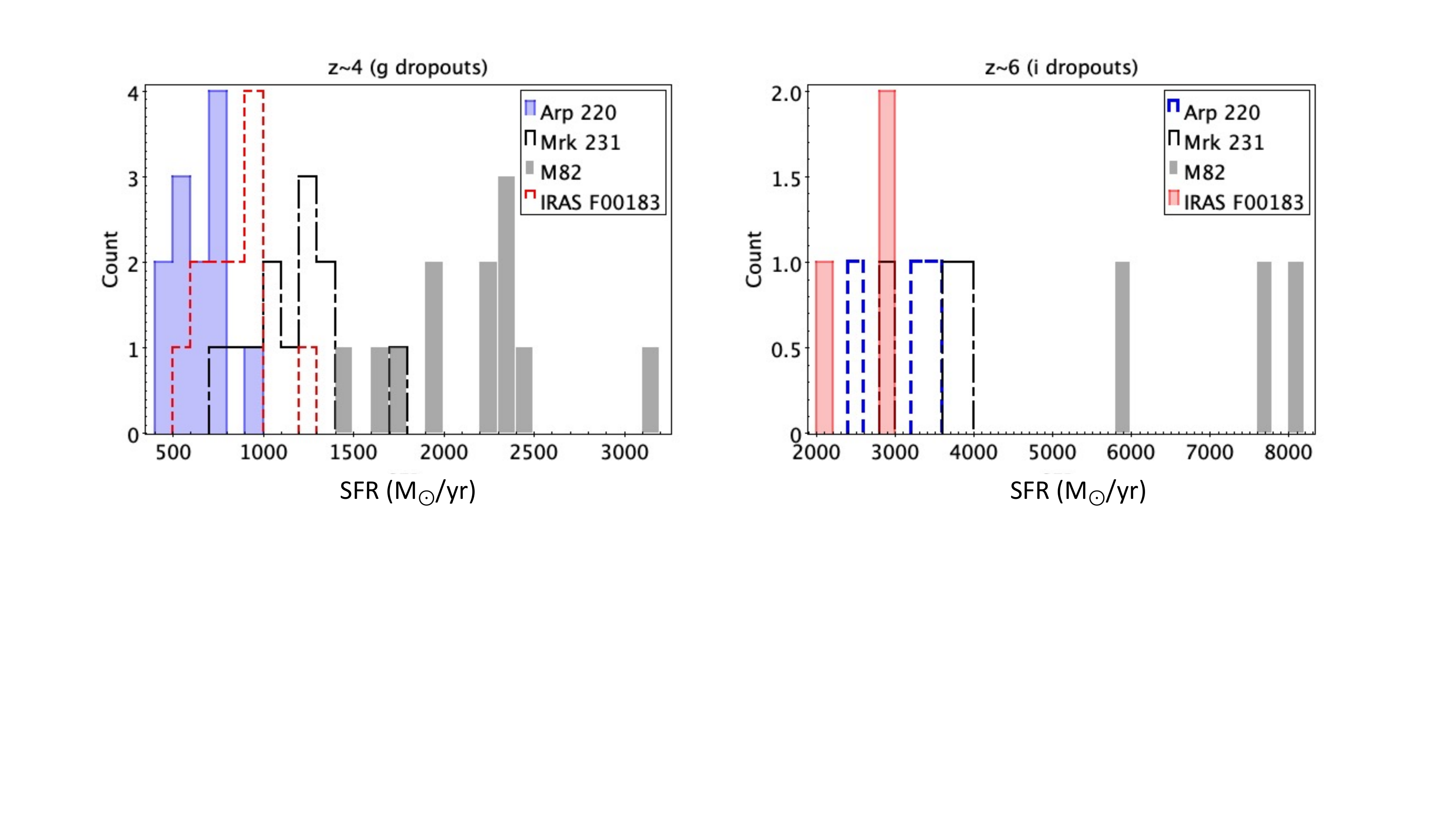}
    \caption{ \ac{SFR} estimated from S\textsubscript{250} by assuming various SED models.}
    \label{fig:sfr}
\end{figure*}
We present the results of this calculation in Figure~\ref{fig:sfr}. The resulting \ac{SFR}s  for $z\sim4$ \& $z\sim6$ samples lie in the range $\sim 400-3500$\, M\textsubscript{$\odot$}/yr \& $\sim 2000-8200$\, M\textsubscript{$\odot$}/yr respectively, which is high but not unphysical, and similar to the \ac{SFR}s reported for \ac{SB}s and SMGs at similar redshifts \citep{barger2014there}. These SFRs would generate a radio power in the order of  $\sim 10^{24}-10^{25}$~W~Hz$^{-1}$, which is typically a small fraction ($\le 10\%$) of the observed radio luminosities.

 We conclude that these galaxies detected by \textit{Herschel}-SPIRE are likely to be composite galaxies containing both a radio-\ac{AGN} and a starburst. We further note that the radio-to-3.4\,$\mu$m flux density ratio of SPIRE detected sources is $<100$. 7 out of 15 250\,$\mu$m detected sources found to have $W1-W2 > 0.8$ WISE colour, suggesting an \ac{AGN} component. At $z\gtrsim4$, W1 WISE band observations probe the optical rest-frame including H$\alpha$ emission, an indicator of SF activity or  quasar light or both. This suggests that a fraction of  high-$z$ radio-\ac{AGN}s ($\log L\textsubscript{1.4}=25$) at $z\gtrsim4$ may have a \ac{SB} host, contrary to low-z radio-\ac{AGN}s which are usually hosted by quiescent galaxies. Similarly, \cite{rees2016radio} 
%ray study on sample of radio-\ac{AGN}s out to redshift $z=2.25$ demonstrated 
found that $z > 1.5$ radio-\ac{AGN}s tend to be hosted by SFGs. Our study suggests that this may be true at even higher redshifts.

%%%%%%%%%%%%%%%%%%%%%%%%%%%%%%%%%%%%%%%%%%%%%%%%%%%%%%%%%%%%%%%%%%%%%%%%%%%
\subsection{WISE colours}
\label{sec:wise_colours}

Mid-IR colour selection criteria are an efficient tool to identify a hot accretion disk, which is an indicator of \ac{AGN} in galaxies. The dust reprocesses the emission from the accretion disk into the IR, which dominates the \ac{AGN} SED at wavelengths from $\sim1$\,$\mu$m to a few tens of microns, the wavelength regime covered by the WISE survey. 

\cite{stern2012mid} demonstrated that a simple WISE colour cut, $W1-W2>0.8$, selects \ac{AGN} with a reliability of 95\% and with a completeness of 78\%. However, this colour cut works efficiently for low-z sources ($z\sim1$) only, given that is defined using observed fluxes not rest-frame ones. \cite{stern2012mid} therefore discussed a number of possible scenarios to interpret W1-W2 colour at higher redshifts.

Following are the possibilities, taken from \cite{stern2012mid}, applicable to our $z\gtrsim4$ sample, 
\begin{enumerate}
 \item At $z\gtrsim3.5$, W1/W2 bands probe optical \& near-IR rest-frame emission
  causing the $\sim1$\,$\mu$m minimum  seen in some starburst galaxies
 shifting to W2 band and H$\alpha$ emission shifting to W1. This results in a blue W1-W2 (<0.8) colour 

 \item A highly obscured \ac{AGN} will have $W1-W2>0.8$ at all redshifts.
 \item For a composite system (\ac{AGN} $+$ \ac{SB}), dilution by host-galaxy light can limit the identification of low luminosity AGNs.
\end{enumerate}

We looked at the W1-W2 colour of our sources, utilizing CATWISE magnitudes. We considered only those sources without any flags and with SNR$\ge$5 in the W1 and W2 bands. This identifies 106 sources, of which only 28 sources satisfy the $W1-W2>0.8$ \ac{AGN} criterion. We further found that all but 2 of the sources satisfying the $W1-W2>0.8$ \ac{AGN} criterion have S\textsubscript{887}$\gtrsim$0.2-2.5\,mJy and  1.4\,GHz-to-3.4\,$\mu$m flux density ratio between 1 and 100. This supports our argument 
that (i) low power radio sources in our sample are powered by a radio-\ac{AGN} (ii) the radio-to-IR flux density ratio of low power radio-\acp{AGN} can extend to lower values (<100). On the other hand, the sources with $W1-W2<0.8$ could be either unobscured quasars, as they have a blue $W1-W2$ colour at high-$z$ as demonstrated in \cite{ross2020near} or the composite galaxies where the SB component results in a blue $W1-W2$ colour.
In addition to this, we verified the WISE magnitudes of a known starbust at $z=4.1$ \citep{ciesla2020hyper} and found that its $W1-W2 = 0.63$ matches our $z\gtrsim4$ sample with S\textsubscript{1.4}$<$1\,mJy.

%%%%%%%%%%%%%%%%%%%%%%%%%%%%%%%%%%%%%%%%%%%%%%%%%%%%%%%%%%%%%%%%%

\subsection{SED modelling}
\label{sec:sed_modelling}

 It is beyond the scope of this paper to perform full SED modelling of these galaxies, but here we make an illustrative comparison of these galaxies with two representative low-redshift galaxies. To do so, we shift two local radio sources to $z\gtrsim4$ for comparison: (i) Cygnus~A, a radio galaxy at $z = 0.05607$, and (ii) IRAS F00183--7111, an ultraluminous infrared galaxy (ULIRG) at $z=0.327$ showing composite emission from a radio loud-\ac{AGN} and a \ac{SB}. Each of these sources is distinct: their radio emission either comes from an \ac{AGN} or a combination of SF and \ac{AGN}. Their broadband observed photometry is obtained from NED.  We added more radio data from  \citet[Table~1]{norris2012radio} to the IRAS F00183--7111 template since the radio coverage is poor and this wavelength regime is critical for this analysis. 

We do not consider galaxies exclusively powered by star formation activity, like M\,82 and Arp~220, as they produce radio flux densities three or four orders of magnitude below the current detection limit when shifted to $z\gtrsim4$, indicating that none of our sources represent typical SFGs. 
 
We create the SED at $z\gtrsim4$ by shifting the observed SED in the frequency space by a factor of \textit{(1+$z$)\textsuperscript{-1}}, and in flux density space by a factor of (\textit{D\textsubscript{obs}}/\textit{D\textsubscript{z})\textsuperscript{2}}$\times {\it (1+z_{\rm {\it shift}}}/{\it 1+z_{\rm {\it obs}}})$.\textit{ D\textsubscript{obs}} and \textit{D\textsubscript{z}} are the luminosity distance to the observed redshift and shifted redshift ($z_{\rm {\it shift}}\sim4$, 5, 6 or 7) respectively, estimated using an online cosmology calculator \citep{wright2006cosmology}.

\begin{figure*}
 \centering
  \begin{subfigure}{.45\textwidth}
 \centering
 \includegraphics[width=\textwidth]{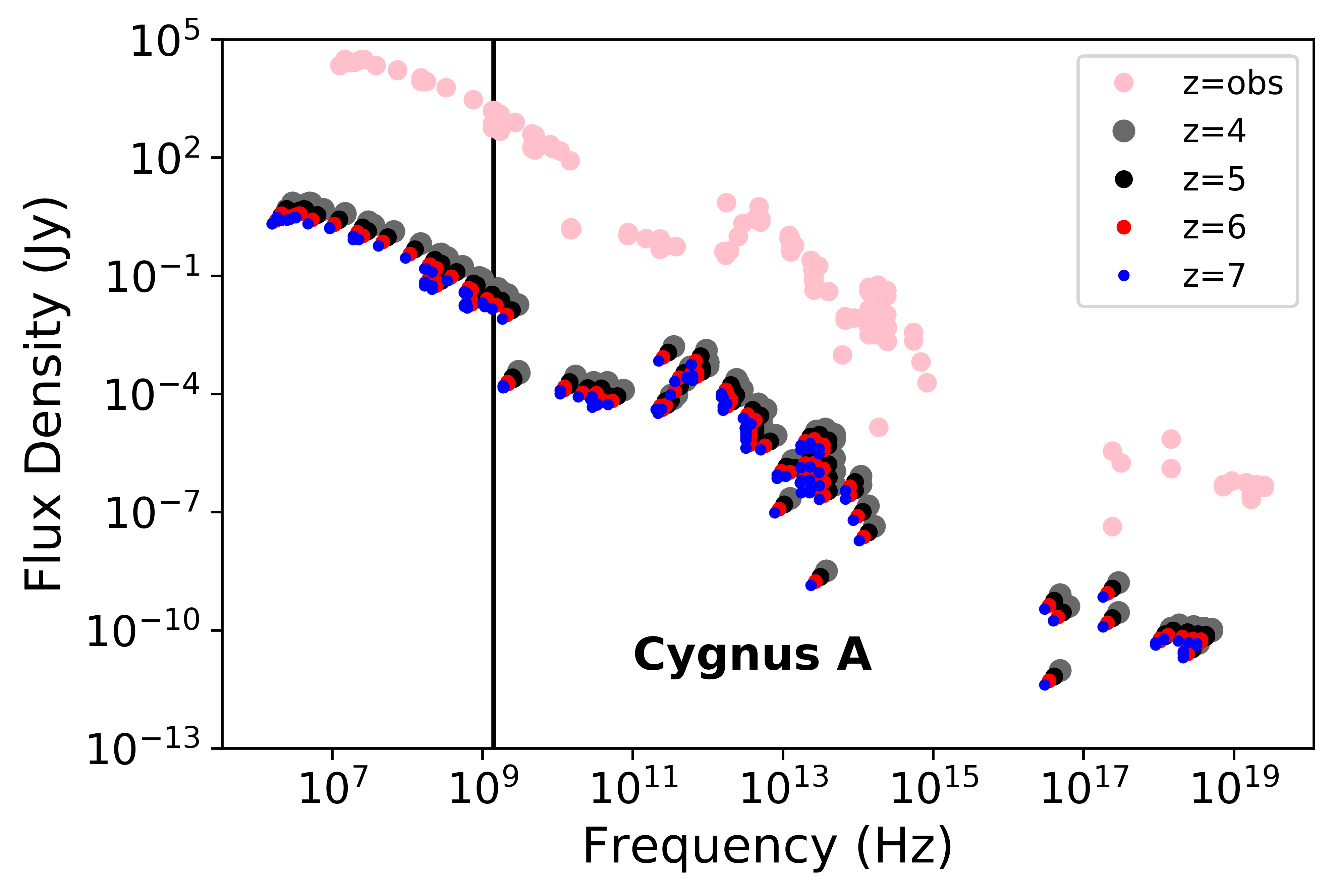} 
\end{subfigure}
\hspace{0.5cm}
 \begin{subfigure}{.45\textwidth}
 \centering
 \includegraphics[width=\textwidth]{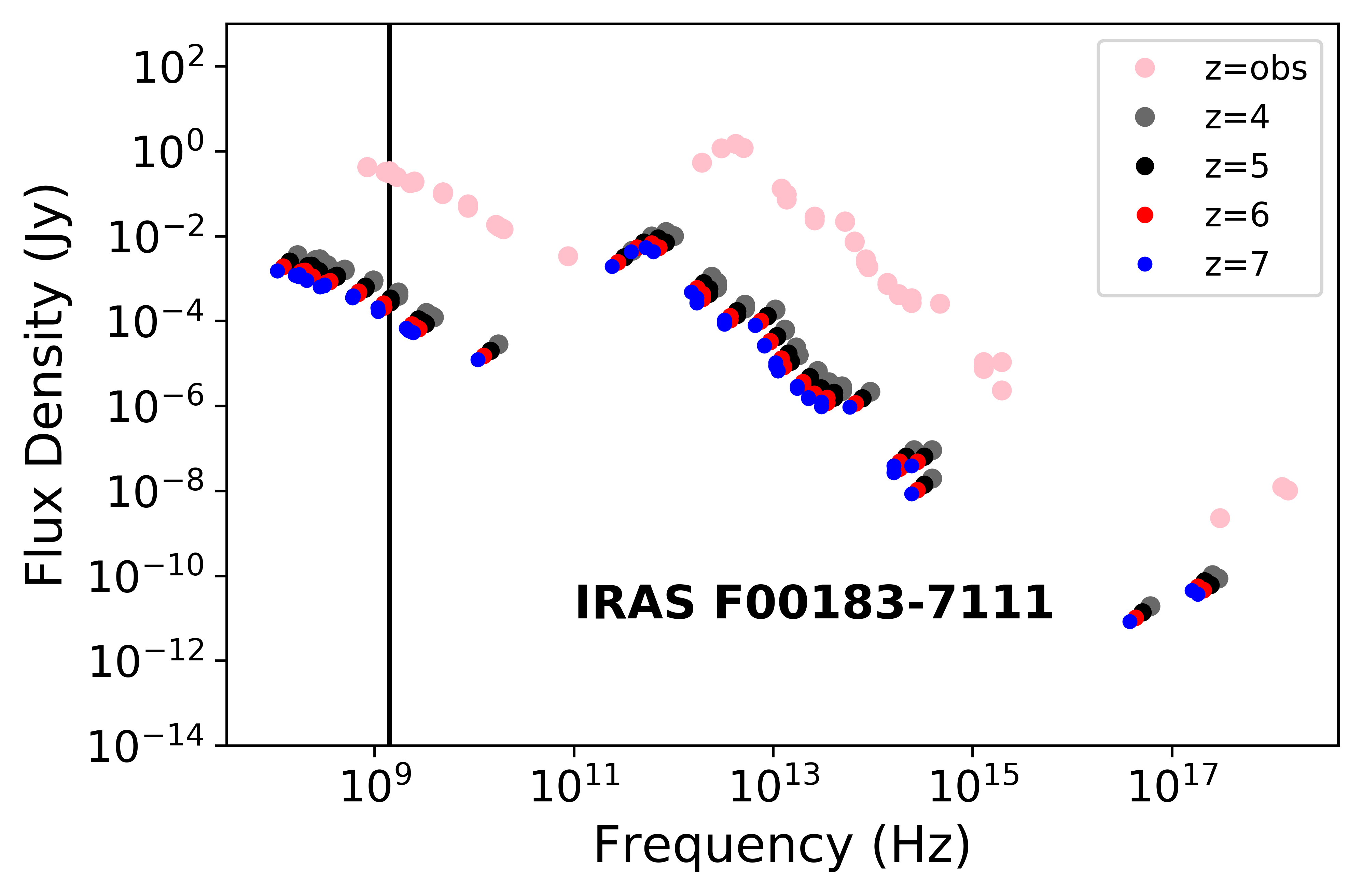} 
\end{subfigure}
 \caption{Template SEDs shifted to redshifts, $z=4, 5, 6, 7$ in grey, black, red and blue colours respectively. SED templates are built by simply connecting the datapoints and performed 1D interpolation using python-scipy (interp1d) package to obtain 1.4\,GHz flux density information from the shifted SEDs. 1.4\,GHz is marked in black solid line. Pink points represent the observed SED, obtained from NED.
 }
 \label{fig:shifted_SED}
\end{figure*}

Figure~\ref{fig:shifted_SED} shows the resulting shifted SEDs. The extracted 1.4\,GHz flux density from the shifted SEDs, shown in Table~\ref{tab:shifted_sed}, demonstrates that (i) active galaxies powered by a radio-\ac{AGN}, such as Cygnus~A, can represent the brighter sources in our sample and (ii) a composite system like IRAS F00183-7111 consisting of a radio-\ac{AGN} and a significant \ac{SB} component, can represent fainter sources in our sample. Thus, our SED modelling shows that a radio-\ac{AGN} and a composite system can reproduce the galaxies in our sample implying that properties of our \acp{HzRS}  are not that extraordinary compared to typical local galaxies. At the same time, we note that IRAS F00183-7111 does not reproduce our sample's observed SPIRE flux densities, as its SFR ($\sim 220$\,$M_{\odot}$/yr; \cite{mao2014star}) is not sufficient.

\begin{table}
 \centering
  \caption{1.4\,GHz flux density extracted from the template SEDs shifted to $z=4,5,6$ by performing 1d interpolation using interp1d function in python-scipy package.}
 \begin{tabular}{c|c|c|c|c}
 \hline
  \multirow{2}{*}{Source} & \multicolumn{3}{c}{S\textsubscript{1.4} (mJy)}\\
  \cline{2-5}
   & $z\sim4$ & $z\sim5$ & $z\sim6$ & $z\sim7$ \\
   \hline
  Cygnus A & 60 & 29 & 20 & 14 \\ 
  IRAS F00183 & 0.6 & 0.3 & 0.2 &  0.1 \\
   \hline
 \end{tabular}
 \label{tab:shifted_sed}
\end{table}
%%%%%%%%%%%%%%%%%%%%%%%%%%%%%%%%%%%%%%%%%%%%%%%%%%%%%%%%%%%%%%%%%%%%%%%%%%%%

\section{Conclusions}
\label{sec:concl}

We have used the Lyman dropout technique to identify a sample of 149 radio sources at $z\gtrsim4-7$, of which one is  a known radio-quasar (VIK J2318−3113; selected as a Z-dropout) at $z=6.44$, and 148 are newly-identified. Our study reveals a new population of high-redshift ($z\sim$ 4--7) radio-\ac{AGN}s, $\sim1-2$ orders of magnitudes fainter than currently known radio-\ac{AGN}s at similar redshifts, but with 1.4\,GHz luminosities typical of lower-redshift radio-loud \ac{AGN}. 

Comparison with spectroscopic redshifts in the COSMOS field indicate that our $z\sim4$  sample is about 88\% reliable, the  $z\sim5$  sample is about 81\% reliable, and the  $z\sim6$  sample is  at least about 62\% reliable. We do not have a reliability estimate for the  $z\sim7$  sample. 

We have explored radio and IR observations to understand  the origin of the radio emission of our  \ac{HzRS} sample.  Our conclusions are as follows.

\begin{enumerate}
 \item The radio (1.4\,GHz) estimated \ac{SFR} of our sample (Table~\ref{tab:sfr}) gives unphysical values, indicating that radio emission in our sources is not solely powered by star formation. 
 
 \item The faint ($\log L\textsubscript{1.4}=25$) and bright ($\log L\textsubscript{1.4}\ge26$)
 end of our sample spans the low and high power radio galaxy luminosity classes respectively, suggesting the presence of a radio-\ac{AGN} component in our sample. This study presents a new population of less powerful radio-\ac{AGN}s candidates at $z\sim4$, 5 and 6 that have been missed by previous surveys.

 \item  $\sim10\%$ of our \ac{HzRS} sample are detected in the {\it Herschel}-SPIRE bands which probe \ac{SB} heated dust emission and \ac{AGN} heated dust torus emission at $z>2$. This suggests that SPIRE detected sources are likely to represent composite systems. 
 
 \item We demonstrate that some high-$z$ radio-\ac{AGN}s tend to have hosts that are \ac{SB} galaxies, in contrast to low-$z$ radio-\ac{AGN}s, which are usually hosted by quiescent elliptical galaxies.
 
  \item  Using the $W1-W2>0.8$ AGN indicator, we identified 28 radio-\acp{AGN},  26 of which are found to be on the faint end of the observed 887.5\,MHz flux density distribution.  We further demonstrate that the 1.4\,GHz-to-3.4\,$\mu$m flux density ratio of these weak radio-\ac{AGN}s extends to lower values (1-100) than previously thought.
 
 \item SED modelling confirms that a composite system (radio-\ac{AGN} $+$ \ac{SB}) and a radio galaxy at $z\gtrsim4$ can produce the radio flux densities similar to ones observed at the faint and bright end respectively.

\end{enumerate}
\section{Future work}

 Spectroscopic follow-up of our sample is essential (i) to confirm the redshifts of these candidate sources identified by the Lyman dropout technique (ii) to verify the reliability of magnitude cut-off introduced in $Z$-band for $i$ dropouts (iii) to establish the criteria to identify interlopers, if present. The above three goals are critical in producing reliable criteria to select \acp{HzRS} in the full \ac{EMU} survey and thus verifying the model \citep{raccanelli2012cosmological} for the redshift distribution of radio sources.

%$$$##############################################

\section*{Acknowledgements}

The Australian SKA Pathfinder is part of the Australia Telescope National Facility which is managed by \ac{CSIRO}. Operation of \ac{ASKAP} is funded by the Australian Government with support from the National Collaborative Research Infrastructure Strategy. \ac{ASKAP} uses the resources of the Pawsey Supercomputing Centre. 
Establishment of \ac{ASKAP}, the Murchison Radio-astronomy Observatory and the Pawsey Supercomputing Centre are initiatives of the Australian Government, with support from the Government of Western Australia and the Science and Industry Endowment Fund. 
We acknowledge the Wajarri Yamatji people as the traditional owners of the Observatory site. 

Based on observations made with ESO Telescopes at the La Silla Paranal Observatory under programme IDs 177.A-3016, 177.A-3017, 177.A-3018 and 179.A-2004, and on data products produced by the KiDS consortium. The KiDS production team acknowledges support from: Deutsche Forschungsgemeinschaft, ERC, NOVA and NWO-M grants; Target; the University of Padova, and the University Federico II (Naples).

Isabella Prandoni acknowledges support from INAF under the PRIN MAIN stream "SAuROS" project, and from CSIRO under its Distinguished Research Visitor Programme. 

 We thank an anonymous referee for helpful comments and some excellent suggestions.
%%%%%%%%%%%%%%%%%%%%%%%%%%%%%%%%%%%%%%%%%%%%%%%%%%
\section*{Data Availability}

%The inclusion of a Data Availability Statement is a requirement for articles published in MNRAS. Data Availability Statements provide a standardised format for readers to understand the availability of data underlying the research results described in the article. The statement may refer to original data generated in the course of the study or to third-party data analysed in the article. The statement should describe and provide means of access, where possible, by linking to the data or providing the required accession numbers for the relevant databases or DOIs.

This study utilised the data available in the following public domains: 
\begin{enumerate}
    \item https://data.csiro.au/domain/casdaObservation
    \item https://kids.strw.leidenuniv.nl/DR4/access.php
    \item http://horus.roe.ac.uk/vsa/index.html
    \item https://irsa.ipac.caltech.edu/
\end{enumerate}

The derived final datasets are available in the article.



\bibliographystyle{mnras}
\bibliography{draft} % if your bibtex file is called example.bib

%\begin{thebibliography}{99}
%\end{thebibliography}

%%%%%%%%%%%%%%%%%%%%%%%%%%%%%%%%%%%%%%%%%%%%%%%%%%
%%%%%%%%%%%%%%%%% APPENDICES %%%%%%%%%%%%%%%%%%%%%
\appendix
%\begin{appendices}

\numberwithin{table}{section}
\section{Tables}

%\Ray{I think its better not to use the Latex appendix as the table numbering falls over. Also I see you have some illegal characters in the table - these are the -'s. I've fixed some but not all}

\label{A}

\begin{table*}
\caption{ Example for the list of published radio-\ac{AGN}s at z$\ge$4.  The full catalogue is available online.}
 \centering
 \begin{tabular}{c|c|c|c|c|c|c|c}
 \hline
  Index & Name & RA & DEC & redshift & S\textsubscript{1.4} & class & Reference\\
   & & (deg) & (deg) & z & (mJy) & \\
  \hline 
1 & NVSS J153050+104932	& 232.70867	& 10.82558	& 5.720	& 7.5 & radio galaxy & \cite{saxena2018discovery}\\ 
2& J085614+022359 & 134.0583 & 2.3997 & 5.55 & 86.5 & radio galaxy & \cite{drouart2020gleaming}\\
3 & TN J0924-2201	& 141.08300	& -22.02819	 &	5.190 &	71.5 	& radio galaxy & \cite{de2001spectroscopy}		\\	
4 & FIRST J163912.1+405236	& 249.80045	&	40.87686 &	4.880 &	22	& radio galaxy &	\cite{jarvis2009discovery}	\\
5 & HSC J083913.17+011308.1 & 8.65366 & 1.21892 & 4.72 & 7.17 & radio galaxy & \cite{yamashita2020wide} \\
$--$ & $--$ & $--$& $--$ & $--$ & $--$ & $--$ & $--$\\
\hline
\end{tabular}
\label{tab:known_radio_AGN}
\end{table*}

\begin{table*}
 \centering
\caption{ A list of  33 new radio sources at $z\gtrsim4$, obtained by cross-matching \acs{SDSS} and \acs{FIRST}/\acs{NVSS}. In each case the SDSS spectrum has been checked for supporting evidence of the redshift, such as the Lyman break or other spectral features.  The full catalogue is available online.} 
 \begin{tabular}{c|c|c|c|c|c|c|c}
 \hline
  Index & Name & RA & DEC & redshift & S\textsubscript{1.4} & class & Reference\\
   & & (deg) & (deg) & z & (mJy) & &\\
    \hline 
1 & SDSS J1148+5251 &	177.06935 &	52.86395 &	6.420 &	0.055 &	qso &\cite{walter2004resolved} \\		
2 & SDSS J100831.57+401910.3 &	152.13150 &	40.31947 &	5.670 & 2.52 &	galaxy &	\\
3 & 2MASX J00300536+2957082 & 7.52242 & 29.95225 & 5.199 & 17.7 & qso & \cite{waldram2007some} \\
4 & SDSS J085826.55+553234.9  & 134.6107 & 55.54291 &  5.076 & 20.6 & qso  & \\ 5 & J142634.86+543622.8 &      216.64524 & 54.60634  &  4.848 &4.36& qso  &     \cite{wang2016survey}       \\                        
$--$ & $--$ & $--$& $--$ & $--$ & $--$ & $--$ & $--$\\
\hline
\end{tabular}
\label{tab:new_radio_AGN}
\end{table*}

%\end{appendices}

% Don't change these lines
\bsp	% typesetting comment
\label{lastpage}
\end{document}